\documentclass[11pt]{article}
\usepackage{a4,bm,epsfig,booktabs}
\usepackage{setspace}
\newcommand{\bdlauto}{\theta}
\usepackage{ifthen,array}
\newcommand{\ams}{\usepackage{amsfonts,amssymb,amsmath}}

\ams
\allowdisplaybreaks[3]
\newlength{\textwidthorig}
\newlength{\oddsidemarginorig}
\newlength{\textheightorig}
\newlength{\topmarginorig}
\def\seitenlaengenabsolut#1 #2 #3 #4 {\setlength{\textwidth}{#1}
                                      \setlength{\oddsidemargin}{#2}
                                      \setlength{\textheight}{#3}
                                      \setlength{\topmargin}{#4}}
\def\seitenlaengenrelzustandard#1 #2 #3 #4 {\setlength{\textwidth}{\textwidthorig+#1}
                                            \setlength{\oddsidemargin}{\oddsidemarginorig+#2}
                                            \setlength{\textheight}{\textheightorig+#3}
                                            \setlength{\topmargin}{\topmarginorig+#4}}
\def\seitenlaengenrelzuvorher#1 #2 #3 #4 {\addtolength{\textwidth}{#1}
                                          \addtolength{\oddsidemargin}{#2}
                                          \addtolength{\textheight}{#3}
                                          \addtolength{\topmargin}{#4}}
\newcommand{\standardseite}{\seitenlaengenrelzuvorher2.2cm -0.8cm 1.8cm -1.5cm }   %
\standardseite
\newcommand{\leerezeile}{\vspace{2ex}}
\newlength{\laengespatium}

\newcommand{\nach}{\longrightarrow}      %Abbildungspfeil

\newcommand{\auf}{\longmapsto}           %Abbildungspfeil fuer Elemente
\newcommand{\txtauf}[1]{\auf}            %Abbildungspfeil mit Opt. "Text dr"uber"
\newcommand{\impliz}{\Longrightarrow}    %Implikationspfeil
\newcommand{\aequ}{\Longleftrightarrow}  %Aequivalenzpfeil (zweifacher Strich)
 
\newcommand{\invimpliz}{\Longleftarrow}  %Implikationspfeil (umgekehrte Richtung
\newcommand{\gegen}{\rightarrow}         %Konvergenzpfeil
\newcommand{\iso}{\cong}                 %Symbol fuer isomorph
\newcommand{\ident}{\equiv}              %Symbol fuer identisch (3 Striche)
\newcommand{\teilmenge}{\subseteq}       %Symbol fuer Teilmenge
\newcommand{\obermenge}{\supseteq}       %Symbol fuer Teilmenge andersherum
\newcommand{\echteteilmenge}{\subset}    %Symbol fuer echte Teilmenge
\newcommand{\aeqrel}{\sim}               %Symbol Aequivalenzrelation
\newcommand{\nichtin}{\not\in}

\newcommand{\senk}{\perp}                %Senkrechtzeichen

\newcommand{\leeremenge}{\varnothing}    %Symbol f"ur leere Menge
\newcommand{\tensor}{\otimes}            %Symbol f"ur Tensor
\newcommand{\kreuz}{\times}              %Symbol f"ur Kreuz

\newcommand{\einschr}[1]{{}\arrowvert_{#1}}      %Einschr"ankung auf ...
\newcommand{\dirsum}{\oplus}           %direkte Summe

\newcommand{\bigtensor}{\bigotimes}      %tensor als grosser Operator
\newcommand{\kp}{\odot}                  %Kreispunkt f"ur Ma"smultiplikation
\newcommand{\betraganpass}[1]%
           {\left| #1 \right|}           %Betragsstriche (variable Gr"o"se) 
\newcommand{\bigbetrag}[1]%
           {\bigl|{#1}\bigr|}            %Betragsstriche (gr"o"ser)
\newcommand{\betrag}[1]%
           {|{#1}|}                      %Betragsstriche  
\newcommand{\betragnichtanpass}[1]%
           {\mid #1 \mid}                %Betragsstriche 
\newcommand{\norm}[1]%
           {{}{\parallel}#1{\parallel}{}}      %Normstriche  
\newcommand{\erww}[1]%
           {\langle #1 \rangle}          %Erwartungwert-Klammern
\newcommand{\skalprod}[2]%
           {\langle #1,#2 \rangle}       %eckiges Skalarprodukt
\newcommand{\supnorm}[1]{{\norm{#1}_\infty}}        %Supremumsnorm allgemein
\newcommand{\quer}{\overline}            %Strich drueber
\newcommand{\dach}{\widehat}             %Dach dr"uber
\newcommand{\inv}[1]{\frac{1}{#1}}       %Liefert 1/#1 als Bruch
\newcommand{\einhalb}{\inv{2}}           %Liefert 1/2 als Bruch
\newcommand{\re}{\text{Re }}                           %Re als Realteil
\newcommand{\im}{\text{im\;}}                          %im als Image (mit platz)
\newcommand{\pr}{{\text{pr}}}                          %pr als Projektion
\newcommand{\tr}{\text{tr}}                           %tr als Spur
\newcommand{\id}{\:\text{id}}                          %id als ident. Abb.
\newcommand{\ido}{\text{id}}                           %id als ident. Abb. (ohne Platz)
\newcommand{\inter}{\text{int}\:}                      %int als Inneres 
\newcommand{\spann}{\text{span}}                       %span als Spann
\newcommand{\rank}{\text{rank }}                       %rank als rank
\newcommand{\dist}{\text{dist}}                        %dist als Distanz
\newcommand{\elanz}{\#}                                %# f"ur Elementanzahl
\newcommand{\del}{\partial}                            %krummes d fuer partielle Abl.
\newcommand{\Hom}{\text{Hom}}                          %Hom als Homomorphismengruppe
\newcommand{\End}{\text{End}}                          %End als Endomorphismengruppe
\newcommand{\Maps}{\text{Maps}}                        %Maps f"ur Set of Maps
\newcommand{\dd}{\text{d}}                             %Differential d
\newcommand{\e}{\text{e}}                              %exp
\newcommand{\I}{\text{i}}                              %i als imagin"are Einheit
\newcommand{\EINS}{{\boldsymbol{1}}}                   %fette 1
\newcommand{\field}[1]{\mathbb{#1}}                    %liefert #1 als mathbb-Zeichen
\newcommand{\C}{{\field{C}}}                           %C fuer komplexe Zahlen
\newcommand{\N}{{\field{N}}}                           %N fuer natuerliche Zahlen
\newcommand{\R}{{\field{R}}}                           %R fuer reelle Zahlen
\newcommand{\Z}{{\field{Z}}}                           %Z fuer ganze Zahlen
\newcommand{\rnkl}[2]{\raisebox{-0.4ex}{$#1$}%
\raisebox{-0.12ex}{{\large$\setminus$}}\,#2}   %0.5/0.2 fr"uher  %Rechtsnebenklassenfaktorraum
\newcommand{\agb}{{\overline{{\cal A}/{\cal G}}}}      %A/G + Strich
\newcommand{\agbfact}[1][]{\text{$\agb/\!\aeqrel$}}    %A/G\~ + Strich
\newcommand{\ag}{{\cal A}/{\cal G}}                    %A/G ohne Strich
\newcommand{\Ab}{{\overline{{\cal A}}}}                %A + Strich
\newcommand{\Gb}{{\overline{{\cal G}}}}                %G + Strich

\newcommand{\AbGb}{{\Ab/\Gb}}                          %A+Strich / G+Strich
\newcommand{\gen}{{\text{gen}}}
\newcommand{\qa}{{\quer{A}}}                           %A als verallg. Zusammenhang 
\newcommand{\qg}{{\quer{g}}}                           %Verallgemeinerte Eichtrf.
\newcommand{\holgr}{{\mathbf H}}                       %Holonomiegruppe
\newcommand{\bz}{{\mathbf B}}                          %Basiszentralisator
\newcommand{\gross}[1]{{\boldsymbol #1}}               %#1 als grosses
\newcommand{\gc}{\gross{\gamma}}                       %gro"ses gamma 
\newcommand{\gd}{\gross{\delta}}                       %gro"ses delta
\newcommand{\Pf}{{\cal P}}                             %Menge aller Pfade
\newcommand{\KG}[1]{\Pf_{#1}}                          %Wege_#1
\newcommand{\BB}{\uparrow\uparrow}                     %Relation gleicher Anfangsweg
\newcommand{\EE}{\downarrow\downarrow}                 %Relation gleicher Endweg
\newcommand{\hyph}{\upsilon}                           %a hyph
\newcommand{\Hyph}{Y}                                  %Y fuer elanz(hyph)
\newcommand{\Haar}{{\text{Haar}}}                      %Index Haar
\newcommand{\LG}{{\mathbf{G}}}                         %Liegruppe G (fett)
\newcommand{\LN}{{\mathbf{N}}}                         %Liegruppe N (fett)
\newcommand{\Lieg}{{\mathfrak{g}}}                            %gotisches g (fuer Liealgebra)
\newcommand{\aeqrelzush}[1][]{\sim}                    %Symbol Aequivalenzrelation
\newcommand{\alg}{\mathfrak{A}}                          %Symbol fuer eine Algebra A
\newcommand{\strat}{{\cal S}}                          %Symbol Stratifizierung
\newcommand{\nklza}[1][]{\ifthenelse{\equal{#1}{}}     %Z(H_\qa) \ G
                                    {\rnkl{Z(\holgr_\qa)}{\LG}}        
                                   {\rnkl{Z(\holgr_{#1})}{\LG}}}       
\newcommand{\nkla}[1][]{\ifthenelse{\equal{#1}{}}      %B(\qa) \ \Gb
                                    {\rnkl{\bz(\qa)}{\Gb}}        
                                    {\rnkl{\bz(#1)}{\Gb}}}       
\newcommand{\he}{{\text{he}}}                          %he f"ur halbeinfach
\newcommand{\ab}{{\text{ab}}}                          %ab f"ur abelsch

\newcommand{\darst}{{\phi}}                            %Darstellung
\newcommand{\charakt}{\chi}                            % Charakter
\newcommand{\matfkt}{{\cal M}}                         %Menge aller Matrixfunktionen
\newcommand{\YM}{{\text{YM}}}                          %YM f"ur Yang-Mills

\newcommand{\ymwirk}[1][]{\ifthenelse{\equal{#1}{}}{S_{\YM}}{S_{\YM,#1}}}
\newcommand{\karte}{\chi}
\renewcommand{\karte}{{\kappa}}             %Karte

\newcommand{\bmat}{\begin{pmatrix}}
\newcommand{\emat}{\end{pmatrix}}

\newcommand{\Homeo}{{\text{Homeo}}}                      % Menge der Homeos
\newcommand{\codim}{\text{codim}}
\newcommand{\ListNullAbstaende}{\setlength{\topsep}{0pt}%
                                \setlength{\parskip}{0pt}%
                                \setlength{\partopsep}{0pt}%
                                \setlength{\itemsep}{0pt}%
                                \setlength{\parsep}{0pt}}
\newcommand{\ListNurAnstrichAbstand}{\setlength{\topsep}{0pt}%
                                     \setlength{\parskip}{0pt}%
                                     \setlength{\partopsep}{0pt}%
                                     \setlength{\parsep}{0pt}}
\newenvironment{StandardListe}[2]%
               {\begin{list}%
                      {#1}%
                      {\settowidth{\leftmargin}{M#1}%
                       \settowidth{\labelwidth}{#1}%
                       \settowidth{\labelsep}{M}%
                       #2%
                      }%
                }%
               {\end{list}}%
\newenvironment{EinfachListe}[1]%
               {\begin{StandardListe}{#1}{\ListNullAbstaende}}%
               {\end{StandardListe}}%
               {\begin{StandardListe}{#1}{\ListNurAnstrichAbstand}}%
               {\end{StandardListe}}%
\newcommand{\labelsatz}[1]{#1}
\newcounter{listennr}                      %
\newlength{\hilfslaenge}
\newlength{\stdlabellaenge}
\newlength{\maximum}
\newcommand{\stdlabel}{}
\newcommand{\Maximum}{}
\newcommand{\iitem}[1][]{\ifthenelse{\equal{#1}{}}%
                           {\item \setlength{\hilfslaenge}{\stdlabellaenge}}%
                           {\item[\labelsatz{#1}\hfill]%
                            \settowidth{\hilfslaenge}{\labelsatz{#1}}}%
                         \ifthenelse{\lengthtest{\maximum < \hilfslaenge}}%
                           {\setlength{\maximum}{\hilfslaenge}%
                            \ifthenelse{\equal{#1}{}}%
                               {\renewcommand{\Maximum}{\stdlabel}}%
                               {\renewcommand{\Maximum}{#1}}}%
                           {}%
                      }      
\makeatletter
\newenvironment{AutoLabelLaengenListe}[2][]%
               {\begin{list}%
                      {\labelsatz{#1}\hfill}%
                      {\stepcounter{listennr}%
                       \settowidth{\leftmargin}{M\labelsatz{\ref{listnr\arabic{listennr}}}}%
                       \settowidth{\labelwidth}{\labelsatz{\ref{listnr\arabic{listennr}}}}%
                       \settowidth{\labelsep}{M}%
                       \settowidth{\stdlabellaenge}{\labelsatz{#1}}%
                       \renewcommand{\stdlabel}{#1}%
                       #2%
                       \renewcommand{\Maximum}{}%
                      }%
                }%
               {\renewcommand{\@currentlabel}{\Maximum}%
                \label{listnr\arabic{listennr}}%
                \end{list}%
                }%
\makeatother
\newenvironment{StandardEinrueckung}[2]%
               {\begin{list}%
                      {#1}%
                      {\settowidth{\leftmargin}{M#1}%
                       \settowidth{\labelwidth}{#1}%
                       \settowidth{\labelsep}{M}%
                       #2%
                      }%
                \item}%
               {\end{list}}%
\newenvironment{Einrueckungpur}[1]%
               {\begin{StandardEinrueckung}{#1}{\ListNullAbstaende}}%
               {\end{StandardEinrueckung}}%
\newenvironment{Einrueckung}[1]%
               {\begin{StandardEinrueckung}{#1}{\setlength{\parsep}{0pt}}}%
               {\end{StandardEinrueckung}}%
\newcommand{\EineZeileGleichung}[2][0.0ex]
           {
            
            \vspace{#1} 
            \noindent
            \hspace*{\fill}
            $\displaystyle{#2}$
            \hspace*{\fill}

            \vspace{#1} 
            
           }

\makeatletter
\newcommand{\EineNumZeileGleichung}[2][0.5ex]
           {
            
            \vspace{#1} 
            \noindent
            \stepcounter{equation}
            \renewcommand{\@currentlabel}{\arabic{equation}}%
            \phantom{(\arabic{equation})}\hspace*{\fill}
            $\displaystyle{#2}$
            \hspace*{\fill}
            (\arabic{equation})

            \vspace{#1} 
            
           }
\makeatother
\makeatletter
\newcommand{\EineErwNumZeileGleichung}[2][0.5ex]
           {
            
            \vspace{#1} 
            \noindent
            \stepcounter{equation}
            \renewcommand{\@currentlabel}{\arabic{equation}}%
            \phantom{(\arabic{equation})}\hspace*{\fill}
            #2 %
            \hspace*{\fill}
            (\arabic{equation})

            \vspace{#1} 
            
           }
\makeatother
\newcommand{\breitrel}[1]{\hspace*{\tabcolsep} #1 \hspace*{\tabcolsep}}
\newlength{\abstaug}              %
\newenvironment{AllgUnnumGleichung}[2][1.0ex]%           %
               {
  
                \setlength{\abstaug}{#1}
                \vspace{\abstaug}
                \hspace*{\fill}
                $\begin{array}[t]{#2}
                }%
               {\end{array}$
                \hspace*{\fill}
  
                \vspace{\abstaug}

                }%
\newenvironment{AllgNumGleichung}[2][0.0ex]%           %
               {
  
                \setlength{\abstaug}{#1}
                \vspace{\abstaug}
                $\begin{tabular*}{\textwidth}[t]{#2}
                }%
               {\end{tabular*}$

                \vspace{\abstaug}

               }%
\newenvironment{StandardUnnumGleichungKlein}[1][0ex]%       %
               {\renewcommand{\s}{\\[#1] }%
                \begin{AllgUnnumGleichung}{rcl}}%
               {\end{AllgUnnumGleichung}}%
\newcommand{\s}{\\[0ex] }             %
\newenvironment{StandardUnnumGleichung}[1][0ex]%       
               {\renewcommand{\s}{\\[#1] }%
                \begin{AllgUnnumGleichung}{>{\displaystyle}rc>{\displaystyle}l}}%
               {\end{AllgUnnumGleichung}}%
\newenvironment{XrelYZNumGleichung}[1][0ex]%       %
               {\renewcommand{\s}{\\[#1] }%
                \begin{AllgNumGleichung}{rcll}}%
               {\end{AllgNumGleichung}}%
\newcommand{\erl}[1]{\hfill\mbox{\hspace*{1.5em}\small (#1)}}

\newcommand{\erllang}[2][0.5\textwidth]%
              {\hfill\hspace*{1.5em}%
               \begin{minipage}[t]{#1}{\small%
                          \begin{list}{(}{\ListNullAbstaende%
                                          \settowidth{\leftmargin}{(}%
                                          \settowidth{\labelwidth}{(}%
                                          \settowidth{\labelsep}{}%
                                         }%
                          \item#2)%
                          \end{list}}%
               \end{minipage}\\[-0.9ex]
              }%         
\newcommand{\DefBemUmgeb}[1]% 
           {\newenvironment{#1}[1][]%
                           {\begin{Einrueckung}{{\bf #1}}%
                            \ifx##1\empty\else{{\bf ##1}
                            
                                                        }\fi%
                            }%
                           {\end{Einrueckung}}}
\newcommand{\DefSBemUmgeb}[2]% %
           {\newenvironment{#1}[1][]%
                           {\begin{Einrueckung}{{\bf #2}}%
                            \ifx##1\empty\else{{\bf ##1}
                            
                                                        }\fi%
                            }%
                           {\end{Einrueckung}}}
\makeatletter
\newcommand{\DefBspUmgeb}[3]% %
           {\newcounter{#2}[#3]%
            \newenvironment{#1}[1][]%
                           {\stepcounter{#2}%
                            \renewcommand{\ZaehlerMarke}{\arabic{#2}}%  
                            \renewcommand{\Einzugsname}{{\bf #1 \ZaehlerMarke}}%
                            \begin{Einrueckung}{\Einzugsname}
                            \ifx##1\empty\else{{\bf ##1}\\}\fi%
                            \renewcommand{\@currentlabel}{\ZaehlerMarke}%
                            }%
                           {\end{Einrueckung}}}
\makeatother
\newcommand{\ZaehlerbisEbene}{section}
\newcommand{\Ebenea}{section}
\newcommand{\Ebeneb}{subsection}

\newcommand{\Abschnittnummer}{%
            \ifx\ZaehlerbisEbene\Ebenea{\arabic{section}}%
             \else{%
              \ifx\ZaehlerbisEbene\Ebeneb{\arabic{section}.\arabic{subsection}}%
               \else{\arabic{section}.\arabic{subsection}.\arabic{subsubsection}}%
              \fi}%     
            \fi}     
\newcommand{\Einzugsname}{}
\newcommand{\ZaehlerMarke}{}
\makeatletter
\newcommand{\DefThmUmgeb}[3]% 
           {\newcounter{#1}[#3]%
            \newenvironment{#1}[1][]%
                           {\stepcounter{#2}%
                            \setcounter{#1}{\value{#2}}%
                            \renewcommand{\ZaehlerMarke}{\Abschnittnummerpunkt\arabic{#1}}%  
                            \renewcommand{\Einzugsname}{{\bf #1 \ZaehlerMarke}}%
                            \begin{Einrueckung}{\Einzugsname}
                            \ifx##1\empty\else{{\bf ##1}
                            
                                                        }\fi%
                            \renewcommand{\@currentlabel}{\ZaehlerMarke}%
                            }%
                           {\end{Einrueckung}}}
\makeatother
\makeatletter
\newcommand{\DefSThmUmgeb}[4]% 
           {\newcounter{#1}[#3]%
            \newenvironment{#1}[1][]%
                           {\stepcounter{#2}%
                            \setcounter{#1}{\value{#2}}%
                            \renewcommand{\ZaehlerMarke}{\Abschnittnummerpunkt\arabic{#1}}%
                            \renewcommand{\Einzugsname}{{\bf #4 \ZaehlerMarke}}
                            \begin{Einrueckung}{\Einzugsname}
                            \ifx##1\empty\else{{\bf ##1}

                                                        }\fi%
                            \renewcommand{\@currentlabel}{\ZaehlerMarke}%
                            }%
                           {\end{Einrueckung}}}
\makeatother
\makeatletter
\newcommand{\DefUnterNumThmUmgeb}[5]% 
           {\newcounter{#1}[#3]%
            \newcounter{#4}%
            \newenvironment{#1}[1][]%
                           {\ifx##1\empty\else{\stepcounter{#2}\setcounter{#4}{0}}\fi%
                            \stepcounter{#4}%
                            \setcounter{#1}{\value{#2}}%
                            \renewcommand{\ZaehlerMarke}{\Abschnittnummerpunkt\arabic{#1}\alph{#4}}%
                            \renewcommand{\Einzugsname}{{\bf #5 \ZaehlerMarke}}
                            \begin{Einrueckung}{\Einzugsname}
                            \renewcommand{\@currentlabel}{\ZaehlerMarke}%
                            }%
                           {\end{Einrueckung}}}
\makeatother
\newenvironment{Beweis}[1][]%
               {\begin{Einrueckung}{{\bf Beweis}}%
                \ifx#1\empty\else{{\bf #1}

                                            }\fi%
                }%
               {\end{Einrueckung}%
                }%
\newenvironment{Proof}[1][]%
               {\begin{Einrueckung}{{\bf Proof}}%
                \ifx#1\empty\else{{\bf #1}

                                            }\fi%
                }%
               {\end{Einrueckung}%
                }%
               {\begin{Einrueckung}{{\bf \glqq Beweis\grqq}}%
                \ifx#1\empty\else{{\bf #1}
                
                                            }\fi%
                }%
               {\end{Einrueckung}%
                }%
               {\begin{Einrueckung}{{\bf Begr"undung}}%
                \ifx#1\empty\else{{\bf #1}
                
                                            }\fi%
                }%
               {\end{Einrueckung}%
                }%
\newenvironment{Hinrichtung}%
               {\begin{Einrueckungpur}{$\impliz$}}%
               {\end{Einrueckungpur}}%
\newenvironment{Rueckrichtung}%
               {\begin{Einrueckungpur}{$\invimpliz$}}%
               {\end{Einrueckungpur}}%
               {\begin{Einrueckungpur}{\glqq$\teilmenge$\grqq}}%
               {\end{Einrueckungpur}}%
               {\begin{Einrueckungpur}{\glqq$\obermenge$\grqq}}%
               {\end{Einrueckungpur}}%
               {\begin{Einrueckungpur}{"$\teilmenge$"}}%
               {\end{Einrueckungpur}}%
               {\begin{Einrueckungpur}{"$\obermenge$"}}%
               {\end{Einrueckungpur}}%
\newcommand{\qed}{\nopagebreak\hspace*{2em}\hspace*{\fill}{\bf qed}}
\newcommand{\ARabic}{\arabic}
\newcommand{\Nummerntypa}{\arabic}   
\newcommand{\Nummerntypb}{\alph}
\newcommand{\Nummerntypc}{\roman}
\newcommand{\Nummerntypd}{\Alph}

\newcommand{\Nra}{\Nummerntypa{Nummera}}            %
\newcommand{\Nrb}{\Nummerntypb{Nummerb}}            %
\newcommand{\Nrc}{\Nummerntypc{Nummerc}}                
\newcommand{\Nrd}{\Nummerntypd{Nummerd}}                
\newcommand{\ZeichenzuNrTyp}[1]%
           {\ifx#1\ARabic {.}\else{)}%
                  \fi}                              %
\newcommand{\NrZeicha}{\ZeichenzuNrTyp{\Nummerntypa}}
\newcommand{\NrZeichb}{\ZeichenzuNrTyp{\Nummerntypb}}
\newcommand{\NrZeichc}{\ZeichenzuNrTyp{\Nummerntypc}}
\newcommand{\NrZeichd}{\ZeichenzuNrTyp{\Nummerntypd}}
\newcommand{\ListMarkea}%
           {\Nra\NrZeicha}
\newcommand{\ListMarkeb}%
           {\Nra\NrZeicha\Nrb\NrZeichb}
\newcommand{\ListMarkec}%
           {\Nra\NrZeicha\Nrb\NrZeichb\Nrc\NrZeichc}
\newcommand{\ListMarked}%
           {\Nra\NrZeicha\Nrb\NrZeichb\Nrc\NrZeichc\Nrd\NrZeichd}
\newcommand{\Anfangszeichen}{}
\newcommand{\Anfangspunkt}{}
\newcounter{Schachtelebene}
\newcounter{Hilfszaehler}
\newcommand{\Hilfsbefehl}{}
\newcommand{\Schachtelebene}{\alph{Schachtelebene}}
\makeatletter
\newenvironment{AllgNumerierteListe}[2][]%      %
               {\addtocounter{Schachtelebene}{1}%
		\setcounter{Hilfszaehler}{#2}%
                \renewcommand{\Anfangszeichen}%
                             {\renewcommand{\Hilfsbefehl}{\csname Nummerntyp\Schachtelebene \endcsname}%
                              \Hilfsbefehl{Hilfszaehler}}%
                \renewcommand{\Anfangspunkt}%
                             {\csname NrZeich\Schachtelebene \endcsname}%
                \begin{list}%
                      {\stepcounter{Nummer\Schachtelebene}%
                       \csname Nr\Schachtelebene \endcsname
                       \csname NrZeich\Schachtelebene \endcsname
                       }%
                      {\settowidth{\leftmargin}{M\Anfangszeichen\Anfangspunkt}%
                       \settowidth{\labelwidth}{\Anfangszeichen\Anfangspunkt}%
                       \settowidth{\labelsep}{M}%
                       \setlength{\topsep}{0pt}%
                       \setlength{\parskip}{0pt}%
                       \setlength{\partopsep}{0pt}%
                       \setlength{\itemsep}{0pt}%
                       \setlength{\parsep}{0pt}%
                      }%
                \renewcommand{\@currentlabel}{\csname ListMarke\Schachtelebene \endcsname}%
                }%      
               {\ifthenelse{\equal{}{}}{\setcounter{Nummer\Schachtelebene}{0}}{}
                \addtocounter{Schachtelebene}{-1}%
                \end{list}}
\makeatother
\newenvironment{NumerierteListe}[1]%      %
               {\begin{AllgNumerierteListe}{#1}}
               {\end{AllgNumerierteListe}}
\newenvironment{WeiterNumerierteListe}[1]%      %
               {\begin{AllgNumerierteListe}[Weiter]{#1}}
               {\end{AllgNumerierteListe}}

\newcommand{\UnnumAnfangszeichen}{}
\newcounter{UnnumSchachtelebene}
\newcommand{\UnnumSchachtelebene}{\alph{UnnumSchachtelebene}}
\makeatletter
\newenvironment{UnnumerierteListe}%          
               {\addtocounter{UnnumSchachtelebene}{1}%
                \renewcommand{\UnnumAnfangszeichen}%
                             {\csname UnnumZeich\UnnumSchachtelebene \endcsname}%
                \begin{list}%
                      {\UnnumAnfangszeichen}%
                      {\settowidth{\leftmargin}{M\UnnumAnfangszeichen}%
                       \settowidth{\labelwidth}{\UnnumAnfangszeichen}%
                       \settowidth{\labelsep}{M}%
                       \setlength{\topsep}{0pt}%
                       \setlength{\parskip}{0pt}%
                       \setlength{\partopsep}{0pt}%
                       \setlength{\itemsep}{0pt}%
                       \setlength{\parsep}{0pt}%
                      }%
                }%
               {\addtocounter{UnnumSchachtelebene}{-1}%
                \end{list}}
\makeatother
\newlength{\fktdefhilfslaenge}
\newcommand{\ohnefktdef}[4]%                 %
           {\hspace*{\fill}
            $\begin{array}[t]{ccc}%
            #1 & \nach & #2 \\
            #3 & \auf  & #4
            \end{array}$
            \hspace*{\fill}}
\newcommand{\fktdef}[5]%                 %
           {\hspace*{\fill}
            $\begin{array}[t]{cccc}%
            #1: & #2 & \nach & #3 \\    
                & #4 & \auf  & #5
            \end{array}$
            \settowidth{\fktdefhilfslaenge}{$#1$:}
            \hspace*{0.6 \fktdefhilfslaenge}  
            \hspace*{\fill}}
\newcommand{\fktdefpur}[5]%                 %
           {$\begin{array}[t]{cccc}%
            #1: & #2 & \nach & #3 \\    
                & #4 & \auf  & #5
            \end{array}$}
\newcommand{\fktdefabgesetztpur}[5]%          %
           {
            
            $\begin{array}[t]{cccc}%
            #1: & #2 & \nach & #3 \\    
                & #4 & \auf  & #5
            \end{array}$
            \settowidth{\fktdefhilfslaenge}{$#1$:}
            \hspace*{0.6 \fktdefhilfslaenge}
            
           }
\newcommand{\fktdefabgesetzt}[5]%                %
           {
           
            \hspace*{\fill}
            $\begin{array}[t]{cccc}%
            #1: & #2 & \nach & #3 \\    
                & #4 & \auf  & #5
            \end{array}$
            \settowidth{\fktdefhilfslaenge}{$#1$:}
            \hspace*{0.6 \fktdefhilfslaenge}  
            \hspace*{\fill}
            
            }
\newcommand{\ohnefktdefabgesetzt}[4]%                %
           {      

            \hspace*{\fill}
            $\begin{array}[t]{ccc}%
            #1 & \nach & #2 \\
            #3 & \auf  & #4
            \end{array}$
            \hspace*{\fill}

            }
\newcommand{\doppelohnefktdefabgesetzt}[6]%                %
           {

            \hspace*{\fill}
            $\begin{array}[t]{ccccc}%
            #1 & \nach & #2 & \nach & #3\\
            #4 & \auf  & #5 & \auf  & #6
            \end{array}$
            \hspace*{\fill}

            }
\newcommand{\anhang}%
           {\appendix
            \sectioninh{Anhang}
            \renewcommand{\Abschnittnummer}{%
                  \ifx\ZaehlerbisEbene\Ebenea{\Alph{section}}%
                  \else{%
                        \ifx\ZaehlerbisEbene\Ebeneb{\Alph{section}.\arabic{subsection}}%
                        \else{\Alph{section}.\arabic{subsection}.\arabic{subsubsection}}%
                        \fi}%     
                  \fi}%
                 
            }            
\newcommand{\anhangengl}%
           {\appendix
            \sectioninh{Appendix}
            \renewcommand{\Abschnittnummer}{%
                  \ifx\ZaehlerbisEbene\Ebenea{\Alph{section}}%
                  \else{%
                        \ifx\ZaehlerbisEbene\Ebeneb{\Alph{section}.\arabic{subsection}}%
                        \else{\Alph{section}.\arabic{subsection}.\arabic{subsubsection}}%
                        \fi}%     
                  \fi}%
                 
            }

\newcounter{wdhlstufe}
\newcommand{\sectioninh}[1]%
           {\section*{#1}%
            \addcontentsline{toc}{section}{#1}}
\newcommand{\bezeichnung}[3]%
           {\begin{Einrueckungpur}{\hbox to 6em{#1}\hbox to 2.4em{\hfill#2}}
            #3
            \end{Einrueckungpur}}

\newcommand{\doppelteinfach}{e}

\newcommand{\ifdoppelt}[1]{\ifthenelse{\equal{\doppelteinfach}{d}}{#1}{}}
\newcommand{\ifeinfach}[1]{\ifthenelse{\equal{\doppelteinfach}{e}}{#1}{}}

\newlength{\querfhilfsl}              %

\newlength{\hll}

\DefThmUmgeb{Theorem}{Theorem}{\ZaehlerbisEbene}
\DefThmUmgeb{Definition}{Definition}{\ZaehlerbisEbene}
\DefThmUmgeb{Satz}{Theorem}{\ZaehlerbisEbene}
\DefThmUmgeb{Proposition}{Theorem}{\ZaehlerbisEbene}
\DefThmUmgeb{Lemma}{Theorem}{\ZaehlerbisEbene}
\DefThmUmgeb{Folgerung}{Theorem}{\ZaehlerbisEbene}
\DefThmUmgeb{Corollary}{Theorem}{\ZaehlerbisEbene}
\DefThmUmgeb{Vorschrift}{Definition}{\ZaehlerbisEbene}
\DefThmUmgeb{Construction}{Definition}{\ZaehlerbisEbene}
\DefSThmUmgeb{FormSatz}{Theorem}{\ZaehlerbisEbene}{\glqq Satz\grqq} 
\DefThmUmgeb{Vermutung}{Theorem}{\ZaehlerbisEbene}
\DefThmUmgeb{Conjecture}{Theorem}{\ZaehlerbisEbene}
\DefThmUmgeb{Konvention}{Definition}{\ZaehlerbisEbene}
\DefThmUmgeb{Feststellung}{Theorem}{\ZaehlerbisEbene}
\DefUnterNumThmUmgeb{DefinitionZusatzNum}{Definition}{\ZaehlerbisEbene}{DefZN}{Definition}
\DefBspUmgeb{Beispiel}{Beispiel}{subsubsection}
\DefBspUmgeb{Example}{Example}{subsubsection}
\DefBspUmgeb{Frage}{Frage}{section}
\DefBspUmgeb{Question}{Question}{section}
\DefBspUmgeb{Aufgabe}{Aufgabe}{section}
\DefBspUmgeb{Ziel}{Ziel}{section}
\DefBemUmgeb{Bemerkung}
\DefBemUmgeb{Remark}
\DefSBemUmgeb{OffeneFrage}{Offene Frage}
\newcommand{\bdf}{\begin{Definition}}
\newcommand{\edf}{\end{Definition}}
\newcommand{\bvorsch}{\begin{Vorschrift}}
\newcommand{\evorsch}{\end{Vorschrift}}
\newcommand{\bconst}{\begin{Construction}}
\newcommand{\econst}{\end{Construction}}
\newcommand{\bthm}{\begin{Theorem}}
\newcommand{\ethm}{\end{Theorem}}
\newcommand{\bsatz}{\begin{Satz}}
\newcommand{\esatz}{\end{Satz}}
\newcommand{\bprop}{\begin{Proposition}}
\newcommand{\eprop}{\end{Proposition}}
\newcommand{\blem}{\begin{Lemma}}
\newcommand{\elem}{\end{Lemma}}
\newcommand{\bfolg}{\begin{Folgerung}}
\newcommand{\efolg}{\end{Folgerung}}
\newcommand{\bcorr}{\begin{Corollary}}
\newcommand{\ecorr}{\end{Corollary}}
\newcommand{\bfest}{\begin{Feststellung}}
\newcommand{\efest}{\end{Feststellung}}
\newcommand{\bbew}{\begin{Beweis}}
\newcommand{\ebew}{\end{Beweis}}
\newcommand{\bpf}{\begin{Proof}}
\newcommand{\epf}{\end{Proof}}
\newcommand{\bwnum}{\begin{WeiterNumerierteListe}}
\newcommand{\ewnum}{\end{WeiterNumerierteListe}}
\newcommand{\bdfzn}{\begin{DefinitionZusatzNum}}
\newcommand{\edfzn}{\end{DefinitionZusatzNum}}
\newcommand{\bbem}{\begin{Bemerkung}}
\newcommand{\ebem}{\end{Bemerkung}}
\newcommand{\brem}{\begin{Remark}}
\newcommand{\erem}{\end{Remark}}
\newcommand{\bnum}{\begin{NumerierteListe}}
\newcommand{\enum}{\end{NumerierteListe}}
\newcommand{\bunum}{\begin{UnnumerierteListe}}
\newcommand{\eunum}{\end{UnnumerierteListe}}
\newcommand{\bbsp}{\begin{Beispiel}}
\newcommand{\ebsp}{\end{Beispiel}}
\newcommand{\bex}{\begin{Example}}
\newcommand{\eex}{\end{Example}}
\newcommand{\bfrag}{\begin{Frage}}
\newcommand{\efrag}{\end{Frage}}
\newcommand{\bquest}{\begin{Question}}
\newcommand{\equest}{\end{Question}}
\newcommand{\baufg}{\begin{Aufgabe}}
\newcommand{\eaufg}{\end{Aufgabe}}
\newcommand{\bof}{\begin{OffeneFrage}}
\newcommand{\eof}{\end{OffeneFrage}}
\newcommand{\bverm}{\begin{Vermutung}}
\newcommand{\everm}{\end{Vermutung}}
\newcommand{\bconj}{\begin{Conjecture}}
\newcommand{\econj}{\end{Conjecture}}
\newcommand{\bkonv}{\begin{Konvention}}
\newcommand{\ekonv}{\end{Konvention}}
\newcommand{\bglklein}{\begin{StandardUnnumGleichungKlein}}
\newcommand{\eglklein}{\end{StandardUnnumGleichungKlein}}
\newcommand{\bgl}{\begin{StandardUnnumGleichung}}
\newcommand{\egl}{\end{StandardUnnumGleichung}}
\newcommand{\bglrtext}{\begin{XrelYZNumGleichung}}
\newcommand{\eglrtext}{\end{XrelYZNumGleichung}}
\newcommand{\zgl}{\EineZeileGleichung}

\newcommand{\berlgl}{\begin{StandardUnnumGleichung}}
\newcommand{\eerlgl}{\end{StandardUnnumGleichung}}
\newcommand{\beinrueck}{\begin{Einrueckungpur}} 
\newcommand{\eeinrueck}{\end{Einrueckungpur}}
\newcommand{\beinflist}{\begin{EinfachListe}} 
\newcommand{\eeinflist}{\end{EinfachListe}}
\newcommand{\beq}{\begin{equation}}
\newcommand{\eeq}{\end{equation}}
\newcommand{\bhin}{\begin{Hinrichtung}}
\newcommand{\ehin}{\end{Hinrichtung}}
\newcommand{\brueck}{\begin{Rueckrichtung}}
\newcommand{\erueck}{\end{Rueckrichtung}}
\newcommand{\bvl}{\begin{AutoLabelLaengenListe}{\ListNullAbstaende}}
\newcommand{\evl}{\end{AutoLabelLaengenListe}}
\newcommand{\df}[1]{{\bf #1}}

\newlength{\adressabstand}
\newcommand{\bigdirsum}{\bigoplus}%
\newcommand{\Bigbetrag}[1]%
           {\Bigl|{#1}\Bigr|}            %Betragsstriche (gr"o"ser)
\newcommand{\erstezeile}[1]{\phantom{\hspace*{#1em}}&&\hspace*{-#1em}\hspace*{-#1em}}
\newcommand{\erz}[1]{\langle#1\rangle}
\newcommand{\casisum}[2]{c_{#1,#2}}
\newcommand{\ons}{\mathbf{E}}

\newcommand{\dwirk}[1]{\alpha_{#1}}
\newcommand{\gwirk}[1]{\beta_{#1}}

\newcommand{\Rho}{{\mathrm{P}}}%
\newcommand{\Nu}{{\mathrm{N}}}%
\newcommand{\tif}{{\mathrm{top}}}%
\newcommand{\dom}{{\mathrm{dom}}}%
\newcommand{\gotd}{{\mathfrak d}}%
\newcommand{\gotso}{{\mathfrak{so}}}%
\newcommand{\confsp}{{\cal C}}%
\newcommand{\qsf}{{\cal S}}%
\renewcommand{\strat}[1]{{\cal #1}}%
\newcommand{\isf}{\Sigma}%
\newcommand{\dsf}{\Delta}%
\newcommand{\mpsg}{{\Lambda}}%  %multiparameter-Subgroup
\newcommand{\mps}{\lambda}%  %multiparameter-Subgroup
\newcommand{\aux}{{\text{aux}}}%
\newcommand{\autos}{{\Diffeo\text{-}\gaugetrfs}}%
\newcommand{\gaugetrfs}{{\cal E}}%
\newcommand{\Diffeo}{{\cal D}}%
\newcommand{\diffeo}{\varphi}%
\newcommand{\qfa}{\Theta}%
\newcommand{\ausl}{-}%
\newcommand{\einl}{+}%
\newcommand{\qwer}{\rho}%
\newcommand{\invers}{{\text{inv}}}%
\newcommand{\winkel}{\alpha}%
\newcommand{\linf}{L_\infty(\Xb,\mu_0)}
\newcommand{\lzwo}{L_2(\Xb,\mu_0)}
\renewcommand{\linf}{L_\infty}
\renewcommand{\lzwo}{L_2}
\newcommand{\vekt}[1]{{\boldsymbol#1}}%
\newcommand{\matfktgsn}{\matfkt_{\mathrm{SN}}}%
\newcommand{\bsns}[1]{{\cal B_{#1}}}%
\newcommand{\algdiff}{{\alg_{\mathrm{Diff}}}}%
\newcommand{\algauto}{{\alg_{\mathrm{Auto}}}}%
\newcommand{\weyl}{{w}}%
\newcommand{\Weyl}{{\cal{W}}}%
\newcommand{\epg}{{\cal{R}}}%
\newcommand{\Weylx}{A}%
\newcommand{\bound}{{\cal{B}}}%
\newcommand{\Xb}{{\cal{X}}}%
\renewcommand{\Xb}{{\Ab}}%
\renewcommand{\he}{{\mathrm{ss}}}%
\newcommand{\home}{{\cal H}}%
\newcommand{\haux}{\hilb_{\mathrm{aux}}}%
\newcommand{\hilb}{\mathfrak{H}}%
\newcommand{\Bignorm}[2][]{\Bigl\lVert#2\Bigr\rVert_{#1}}
\newcommand{\bignorm}[2][]{\bigl\lVert#2\bigr\rVert_{#1}}

\newcommand{\Pfgen}{\Pf_\gen}
\newcommand{\adm}{{\cal Q}}

\newcommand{\qwert}{\lambda}%
\newcommand{\rrr}{\kappa}%
\newcommand{\Germ}{{\mathrm{Germ}}}%
\newcommand{\qgerm}{\rho}%
\newcommand{\neuzh}[1]{\dach{#1}}
\newcommand{\neueseite}{\newpage}
\newcommand{\zusatzzeile}[1][]{\enlargethispage{#1\baselineskip}}
\begin{document}
\title{Representations of the Weyl Algebra in Quantum Geometry}
\author{Christian Fleischhack\thanks{e-mail: 
            {\tt christian.fleischhack@math.uni-hamburg.de}} \\   
        \\
        {\normalsize\em Max-Planck-Institut f\"ur Mathematik in den
                        Naturwissenschaften$^{\phantom{2}}$}\\[\adressabstand]
        {\normalsize\em Inselstra\ss e 22--26}\\[\adressabstand]
        {\normalsize\em 04103 Leipzig, Germany}
        \\[-25\adressabstand]      
        {\normalsize\em Department Mathematik}$^{\phantom{1}}$\\[\adressabstand]
        {\normalsize\em Universit\"at Hamburg}\\[\adressabstand]
        {\normalsize\em Bundesstra\ss e 55}\\[\adressabstand]
        {\normalsize\em 20146 Hamburg, Germany}
        \\[-25\adressabstand]}      
\date{May 5, 2009} %cmp in memoriam hf 20a
\maketitle
\begin{abstract}
The Weyl algebra $\alg$
of continuous functions and exponentiated fluxes,
introduced by Ashtekar, Lewandowski and others,
in quantum geometry is studied.
It is shown that, in the piecewise analytic category,
every regular representation of $\alg$ having a 
cyclic and diffeomorphism invariant vector, is already 
unitarily equivalent to the fundamental representation.
Additional assumptions concern the dimension of the underlying 
analytic manifold (at least three), the finite wide triangulizability of 
surfaces in it to be used for the fluxes and the naturality 
of the action of diffeomorphisms -- 
but neither any domain properties of the represented
Weyl operators nor the requirement that the diffeomorphisms act by pull-backs.
For this, the general behaviour of $C^\ast$-algebras
generated by continuous functions and pull-backs of homeomorphisms,
as well as the properties of stratified analytic diffeomorphisms are studied.
Additionally, the paper includes also a 
short and direct proof of the irreducibility of $\alg$.
\end{abstract}

%------------------------------------------------------------------------%
%            Abschnitt: Introduction                                     %
%------------------------------------------------------------------------%
\section{Introduction}
Every physical theory requires fundamental mathematical assumptions 
at the very beginning.
It is highly desirable to justify them by even more fundamental axioms
that are both mathematically and physically as plausible as possible. 
In loop quantum gravity, there are a few of such technical
prerequisites. First of all, of course, one assumes that all 
objects are constructed out of parallel transports along graphs 
in a base manifold of an $SU(2)$ principal fibre bundle 
(or maybe also using higher dimensional objects like in spin foam theory).
This is reasonable by the fact that classical (canonical) gravity
is an $SU(2)$ gauge field theory with constraints as discovered
by Ashtekar in the mid-80s \cite{a117}.
Secondly, one needs inputs about the quantization of this classical
system. For this, 
at least the structure of the configuration space $\confsp$ of all those 
parallel transports (modulo gauge transforms) has to be fixed. 
If one wants to use functional integrals for quantization, one
is forced to study measures on that space.
The usage of parallel transports corresponding
to smooth connections only, however,
has lead to enormous mathematical problems. These
could be widely avoided only 
by including distributional connections as well \cite{a72}. 
Namely, by the assumption that the reductions of the full theory 
to finitely many degrees of freedom 
(i.e.\ parallel transports on a finite graph)
are continuous, one finds that
the topology of $\confsp$ is a projective limit topology%
\footnote{Of course, any refinement of this topology 
leads to continuous reductions again. However, for simplicity, one
ignores this possibility.} making $\confsp$ a compact space.
Here, the compactness is induced 
by that of the underlying structure group $SU(2)$
comprising the values of the parallel transports.
This strategy can be reused to find natural measures on $\confsp$ -- 
one simply uses the assumption
that the restrictions of the theory to finite graphs 
push forward the measure on $\confsp$
to the Haar measures on the finite powers of $SU(2)$. This
leads to the Ashtekar-Lewandowski measure $\mu_0$ \cite{a48}.
Of course, this measure is ``natural'', since the Haar measure
on a Lie group is ``natural'' as well. However, this is at most
a mathematical statement or a statement of beauty. The deeper
question behind is how one can justify this choice  by
mathematical physics arguments.

%------------------------------------------------------------------------%
%            Abschnitt: Introduction                                     %
%------------------------------------------------------------------------%
\subsection{Early Attempts}

For the first time, this problem has been raised by Sahlmann \cite{d64}.
He considered the class of measures on $\confsp$
that are absolutely continuous w.r.t.\ $\mu_0$,
and realized \cite{d64,d65} that (up to some additional technical assumptions)
only $\mu_0$ allows
for a diffeomorphism invariant measure such that the flux variables
are represented as operators on the corresponding $L_2$ space.
Although these results were proven for the case of a $U(1)$ gauge
theory, they have been expected to hold also for the case of a general
compact structure Lie group $\LG$. Moreover, it suggests that 
the diffeomorphism invariance of gravity together with its
full phase space description could be responsible for
the uniqueness of $\mu_0$. 
The situation is similar to ordinary
quantum mechanics. There, the Stone-von Neumann theorem \cite{BratRob2}
tells us that there is (up to equivalence) 
precisely one irreducible regular representation of the Weyl algebra 
generated by the exponentiated position and momentum operators
together with their Poisson relations. In the
standard Schr\"odinger representation on $L_2(\R,\dd x)$, 
these unitary operators are given by 
\bgl\
[\e^{\I \pi \dach x} \psi] (x) \breitrel= \e^{\I \pi x} \psi (x)
& \breitrel{\text{ and }} & 
[\e^{\I \xi \dach p} \psi] (x) \breitrel= \psi(x + \xi).
\egl\noindent
In loop quantum gravity, on the other hand, 
the connections are the generalized positions and the 
densitized dreibein fields are the generalized momenta. 
Exponentiation here includes also smearing: Connections are smeared
along one-dimensional objects (i.e.\ paths) and exponentiated
to give parallel transports --
dreibeine along one-codimensional objects (i.e.\ hypersurfaces)
to give flux variables. Now, one possible (even irreducible and regular) 
representation for the
corresponding Weyl algebra $\alg$ is given by multiplication
and translation operators, respectively,
on $L_2$ functions on $\confsp$ w.r.t.\
the Ashtekar-Lewandowski measure. 
All that suggests that maybe this representation $\pi_0$ 
is even uniquely determined as well by certain reasonable assumptions.
Sahlmann and Thiemann \cite{d60,d63}, supported by results
of Lewandowski and Oko\l\'ow \cite{d61} (see also \cite{d66} for further
discussion), had argued that 
$\pi_0$ may be the only irreducible, regular and diffeomorphism invariant
representation of $\alg$. 
Despite the progress
given by these papers,
there had remained many points open, both technically and conceptually. 
A conceptual one concerned the domain properties of the represented 
operators. In fact, all results for non-abelian structure groups in \cite{d60}
relied crucially on the fact that the
self-adjoint generators of both the represented and the non-represented 
unitary operators share a certain, but not naturally given common dense domain.
Another issue regarding the smoothness properties of the diffeomorphisms
will be discussed below.

%------------------------------------------------------------------------%
%            Abschnitt: Introduction                                     %
%------------------------------------------------------------------------%
\subsection{Achievements of the Present Paper}

The situation above has described the status some five years ago. 
The goal of our present paper is now to give a complete and rigorous proof
of a Stone-von Neumann-like theorem in quantum geometry avoiding most of these
problems. More precisely, we will show that every regular representation
of $\alg$ that has a cyclic and diffeomorphism invariant vector,
is unitarily equivalent to the fundamental representation $\pi_0$,
provided the action of diffeomorphisms satisfies some rather
mild condition.
The main conceptual achievements of our theorem, in comparison to \cite{d60},
are the following:
\bunum
\item
There are no longer any requirements concerning the domains of the operators in
the game. This will be possible, since we consequently, from the very beginning,
work with the exponentiated fluxes only. At no point, will we
use their self-adjoint generators. There is only one issue, where we
use the relation between operators and their generators. This will concern
one-parameter subgroups in a compact Lie group in order to get some
estimate for certain products in it. However, we will completely leave this 
infinitesimal arena before going back to the Weyl algebra level.
\item
The requirements concerning the representations
of the diffeomorphisms are drastically weakened. 
In \cite{d60}, it had to be assumed that
these are represented via pull-backs and respect
the decomposition of the representation restricted to $C(\confsp)$ 
into cyclic generators. 
In particular, one had to assume that each of these components 
contains a diffeomorphism invariant cyclic vector.
As to be discussed at the end of the paper, a priori
these requirements drastically reduce the measures 
allowed in these decompositions.
We will now be able to show that this assumption can be replaced
by a weaker one. We only require that coinciding addends in the
decomposition share the same representation of diffeomorphisms
if at least one addend is diffeomorphism invariant.
\item
Moreover, we will be able to clarify the particular class of diffeomorphisms
to be used. Analytic diffeomorphisms are unsatisfactory from two points of 
view: Physically, they contradict the notion of locality, i.e., if we
transform some set in the space(-time) manifold locally, then we transform 
this manifold even globally. Mathematically, they are not flexible enough
as well, i.e., it will often be very difficult, if not impossible, 
to locally map objects onto each other under very rigid conditions, as we will
see below. Therefore, we are forced to extend the class of
isomorphisms. In fact, it will be manageable to 
use stratified analytic diffeomorphisms,
slightly modifying the similar structures in, e.g., \cite{m5,m4,m6}.
This, at the same time, leads to a natural extension of the 
surfaces used to define the Weyl operators, from analytic submanifolds
to semianalytic sets. However, this is not a severe extension, 
since every semianalytic set can be stratified into a locally finite set of 
analytic submanifolds being mutually disjoint, i.e., having commuting
Weyl operators.
\eunum

%------------------------------------------------------------------------%
%            Abschnitt: Introduction                                     %
%------------------------------------------------------------------------%
\subsection{Idea of the Proof}
Let us very shortly outline the proof of the uniqueness theorem.
As usual (see, e.g., \cite{d60}), the restriction of 
any representation $\pi$ of a Weyl-like algebra to the continuous functions,
can be decomposed into (w.r.t.\ $C(\confsp)$) cyclic ones. 
These are always
the canonical representations on some $L_2(\confsp,\mu_\nu)$
with appropriate measures $\mu_\nu$ on $\confsp$. 
Assuming that $\pi$ contained a cyclic
vector having some invariance property, 
we may find such a decomposition, such that one of the
constant vectors $\EINS_\nu\in L_2(\confsp,\mu_\nu)$ 
has these properties as well. 
Then, being the first step where we use the particular structures of quantum
geometry, regularity and diffeomorphism invariance imply
that this $\mu_\nu$ is the Ashtekar-Lewandowski measure.
Now, being the second step relying on quantum geometry, 
we may show that certain Weyl operators
are diffeomorphism conjugate to their adjoints.
By general arguments, using the two properties above and 
adding invariance and cyclicity of $\EINS_\nu$,
we prove that $\pi$ equals (up to unitary equivalence) 
the fundamental representation of $\alg$.

\makeatletter
\newcounter{tab}
\newcommand{\fusszeile}{}
\newenvironment{tabelleengl}[1][]
                        {\stepcounter{tab}%
                         \renewcommand{\ZaehlerMarke}{\arabic{tab}}%  
                         \renewcommand{\Einzugsname}{{Table \ZaehlerMarke}}%
                         \renewcommand{\fusszeile}{\Einzugsname\ifthenelse{\equal{#1}{}}{}{: #1}}
                         \begin{center}
                         \renewcommand{\@currentlabel}{\ZaehlerMarke}%
                        }%
                        {\par\noindent\fusszeile
                         \end{center}}
\makeatother
\newcommand{\mdf}[1]{\text{\boldmath $#1$}}
\newcommand{\LGG}{G}
\newcommand{\ansleer}{$\phantom\cdot$\:\:}
\newcommand{\ans}{$\cdot$\:\:}
%------------------------------------------------------------------------%
%            Konstruktion: neue Zusammenh"ange                           %
%------------------------------------------------------------------------%
\subsection{Comparison with LOST Paper}
\label{subsect:vgl_lost}
\label{addendum:start}
While this paper was prepared, Lewandowski, Oko\l\'ow, Sahlmann
and Thiemann (LOST) were working on a similar problem for the 
holonomy-flux $\ast$-algebra. This algebra is given if the 
fluxes themselves are considered 
together with the continuous functions on $\confsp$.
Some time after the present article had been sent to the {\tt arxiv},
the four-men paper \cite{lost} has been finished and 
appeared there as well. In this 
subsection,
we are going to compare the corresponding results.

As already mentioned, 
the most striking difference between the two approaches lies in the algebra:
We use both exponentiated positions and momenta, but LOST exponentiate
positions only and keep the fluxes non-exponentiated. 
Consequently, LOST investigate the holonomy-flux algebra, a $\ast$-algebra, 
but we consider the Weyl algebra --
a $C^\ast$-algebra. Here the exponentiated fluxes are 
implemented as unitaries, whereas LOST study 
implicitly their self-adjoint generators being, of course, unbounded. 
The price to pay is that,
in contrast to our case, LOST have to get rid of
the persistent domain problems. 
This is done very directly using a state, since that --via GNS-- 
guarantees the existence of a common dense domain for all the operators.
By construction, this domain is spanned by the cylindrical functions on $\confsp$.
On the other hand, 
we only assume that the Weyl operators are continuously represented
w.r.t.\ their smearing. This means that each corresponding one-parameter
subgroup has some self-adjoint generator.
If this was not the case, it is expected that then 
there exist other diffeoinvariant representations of the Weyl algebra.
Nevertheless, note that our regularity assumption for each {\em single}\/
one-parameter subgroup is much weaker than
that of the existence of a certain {\em common}\/ dense domain for {\em all}\/
generators as in the LOST case. Indeed, our assumption
follows from the LOST requirements: The GNS construction
implies that, given a state, the $\ast$-invariant fluxes become symmetric
operators. As it turns out, they are even self-adjoint.
Hence they generate weakly continuous one-parameter subgroups.

All that seems to show that our result is much stronger 
than that of LOST. However,
there will be an additional assumption made in our paper only:
the diffeomorphisms are implemented naturally. 
Until now, by no means, neither the relevance of this requirement 
nor its possible counterpart in the LOST paper is clear. 
However, while, as a matter of principle, it cannot be expected that the
domain assumptions above can be dropped by LOST, we do hope that
the naturality condition can be shown obsolete sometime. 

The remaining differences are, from our point of view, secondary.
Let us only sketch a few of them.
The technical advantage of the $\ast$-algebra case is
the linearity of the fluxes w.r.t.\ the smearing, which enables LOST to
use the scalar-product trick by Oko\l\'ow. At the same time, LOST 
have to use compactly supported smearing functions. We, on the other hand,
are confined to (up-to-gauge) constant smearings, although there is some hope
to relax that. Since compactly supported smearings
mean that one can restrict oneself to ``nice'' parts of the surfaces
and forget about near-boundary regions, LOST --in contrast to us-- did
not have to assume that the surfaces are (widely) triangulizable.

Rather similar are the general assumptions concerning smoothness. The striking
idea that underlies both investigations is that stratified analytic
objects comprise both the advantages of analyticity and those of locality.
Only the implementation somewhat differs. Both are influenced by the notion
of semianalyticity introduced mainly by \L ojasiewicz, but 
-- for simplicity --
we mostly study these
structures on a given analytic manifold, whereas LOST define semianalytic 
structures in a more categorical way. Nevertheless,
essentially all of our considerations
should be directly transferable to the LOST framework and vice versa. 
There should also be no
significant changes if we required semianalyticity to include not only 
continuity at the boundaries, but also $C^k$ as in the LOST regime. 
Only in the $C^\infty$ case, this is not completely clear. 

Finally, we summarize our comparison 
in Table \ref{tab:vgl} on page \pageref{tab:vgl}.
Note that there we slightly modify the notions used in the respective
article to better explain coincidences and differences. 
\begin{figure}
\begin{tabelleengl}[Comparison between LOST and Fleischhack]
\label{tab:vgl}
{
\small
\begin{tabular}{p{0.27\textwidth}p{0.335\textwidth}p{0.314\textwidth}}
\toprule
       & LOST & Fleischhack  \\
\midrule\midrule
\df{theory} & gauge field theory & gauge field theory \\\midrule
geometric ingredients       & principal fibre bundle $P$ & principal fibre bundle $P$ \\
       & \ans structure group $\LGG$ & \ans structure group $\LGG$ \\
       & \ans base manifold $M$ & \ans base manifold $M$ \\\midrule       
smoothness & stratified analytic & stratified analytic \\
           & \ans$C^{\mdf{k}}$ & \ans$C^{\mdf{0}}$ \\
           & \ans semianalytic & \ans semi- \df{or subanalytic} \\\midrule
basic assumptions & \ans$\LGG$ compact connected Lie & \ans$\LGG$ compact connected Lie \\
                  & \ans$M$ stratified analytic & \ans$M$ analytic \\
                  & \ans$\dim M \geq \mdf{2}$ & \ans$\dim M \geq \mdf{3}$ \\\midrule
diffeomorphisms & stratified analytic & stratified analytic \\\midrule
\midrule
\df{positions} & connections & connections \\
\ans exponentiated & \ans yes & \ans yes \\
\ans smeared along & \ans paths & \ans paths \\\midrule
paths & stratified analytic & stratified analytic \\\midrule
\midrule
\df{momenta} & fluxes & fluxes \\
\ans exponentiated & \ans\df{no} & \ans\df{yes} \\
\ans smeared along & \ans surfaces & \ans surfaces \\\midrule
surfaces & stratified analytic & stratified analytic  \\
     & \ans open & \ans open \\
     & \ans codimension $1$ & \ans codimension $1\mdf{+}$ \\
     & \ans---    & \ans\df{widely triangulizable} \\\midrule
smearing functions & stratified analytic & stratified analytic \\
& \df{compactly supported} & \df{constant on strata} \\
\midrule
\midrule
\df{algebra} & holonomy-flux algebra & Weyl algebra \\\midrule
type    & $\mdf{\ast}$-algebra & $\mdf{C^\ast}$-algebra \\\midrule
generators & positions & positions \\ 
          & \ans\df{cylindrical} functions on $\confsp$ & \ans\df{continuous} functions on $\confsp$ \\\cmidrule{2-3}
          & momenta & momenta \df{(unitary)} \\
          & \ans \df{weak derivatives of} &    \\
          & \ansleer pull-backs of left/right & \ans pull-backs of left/right  \\
          & \ansleer translations on $\confsp$ & \ansleer translations on $\confsp$ \\
\midrule
\midrule
\df{uniqueness} & \df{state} & \df{representation} \\\midrule
assumed cyclicity & cyclic invariant vector 
 & cyclic invariant vector \\\midrule
domain assumptions & \df{common dense domain:} & --- \\
                   & \df{cylindrical functions} & \\\midrule
regularity assumptions & ---  & \df{regularity w.r.t.\ smearing} \\\midrule
add'l assumptions & ---  & \df{natural diffeo action} \\\midrule
required invariance & \df{all} bundle automorphisms & some bundle automorphisms \\
                    & \ans diffeomorphisms & \ans some diffeomorphisms\\
                    & \ans\df{gauge transformations} & \ans---
\\\bottomrule
\end{tabular}
}
\leerezeile
\end{tabelleengl}
\end{figure}

\label{addendum:end}

%------------------------------------------------------------------------%
%            Abschnitt: Introduction                                     %
%------------------------------------------------------------------------%
\subsection{Further Developments}
Both the LOST and the present paper originate from the quest for a quantum gravity
theory. Therefore, as said above, its main application concerns an $SU(2)$-gauge field theory
over a three-dimensional manifold $M$ (i.e., some Cauchy surface) with 
diffeomorphism invariance as a fundamental symmetry. All the results
contain, of course, this case, but go much beyond. Nevertheless, 
some related questions are still unsolved. For instance, what about theories with
other symmetries or another field content? First results have been obtained
for homeomorphism invariant scalar field 
theories \cite{d67,d68}. Here, it turned out, that there are indeed other states, labelled by
the Euler characteristics, i.e.\ algebraic-topological properties of the hypersurfaces.
Another approach currently under investigation, 
has been taken by Bahr and Thiemann \cite{d70} extending the diffeomorphism group symmetry
to general automorphisms of the path groupoid.

%------------------------------------------------------------------------%
%            Abschnitt: Introduction                                     %
%------------------------------------------------------------------------%
\subsection{Structure of the Article}
To finish the introduction, let us briefly 
outline the present paper. In Section \ref{sect:gen_set}
we start with a general investigation of $C^\ast$-algebras
that are generated by the continuous functions on a compact Hausdorff
space $X$ and by pull-backs of homeomorphisms of $X$. 
Afterwards, we switch over to quantum geometry.
Since we would like to make the theory applicable to weaker
smoothness classes the paths are required to belong to, 
we generalize the notion of oriented surfaces introducing quasi-surfaces and
intersection functions in Section \ref{sect:qgeo_backgr}. 
Then, in Section \ref{sect:weyl_alg}, the Weyl algebra 
of quantum geometry is defined and the
assumed structures regarding paths, hypersurfaces, diffeomorphisms etc.\
are fixed.
After presenting a pretty short and direct proof for
the irreducibility of the Weyl algebra in Section \ref{sect:irred},
we study the theory of stratified diffeomorphisms
in detail in Section \ref{sect:strat_diffeo}.
The main result on the uniqueness of representations is then contained
in Section \ref{sect:repr}, including a discussion
of the assumptions made and the extensions possible.

%------------------------------------------------------------------------%
%            Abschnitt: Introduction                                     %
%------------------------------------------------------------------------%
\section{General Setting}
\label{sect:gen_set}
Let $X$ be a compact 
Hausdorff space and $\Homeo(X)$ be the set of all homeomorphisms of $X$.
Given some $\xi\in\Homeo(X)$, its pull-back to $C(X)$ is denoted
by $\weyl_\xi$ or, as usual, $\xi^\ast$. Correspondingly, for every 
$\home \teilmenge \Homeo(X)$, the set $\Weyl_\home \ident \home^\ast
\teilmenge \Homeo^\ast(X)$
contains precisely the pull-backs of all elements in $\home$.
The other way round, given some pull-back $\weyl \in \Homeo^\ast(X)$, 
the corresponding homeomorphism is denoted by $\xi_\weyl$, i.e., we
have $\xi_\weyl^\ast = \weyl$. Analogously, $\home_\Weyl\teilmenge \Homeo(X)$ 
is defined for all $\Weyl \teilmenge \Homeo^\ast(X)$. Moreover, we denote
by $\erz\Weyl$ the (abstract) subgroup of $\Homeo^\ast(X)$ generated by $\Weyl$
and define, analogously, $\erz\home$. Obviously, 
$\erz{\home_\Weyl} = \home_{\erz\Weyl}$ and
$\erz{\Weyl_\home} = \Weyl_{\erz\home}$.   
Next, for every measure%
\footnote{If not stated otherwise, by a measure we always mean a normalized
regular Borel measure.}
$\mu$ on $X$,
we denote by $\home(\mu)$
the set of all homeomorphisms on $X$ leaving $\mu$ invariant.
Clearly, $\erz{\home(\mu)} = \home(\mu)$.
Moreover, every $\weyl\in\Weyl_{\home(\mu)}$ 
extends naturally to a unitary operator on $L_2(X,\mu)$,
again denoted by $\weyl$. 
By $\weyl(f\psi) = \weyl(f) \weyl(\psi)$ for all $f\in C(X)$, 
$\psi\in L_2(X,\mu)$ and $\weyl\in\Weyl_{\home(\mu)}$,
we have $\weyl \circ f \circ \weyl^{-1} = \weyl(f)$
as operators in $\bound(L_2(X,\mu))$.
Sometimes, we will extend the notion to operators:
$\weyl_1(\weyl_2) := \weyl_1 \circ \weyl_2 \circ \weyl_1^{-1}$
for $\weyl_1,\weyl_2\in\Weyl_{\home(\mu)}$. 
Finally, let $\alg(\Weyl,\mu)$ denote 
the $C^\ast$-subalgebra in $\bound(L_2(X,\mu))$
generated by $C(X)$ and $\Weyl \teilmenge \Weyl_{\home(\mu)}$,
and let $\pi_0$ be the identical (or fundamental) 
representation of $\alg(\Weyl,\mu)$
on $L_2(X,\mu)$.

\blem
\label{lem:erz(cont*weyl)}
For every $\Weyl\teilmenge\Weyl_{\home(\mu)}$,
the subalgebra spanned by all products $f \circ w$ with $f\in C(X)$ and
$\weyl\in\erz\Weyl$ is dense in $\alg(\Weyl,\mu)$.
\elem
\bpf
Since 
$\weyl \circ f = \weyl(f) \circ \weyl$ for all $\weyl\in\erz\Weyl$
and $f\in C(X)$, 
\bgl
\erstezeile3
f_1 \circ \weyl_1 \circ f_2 \circ \weyl_2 \circ \cdots \circ \weyl_k \circ f_{k+1} \\
 & = & \bigl(f_1 \cdot \weyl_1(f_2) \cdot \cdots \cdot 
                \weyl_1(\weyl_2(\ldots(\weyl_k(f_{k+1}))\ldots))\bigr) \circ 
       \bigl(\weyl_1 \circ \cdots \circ \weyl_k \bigr) 
\egl
is always in $C(X) \circ \erz\Weyl$.
Moreover, with $f$, also $f^\ast \ident \quer f$ is in $C(X)$, and
with $\weyl$, also $\weyl^\ast = \weyl^{-1}$ is in $\erz\Weyl$. 
Therefore, the span of $C(X) \circ \erz\Weyl$ equals the
$\ast$-subalgebra of $\bound(L_2(X,\mu))$ generated by $C(X)$ and $\Weyl$.
\qed
\epf

Throughout the whole section, let $\mu$ be some arbitrary, but fixed measure
on $X$.

%------------------------------------------------------------------------%
%            Abschnitt: Introduction                                     %
%------------------------------------------------------------------------%
\subsection{First-Step Decomposition}
Since every representation of a $C^\ast$-algebra is the direct sum of
a zero representation and a non-degenerate one, we may
restrict ourselves to non-degenerate representations in the
following. 

\blem
\label{lem:decomp(nondeg)}
Fix some $\Weyl \teilmenge \Weyl_{\home(\mu)}$ and
let $\pi$ be a non-degenerate 
representation of $\alg(\Weyl,\mu)$ on some Hilbert space 
$\hilb$. 

Then there are measures $\mu_\nu$ on $X$ with $\nu$ running 
over some (not necessarily countable) index set $\Nu$, such that
$\pi\einschr{C(X)}$ is unitarily equivalent to the direct-sum 
representation
$\bigdirsum_\nu \pi_{\mu_\nu}$, where $\pi_{\mu_\nu}$ 
denotes the canonical representation of $C(X)$ on $L_2(X,\mu_\nu)$
by multiplication operators.
Moreover, these measures may be chosen, such that two of them
are equal if they are equivalent (w.r.t.\ absolute continuity).
\elem
\bpf
Every non-degenerate representation of a $C^\ast$-algebra
is (up to unitary equivalence) the direct sum of cyclic 
representations \cite{BratRob1}. The first assertion now follows,
because every cyclic representation of $C(X)$ is equivalent 
to the canonical representation on $L_2(X,\mu_\nu)$ by
multiplication operators for some regular Borel measure $\mu_\nu$ \cite{EMS124}.
Note that $\pi\einschr{C(X)}$ is non-degenerate by $\EINS\in C(X)$.
Since measures on $X$ are equivalent w.r.t.\
absolute continuity iff the corresponding canonical representations
are equivalent \cite{EMS124}, we get the proof.
\qed
\epf

\bdf
A decomposition $\bigdirsum_\nu \pi_{\mu_\nu}$
as given in Lemma \ref{lem:decomp(nondeg)}
is called
\df{first-step decomposition} of $\pi$. 
\edf
Sometimes we write $(\mu_\nu)_{\nu\in\Nu}$ or shortly $\gross\mu$
to characterize such a decomposition. Moreover, if the particular
$\Weyl$ is not important, we will consider first-step decompositions without
any reference to some $\pi$.

\bdf
A first-step decomposition is called
\df{short} iff $\Nu$ consists of a single element.
\edf
\brem
First-step decompositions are not at all unique.
In fact, consider a short one with $\mu_\nu = \mu$
and choose $U\teilmenge X$ with $0 < \mu(U) < 1$.
Decomposing
 any $\psi\in\hilb$ into $\psi = 1_U \psi + 1_{X\setminus U} \psi$
with $1_U$ being the characteristic function on $U$,
we get a first-step decomposition 
$\pi_{\mu_U} \dirsum \pi_{\mu_{X \setminus U}}$. 
Here, $\mu_U$ is the normalization of $1_U \kp \mu$.
\erem

In the following, 
given some representation $\pi$ of $\alg(\Weyl,\mu)$ on $\hilb$,
we will usually assume that 
$\pi\einschr{C(X)}$ equals (one of) its first-step decomposition(s).
Moreover,
we usually write shortly $\pi_\nu$ instead of $\pi_{\mu_\nu}$.
By $\norm\cdot_{\mu_\nu}$ we denote the norm on 
$L_2(X,\mu_\nu) =: \hilb_\nu$ 
and by $P_\nu$ the respective orthogonal projector mapping $\hilb$ to 
$\hilb_\nu$. 
In particular, we have 
$\norm{\pi(f)\psi}_{\hilb}^2 = \sum_\nu \norm{f \cdot P_\nu \psi}_{\mu_\nu}^2$
for all $f\in C(X)$ and $\psi\in\hilb$.
Next, let $I_\nu : \hilb_\nu \nach \hilb$ denote
the (norm-preserving) canonical embedding of $\hilb_\nu$ into $\hilb$ and set 
$\EINS_\nu := I_\nu (\EINS)$, where $\EINS$ is seen not only as an element 
in $C(X)$, but in $\hilb_\nu$ as well. 
Anyway, often we will simply drop $I_\nu$.
Analogously, we do not explicitly mark the transition from
continuous functions to their classes in $L_2$, when calculating
scalar products.
Note, however, that $C(X)$ is, in general, 
{\em not embedded}\/ into $L_2(X,\mu_\nu)$.
Let, e.g., $\mu_\nu$ be the Dirac measure at some point in $X$, then
the image of $C(X)$ is isomorphic to $\C$. Therefore, one has to be careful
when operating with pull-backs of homeomorphisms that do not
leave $\mu_\nu$ invariant.
Finally, for $\mu_{\nu_1} = \mu_{\nu_2}$ we denote the canonical isomorphism
mapping $I_{\nu_1}(\hilb_{\nu_1})$ to $I_{\nu_2}(\hilb_{\nu_2})$
by $I^{\nu_1}_{\nu_2}$.

\bdf
Let $\Weyl$ be a subset of $\Weyl_{\home(\mu)}$
and let $\pi$ be some representation of $\alg(\Weyl,\mu)$ on some
Hilbert space $\hilb$.

A vector $\psi \in \hilb$ is called \df{$\Weyl$-invariant}
iff $\pi(\weyl) \psi = \psi$ for all $\weyl\in\Weyl$.
\edf
Note that we tacitly assume some information about $\pi$ to be given
when we speak on invariance w.r.t.\ some $\Weyl$.
This will avoid some cumbersome notation when we study equivalent
representations.
\blem
\label{lem:canon_decomp:inv+cycl}
Let $\Weyl$ and $\Weyl'$ be subsets of $\Weyl_{\home(\mu)}$, 
let $\pi'$ be a representation of $\alg(\Weyl\cup\Weyl',\mu)$ on some
Hilbert space $\hilb$, and let
$\psi\in\hilb$ be a $\Weyl'$-invariant vector. 

Then there is a first-step decomposition $\bigdirsum_{\nu\in\Nu} \pi_{\mu_\nu}$
of $\pi'$ and 
some $\nu\in\Nu$,
such that $\EINS_\nu$ is a $\Weyl'$-invariant vector.
If, moreover, $\psi$ is cyclic for $\pi'\einschr{\alg(\Weyl,\mu)}$, 
then $\EINS_\nu$ may be chosen cyclic as well.
\elem
\bpf
Define $\hilb_\nu := \quer{\pi'(C(X)) \psi} \teilmenge \hilb$.
Then both $\hilb_\nu$ and $\hilb_\nu^\senk$ are invariant w.r.t.\ $\pi'(C(X))$.
Since $\hilb_\nu^\senk$ is non-degenerate (if not zero), 
the projection of $\pi'\einschr{C(X)}$ to $\hilb_\nu^\senk$
is (up to equivalence) some direct sum 
$\bigdirsum_{\nu'\in\Nu'} \pi_{\mu_{\nu'}}$ of cyclic representations
of $C(X)$. 
Since, on the other hand,
$\pi'\einschr{C(X)}$ is cyclic on $\hilb_\nu$, it is equivalent
to the canonical representation $\pi_{\mu_\nu}$ of $C(X)$ on some 
$L_2(X,\mu_\nu)$, whereas the corresponding intertwiner maps $\psi$ to 
$\EINS_\nu$. 
Now, by construction,
$\pi_{\mu_\nu} \dirsum \bigdirsum_{\nu'\in\Nu'} \pi_{\mu_{\nu'}}$ 
is a first-step decomposition of $\pi'$. 
Moreover, the $\Weyl'$-invariance of $\psi$ translates into that 
of $\EINS_\nu$ and the cyclicity, if given, as well.
\qed
\epf

Now, throughout the whole Section \ref{sect:gen_set}, we let 
$\Weyl$ and $\Weyl'$ be some arbitrary subsets of $\Weyl_{\home(\mu)}$,
whereas $\weyl'(\Weyl) \teilmenge \Weyl$ for all $\weyl'\in\Weyl'$.
Note that we do not assume that they are fixed once and for all,
i.e., they may be changed from one statement to the other.
Next, $\pi$ and $\pi'$ are always non-degenerate representations
of $\alg(\Weyl,\mu)$ and $\alg(\Weyl \cup \Weyl',\mu)$,
respectively, on some Hilbert space $\hilb$,
where $\pi$ is the restriction of $\pi'$ to $\alg(\Weyl,\mu)$.%
\footnote{Often, we will not refer to $\pi'$ at all. Then,
in general, we tacitly set $\Weyl' = \leeremenge$ and $\pi' = \pi$.}
We let $\bigdirsum_\nu \pi_{\mu_\nu}$ 
be a fixed first-step decomposition of $\pi$ on
$\hilb = \bigdirsum_\nu \hilb_\nu = \bigdirsum_\nu L_2(X,\mu_\nu)$ 
and usually set 
$\pi_\nu := \pi_{\mu_\nu}$ for simplicity. 
Note that every first-step decomposition
of $\pi$ is also some for $\pi'$ and vice versa, since 
$\pi$ and $\pi'$ coincide on $\alg(\Weyl,\mu)$ containing $C(X)$.
Moreover, if there is some $\Weyl'$-invariant
(and $\pi$-cyclic) vector, then we assume that there is some
$\nu\in\Nu$, such that $\EINS_\nu$ is $\Weyl'$-invariant (and $\pi$-cyclic).
Note that this does not contradict the assumption above that measures in a
first-step decomposition are equal if they are equivalent.
Finally, in order to fix a home for the one-parameter subgroups in $\Weyl$
introduced later,
we fix some subset $\epg$ in the set $\Hom(\R,\Weyl)$ of
homomorphisms from $\R$ to $\Weyl$.

%------------------------------------------------------------------------%
%            Abschnitt: Introduction                                     %
%------------------------------------------------------------------------%
\subsection{$\pi_\nu$-Scalars and $\pi_\nu$-Units}

\bdf
An element $\weyl\in\Weyl$ is called 
\bunum
\item
\df{$\pi_\nu$-scalar} iff
$P_\nu \pi(\weyl) \EINS_\nu =  c \:\EINS$ for some $c\in\C$;
\item
\df{$\pi_\nu$-unit} iff 
$P_\nu \pi(\weyl) \EINS_\nu =  \EINS$.
\eunum
\edf
Analogously, we define these properties for $\weyl'\in\Weyl'$.
Since $\weyl$ is unitary, we have
\blem
\label{lem:inv=pi_nu-unit}
$\EINS_\nu$ is $\pi(\weyl)$-invariant 
 $\aequ$  $\weyl$ is a $\pi_\nu$-unit
 $\aequ$  $\weyl^\ast$ is a $\pi_\nu$-unit.
\elem

\bcorr
\label{corr:prod(pi_nu-units)}
Any finite product of $\pi_\nu$-units is a $\pi_\nu$-unit.
\ecorr

\blem
\label{lem:inv(eins)->inv(hilb+hilbsenk)}
If $\EINS_\nu$ is $\pi(\weyl)$-invariant, then $\hilb_\nu$ and
$\hilb_\nu^\senk$ are $\pi(\weyl)$-invariant.
\elem
\bpf
Fix some $\psi_\nu\in\hilb_\nu$ and 
recall that $\EINS_\nu$ is cyclic for $\pi_\nu$, i.e.,
for every $\varepsilon > 0$ there is some 
$f\in C(X)$ with $\norm{\pi(f) \EINS_\nu - \psi_\nu} < \varepsilon$.

Since $\weyl$ is a $\pi_\nu$-unit, we have 
$\pi(\weyl) \pi(f) \EINS_\nu = \pi(\weyl(f)) \EINS_\nu \in \hilb_\nu$.
By unitarity of $\weyl$, we get
$\norm{\pi(\weyl)\psi_\nu - \pi(\weyl) \pi(f) \EINS_\nu} < \varepsilon$,
i.e.\ $\dist(\pi(\weyl) \psi_\nu,\hilb_\nu) < \varepsilon$
for all $\varepsilon > 0$. Hence,
$\pi(\weyl)\psi_\nu \in \hilb_\nu$.

The invariance of $\hilb_\nu^\senk$ follows from the unitarity of $\pi(\weyl)$.
\qed
\epf

\bcorr
\label{corr:unit-nu->cyclic}
If each $\weyl\in\Weyl$ is a $\pi_\nu$-unit, then the restriction
of $\pi$ to $\hilb_\nu$ is cyclic.
\ecorr
\bpf
Since every $\pi_\nu$-unit leaves $\hilb_\nu$ invariant, $\pi(\alg(\Weyl,\mu))$
leaves $\hilb_\nu$ invariant. Since $\EINS_\nu$ is already 
cyclic on $\hilb_\nu$ for $\pi$ restricted to $C(X) \teilmenge\alg(\Weyl,\mu)$, 
we get the assertion.

\qed
\epf

\blem
\label{lem:divide_into_inv_subsp}
Let $\weyl\in\Weyl$ be a $\pi_\nu$-scalar  
and assume $\psi_0 := (\EINS - I_\nu P_\nu) \pi(\weyl) \EINS_\nu \neq 0$.
Define $\psi$ to be the normalization of $\psi_0$, and let $\hilb_\psi$
be the completion of $\pi(C(X)) \psi$. Finally, assume that
$\mu_\nu$ is $\xi_\weyl$-invariant, i.e.\ $\weyl \in \Weyl_{\home(\mu_\nu)}$.

Then the restriction of $\pi\einschr{C(X)}$ to 
its invariant subspace $\hilb_\psi$ is 
equivalent to the canonical representation of $C(X)$ on $L_2(X,\mu_\nu)$.
Moreover, $\hilb_\nu$ and $\hilb_\psi$ are orthogonal.
\elem
\bpf
Of course, by definition, $\hilb_\psi$ is invariant w.r.t.\ $\pi(C(X))$.
Let now $f_1$ and $f_2$ be in $C(X)$. Then, by unitarity of $\pi(\weyl)$
and $\xi_\weyl$-invariance of $\mu_\nu$, we have
\bglklein
 &   & \skalprod{\pi(f_1) \psi_0}{\pi(f_2) \psi_0}_\hilb \\
 & = & \skalprod{\pi(f_1) (\EINS - I_\nu P_\nu) \pi(\weyl) \EINS_\nu}{\pi(f_2) (\EINS - I_\nu P_\nu) \pi(\weyl) \EINS_\nu}_\hilb \\
 & = & \skalprod{(\EINS - I_\nu P_\nu) \pi(f_1) \pi(\weyl) \EINS_\nu}{(\EINS - I_\nu P_\nu) \pi(f_2) \pi(\weyl) \EINS_\nu}_\hilb \\
 & = & \skalprod{\pi(f_1) \pi(\weyl) \EINS_\nu}{\pi(f_2) \pi(\weyl) \EINS_\nu}_\hilb  - 
       \skalprod{I_\nu P_\nu \pi(f_1) \pi(\weyl) \EINS_\nu}{I_\nu P_\nu \pi(f_2) \pi(\weyl) \EINS_\nu}_\hilb \\
 & = & \skalprod{\pi(\weyl) \pi(\weyl^\ast(f_1)) \EINS_\nu}{\pi(\weyl) \pi(\weyl^\ast(f_2))\EINS_\nu}_\hilb  - 
       \skalprod{f_1 \cdot P_\nu \pi(\weyl) \EINS_\nu}{f_2 \cdot P_\nu \pi(\weyl) \EINS_\nu}_{\mu_\nu} \\
 & = & \skalprod{\weyl^\ast(f_1)}{\weyl^\ast(f_2)}_{\mu_\nu} -
       \betrag{c}^2 \: \skalprod{f_1}{f_2}_{\mu_\nu} \\
 & = & (1 - \betrag{c}^2) \: \skalprod{f_1}{f_2}_{\mu_\nu},
\eglklein
where $c$ is given by $P_\nu\pi(\weyl)\EINS_\nu = c \EINS$.
By the arguments above, $\norm{\psi_0}^2 = 1 - \betrag{c}^2$,
implying
\bglklein
\skalprod{\pi(f_1) \psi}{\pi(f_2) \psi}_\hilb 
 & = & \skalprod{f_1}{f_2}_{\mu_\nu}
 \breitrel= \skalprod{\pi(f_1) \EINS_\nu}{\pi(f_2) \EINS_\nu}_\hilb 
\eglklein
for all $f_1, f_2 \in C(X)$. The orthogonality of $\hilb_\nu$ and
$\hilb_\psi$ follows directly from that of $\EINS_\nu$ and $\psi$.
\qed
\epf

\bdf
\label{def:natural}
$\pi'$ is called \df{$\Weyl'$-natural} iff, for any first-step decomposition
$\bigdirsum_{\nu\in\Nu} \pi_{\mu_\nu}$ and for all $\nu_1,\nu_2\in\Nu$
with $\mu_{\nu_1} = \mu_{\nu_2}$, an appearing $\pi'(\Weyl')$-invariance
of $\hilb_{\nu_1}$ implies that of $\hilb_{\nu_2}$ and 
\bgl
 I^{\nu_1}_{\nu_2} \circ I_{\nu_1} P_{\nu_1} \bigl(\pi'\einschr{\alg(\Weyl',\mu)}) & = & 
  I_{\nu_2} P_{\nu_2} \bigl(\pi'\einschr{\alg(\Weyl',\mu)}) \circ I^{\nu_1}_{\nu_2}.
\egl
\edf
Obviously, $\pi'$ is $\Weyl'$-natural,
if the respective first-step decomposition is short.
Moreover, if $\pi'$ is $\Weyl'$-natural and if $\mu_{\nu_1} = \mu_{\nu_2}$,
then 
$\EINS_{\nu_1}$ is $\pi'(\weyl')$-invariant iff
$\EINS_{\nu_2}$ is $\pi'(\weyl')$-invariant.

\bcorr
\label{corr:preserve-invar}
Let $\weyl\in\Weyl \cap \Weyl_{\home(\mu_\nu)}$ 
be a $\pi_\nu$-scalar. Moreover, 
let $\pi'$ be $\Weyl'$-natural.

If $\EINS_\nu$ is $\pi'(\weyl')$-invariant, then
$\pi(\weyl) \EINS_\nu$ is $\pi'(\weyl')$-invariant.
\ecorr
\bpf
Since $\weyl'$ is a $\pi_\nu$-unit, $\hilb_\nu$ is $\pi'(\weyl')$-invariant.
Hence, 
\bglklein
\pi'(\weyl') I_\nu P_\nu \pi(\weyl) \EINS_\nu
 & = & \pi'(\weyl') \: c \EINS_\nu
 \breitrel=  c \EINS_\nu
 \breitrel=  I_\nu P_\nu \pi(\weyl) \EINS_\nu.
\eglklein
If $\psi_0 := (\EINS - I_\nu P_\nu) \pi(\weyl) \EINS_\nu = 0$,
the statement is trivial.
Otherwise, we know from the lemma above and the notations there 
that $\hilb_\nu$ and $\hilb_\psi$ are orthogonal. 
Choose a first-step decomposition of $\pi$ containing
the representation $\pi_{\mu_\nu}$ for $\hilb_\nu$ and for $\hilb_\psi$.
In fact, simply construct a first-step decomposition of
the orthogonal complement of $\hilb_\nu \dirsum \hilb_\psi$ in 
$\hilb$. Now, since $\pi'$ is $\Weyl'$-natural, 
$\psi$ is $\pi'(\weyl')$-invariant 
as well, by the $\pi'(\weyl')$-invariance of $\EINS_\nu$ 
and the lemma above.
The proof follows from 
$\pi(\weyl)\EINS_\nu = \psi_0 + I_\nu P_\nu \pi(\weyl)\EINS_\nu$.
\qed
\epf

\bcorr
\label{corr:scal-krit-nu}
Let $\weyl\in\Weyl \cap \Weyl_{\home(\mu_\nu)}$. 
Additionally, let $\pi'$ be $\Weyl'$-natural,
and let $\EINS_\nu$ be $\pi'(\weyl')$-invariant
for some $\weyl'\in\Weyl'$.

If $\weyl$ is a $\pi_\nu$-scalar, then  
$\pi(\weyl'(\weyl)) \EINS_\nu 
   = \pi'(\weyl') \pi(\weyl) \EINS_\nu = \pi(\weyl) \EINS_\nu$.

This means, in particular,
\bgl
\text{$\weyl$ is a $\pi_\nu$-scalar.} 
 & \aequ & \text{$\weyl'(\weyl)$ is a $\pi_\nu$-scalar.} \\
\text{$\weyl$ is a $\pi_\nu$-unit.} \hspace*{\fill}
 & \aequ & \text{$\weyl'(\weyl)$ is a $\pi_\nu$-unit.} \\
\egl

\ecorr

\bcorr
\label{corr:unitaer:scalar-nu->unit}
Assume 
that $\pi'$ is $\Weyl'$-natural and
that $\EINS_\nu$ is $\pi'(\Weyl')$-invariant. 

Then, for all $\pi_\nu$-scalars 
$\weyl\in\Weyl \cap \Weyl_{\home(\mu_\nu)}$ and all
$\weyl',\weyl'_1,\weyl'_2\in\Weyl'$, we have
\bgl
\weyl'(\weyl) = \weyl'_1(\weyl) \circ \weyl'_2(\weyl)
 & \impliz & \text{$\weyl^{\phantom2}$ is a $\pi_\nu$-unit,} \\
\weyl'(\weyl) = \weyl^\ast \hspace*{\fill}
 & \impliz & \hspace*{\fill} \text{$\weyl^2$ is a $\pi_\nu$-unit.} \\
\egl
\ecorr
\bpf
Using Corollary \ref{corr:preserve-invar} we have in the first case 
\bglklein
       \pi(\weyl) \EINS_\nu 
 & = & \pi'(\weyl'_1)^\ast \pi(\weyl) \EINS_\nu 
 \breitrel= \pi'(\weyl'_1)^\ast \pi(\weyl'(\weyl)) \EINS_\nu  \\
 & = & \pi'(\weyl'_1)^\ast \pi(\weyl'_1(\weyl) \circ \weyl'_2(\weyl)) \EINS_\nu 
 \breitrel= \pi'(\weyl'_1)^\ast \pi(\weyl'_1(\weyl)) \pi(\weyl'_2(\weyl)) \EINS_\nu \\
 & = & \pi(\weyl) \pi'(\weyl'_1)^\ast \pi'(\weyl'_2) \pi(\weyl) \pi'(\weyl'_2)^\ast \EINS_\nu 
 \breitrel= \pi(\weyl) \pi(\weyl) \EINS_\nu
\eglklein
and, in the second one, 
\bglklein
       \pi(\weyl) \EINS_\nu 
 & = & \pi(\weyl'(\weyl)) \EINS_\nu 
 \breitrel= \pi(\weyl^\ast) \EINS_\nu.
\eglklein
\qed
\epf

\blem
\label{lem:pi-unit->pi=pi_0}
If $\weyl$ is a $\pi_\nu$-unit and $\mu_\nu$ equals $\mu$,
then $P_\nu \pi(\weyl) = \pi_0(\weyl) P_\nu$.
\elem
\bpf
We have 
\bgl
\pi(\weyl) f \EINS_\nu 
  & = & \pi(\weyl) \pi(f) \EINS_\nu 
  \breitrel= \pi(\weyl(f)) \pi(\weyl) \EINS_\nu 
  \breitrel= \pi(\weyl(f)) \EINS_\nu 
  \breitrel= \weyl(f)  \EINS_\nu,
\egl
hence $P_\nu \pi(\weyl) f \EINS_\nu = \pi_0(\weyl) P_\nu f \EINS_\nu$
for all $f\in C(X)$. By continuity of $\pi_0(\weyl)$ on $L_2(X,\mu)$
and by cyclicity of $\EINS_\nu$ w.r.t.\ $C(X)$, we get
$P_\nu \pi(\weyl) = \pi_0(\weyl) P_\nu$ on $\hilb_\nu$.
Finally, both $P_\nu \pi(\weyl)$ 
(by Lemma \ref{lem:inv(eins)->inv(hilb+hilbsenk)})
and $\pi_0(\weyl) P_\nu$ are zero on $\hilb_\nu^\senk$.
\qed
\epf

\bcorr
\label{corr:commut+pi-unit-nu->inv}
Let $\weyl$ and $\weyl_0$ be commuting elements in $\Weyl$.
Moreover, let $\mu_\nu = \mu$. 

If $\weyl_0$ is a $\pi_\nu$-unit, then it
leaves $P_\nu \pi(\weyl) \EINS_\nu$ invariant.
\ecorr
\bpf
\bgl
\weyl_0 \bigl(P_{\nu} \pi(\weyl) \EINS_{\nu}\bigr)
 & \ident & \pi_0(\weyl_0) \bigl(P_{\nu} \pi(\weyl) \EINS_{\nu}\bigr) \\
 & = & P_{\nu} \pi(\weyl_0) \pi(\weyl) \EINS_{\nu} \erl{Lemma \ref{lem:pi-unit->pi=pi_0}} \\
 & = & P_{\nu} \pi(\weyl) \pi(\weyl_0) \EINS_{\nu} \\
 & = & P_{\nu} \pi(\weyl) \EINS_{\nu}.
\egl
\qed
\epf

%------------------------------------------------------------------------%
%            Abschnitt: Introduction                                     %
%------------------------------------------------------------------------%
\subsection{Continuous $\mu_0$-Generating Systems}

Until the end of this subsection, let $\mu_0$ be some measure on $X$.
\bdf
A subset $\ons$ of $C(X)$ is called 
\df{continuous $\mu_0$-generating system}
iff 
\bunum
\item
$\EINS\in\ons$ is orthogonal in $L_2(X,\mu_0)$ to each other element in $\ons$ and
\item
$\spann_\C \ons$ is dense both in $C(X)$ and in $L_2(X,\mu_0)$.
\eunum
\edf
\blem
\label{lem:cont_gen_syst:scalar}
Let $\ons\teilmenge C(X)$ be a continuous $\mu_0$-generating system for some
measure $\mu_0$, and let $\psi$ be a vector in $L_2(X,\mu_0)$. 

Then $\skalprod{f}{\psi}_{\mu_0} = 0$ for all $\EINS \neq f \in \ons$
implies that $\psi = c \norm{\psi} \: \EINS$ for some $c\in U(1)$.
\elem
\bpf
Use 
$L_2(X,\mu_0) 
  = \quer{\spann_\C \ons} 
  = \quer{\spann_\C \{\EINS\} \dirsum \spann_\C (\ons\setminus\{\EINS\})}
  = \C \EINS \dirsum \quer{\spann_\C (\ons\setminus\{\EINS\})}.$

\qed
\epf

\blem
\label{lem:eq(mu-gen_syst)}
If $\ons \teilmenge C(X)$ is a continuous generating system w.r.t.\ 
two measures $\mu_1$ and $\mu_2$, then $\mu_1$ equals $\mu_2$. 
\elem
\bpf
We have 
$\int_X f \: \dd\mu_1 = \skalprod{\EINS}{f}_{\mu_1} = 0 =
 \skalprod{\EINS}{f}_{\mu_2} = \int_X f \: \dd\mu_2$ 
for all $\EINS \neq f \in \ons$. Since 
$\spann_\C \ons$ is dense in $C(X)$,
the assertion follows from the regularity of the measures.
\qed
\epf
\blem
Continuous $\mu_0$-generating systems always exist.
\elem
\bpf
$C(X)$ always spans a dense subset in $L_2(X,\mu_0)$.
Let now $\ons$ contain $\EINS$ and all $f - \skalprod{\EINS}{f}_{\mu_0}\EINS$
with $f$ in $C(X)$. 
\qed
\epf

\blem
\label{lem:pi(w)-orth}
Let $\weyl\in\Weyl$ be some element. 
Assume that $\pi'$ is $\Weyl'$-natural and 
that $\EINS_\nu$ is $\Weyl'$-invariant.
Moreover, let $\ons_0 \teilmenge C(X)$ be some subset, such that 
for every non-constant $f\in\ons_0$ there are infinitely many elements
$\{\weyl'_\iota\}$ in $\erz{\Weyl'}$ commuting with $\weyl$, such that
$\{\weyl'_\iota(f)\} \teilmenge C(X)$ 
forms an orthonormal system in $L_2(X,\mu_\nu)$. 

Then $P_\nu\pi(\weyl)\EINS_{\nu'}$ is orthogonal to the
span of $\ons_0$ for all $\nu'\in\Nu$ with $\mu_{\nu'} = \mu_\nu$.
Here, $\ons_0$ is seen as a subset in $\hilb_\nu$.
\elem
\bpf
Let $f\in\ons_0$. Then there are 
infinitely many $\{\weyl'_\iota\}$ in $\erz{\Weyl'}$ commuting with
$\weyl$ and fulfilling
\bgl
\skalprod{\pi(\weyl'_{\iota_1}(f)) \EINS_\nu}{\pi(\weyl'_{\iota_2}(f)) \EINS_\nu}_\hilb
 & = & \skalprod{\weyl'_{\iota_1}(f)}{\weyl'_{\iota_2}(f)}_{\mu_\nu}
 \breitrel= \delta_{\iota_1 \iota_2}.
\egl
By naturality, $\EINS_{\nu'}$ is $\Weyl'$-invariant as well. Hence,
\bgl
\skalprod{\pi(\weyl'_\iota(f)) \EINS_\nu}{\pi(\weyl) \EINS_{\nu'}}_\hilb
 & = & \skalprod{\pi'(\weyl'_\iota) \pi(f) \pi'(\weyl'_\iota)^\ast \EINS_\nu}{\pi(\weyl) \EINS_{\nu'}}_\hilb \\
 & = & \skalprod{\pi(f) \EINS_\nu}{\pi'(\weyl'_\iota)^\ast \pi(\weyl) \EINS_{\nu'}}_\hilb \\
 & = & \skalprod{\pi(f) \EINS_\nu}{\pi(\weyl) \pi'(\weyl'_\iota)^\ast \EINS_{\nu'}}_\hilb \\
 & = & \skalprod{\pi(f) \EINS_\nu}{\pi(\weyl) \EINS_{\nu'}}_\hilb
\egl
for all $\iota$. Consequently, 
\bgl
\skalprod{f}{P_\nu \pi(\weyl) \EINS_{\nu'}}_{\mu_\nu} 
 & = & \skalprod{\pi(f) \EINS_\nu}{\pi(\weyl) \EINS_{\nu'}}_\hilb 
 \breitrel= 0.
\egl
\qed
\epf

%------------------------------------------------------------------------%
%            Abschnitt: Preliminaries                                    %
%------------------------------------------------------------------------%
\subsection{Regularity}

\bdf
Precisely the elements of $\Hom(\R,\Weyl)$
are called one-parameter subgroups in $\Weyl$,
those in $\epg \teilmenge \Hom(\R,\Weyl)$ 
\df{one-parameter $\epg$-subgroups in $\Weyl$}.
\edf

\bdf
\label{def:reg-1ps}
A one-parameter subgroup is called \df{regular}
iff it is weakly continuous.
\edf

\bdf
A representation $\pi$ of $\alg(\Weyl,\mu)$ 
is called \df{regular w.r.t.\ $\epg$}
iff $\pi$ maps regular one-parameter $\epg$-subgroups in $\Weyl$ to 
weakly continuous one-parameter subgroups in $\pi(\Weyl)$.
\edf
If $\epg$ is clear from the context, we will simply speak about regular
representations.

\bdf
\bunum
\item
Two one-parameter subgroups $t \auf \weyl_{1,t}$ and $t \auf \weyl_{2,t}$ 
in $\Weyl$ are called \df{commuting} iff $\weyl_{1,t_1}$ and 
$\weyl_{2,t_2}$ commute for all $t_1,t_2\in\R$.
\item
The set given by all finite (pointwise) products of mutually commuting 
one-para\-meter $\epg$-subgroups in $\Weyl$ is denoted by
$\erz\epg$.
\eunum
\edf
\blem
The product of finitely many, mutually commuting one-parameter 
$\epg$-sub\-groups in $\Weyl$ is a one-parameter $\erz\epg$-subgroup
in $\erz\Weyl$. Moreover, if $\pi$ is regular w.r.t.\ $\epg$,
then $\pi$ is regular w.r.t.\ $\erz\epg$.
\elem
\bpf
The first part is clear. 
For the second one use 
$\norm{\pi(\weyl_t)}_{\bound(\hilb)} \leq \norm{\weyl_t}_{\alg(\Weyl,\mu)} = 1$
for all $t$ to show 
\bglklein
\norm{\bigl(\prod_i \pi(\weyl_{i,t})\bigr) \psi - \psi}
 & \leq & \sum_j \norm{\bigl(\prod_{i<j} \pi(\weyl_{i,t})\bigr) (\pi(\weyl_{j,t})\psi - \psi)} \\
 & \leq & \sum_j \norm{\pi(\weyl_{j,t})\psi - \psi} \\
 & \gegen & 0
\eglklein
for $t\gegen 0$.
\qed
\epf
Therefore, in what follows, we will often assume that 
$\epg$ is replaced tacitly by $\erz\epg$.

%------------------------------------------------------------------------%
%            Abschnitt: Preliminaries                                    %
%------------------------------------------------------------------------%
\subsection{Splitting}
\blem
\label{lem:norm(diffeo_f)}
We have $\norm{\pi(\weyl'(f))}_{\bound(\hilb)} \leq \supnorm{f}$
for all $\weyl'\in\Weyl_{\home(\mu)}$ and all $f \in C(X)$.

Here, the equality holds if $\pi$ is faithful.
\elem
\bpf
We get
$\norm{\pi(\weyl'(f))}_{\bound(\hilb)}  
   \leq \norm{\weyl'(f)}_{\alg(\Weyl,\mu)} 
   = \supnorm{\weyl'(f)} = \supnorm{f}$,
since $\weyl'$ is a pull-back of a homeomorphism.
If $\pi$ is faithful, then even 
$\norm{\pi(\weyl'(f))}_{\bound(\hilb)} = \norm{\weyl'(f)}_{\alg(\Weyl,\mu)}$.
\qed
\epf

\blem
\label{lem:absch(w-1_diffeo_f)}

We have 
\bgl
 \sup_{\weyl'\in\Weyl_{\home(\mu)}}
 \betrag{\skalprod{\psi}{\pi((\weyl - \EINS)(w'(f))) \psi}_\hilb} 
 & \leq & 2 \: \supnorm{f} \: 
               \norm{\psi}_\hilb \:
               \norm{(\pi(\weyl) - \EINS) \psi}_\hilb     \\
\egl
for all $\psi\in\hilb$, $\weyl\in\Weyl$ and $f \in C(X)$.
\elem
\bpf
We have
\bgl
 &      & \betrag{\skalprod{\psi}{\pi((\weyl - \EINS)(f)) \psi}_\hilb} \\
 &  =   & \betrag{\skalprod{\psi}{\pi(\weyl(f)) \psi}_\hilb - 
                  \skalprod{\psi}{\pi(f) \psi}_\hilb} \\
 &  =   & \betrag{\skalprod{\psi}{\pi(\weyl) \pi(f) \pi(\weyl)^\ast \psi}_\hilb - 
                  \skalprod{\psi}{\pi(f) \psi}_\hilb} \\
 &  =   & \betrag{\skalprod{\pi(\weyl)^\ast \psi}{\pi(f) \pi(\weyl)^\ast \psi}_\hilb - 
                  \skalprod{\psi}{\pi(f) \psi}_\hilb} \\
 & \leq & \betrag{\skalprod{(\pi(\weyl)^\ast - \EINS) \psi}{\pi(f) \pi(\weyl)^\ast \psi}_\hilb} +
          \betrag{\skalprod{\psi}{\pi(f) (\pi(\weyl)^\ast - \EINS) \psi}_\hilb} \\
 & \leq & \norm{\pi(f)}_{\bound(\hilb)} \: 
               (\norm{\pi(\weyl)^\ast}_{\bound(\hilb)} + 1) \: \norm{\psi}_\hilb \:
               \norm{(\pi(\weyl)^\ast - \EINS) \psi}_\hilb      \\
 & \leq & 2 \: \norm{\pi(f)}_{\bound(\hilb)} \: 
               \norm{\psi}_\hilb \:
               \norm{(\pi(\weyl) - \EINS) \psi}_\hilb,     \erl{Unitarity of $\weyl$} \\
\egl
hence, for all $\weyl'\in\Weyl_{\home(\mu)}$,
\bgl
         \betrag{\skalprod{\psi}{\pi((\weyl - \EINS)(\weyl'(f))) \psi}_\hilb}   
 & \leq & 2 \: \norm{\pi(\weyl'(f))}_{\bound(\hilb)} \: 
               \norm{\psi}_\hilb \:
               \norm{(\pi(\weyl) - \EINS) \psi}_\hilb    \\
 & \leq & 2 \: \supnorm{f} \: 
               \norm{\psi}_\hilb \:
               \norm{(\pi(\weyl) - \EINS) \psi}_\hilb   
\egl
with Lemma \ref{lem:norm(diffeo_f)}.
\qed
\epf

\bdf
Let $\psi\in\hilb$ be some vector. 
\bunum
\item
Let $f\in C(X)$.

We say \df{$\Weyl'$ splits $\Weyl$ at $\psi$ for $f$} 
iff there is a one-parameter $\epg$-sub\-group
$\weyl_t$ in $\Weyl$, some $\varepsilon > 0$ and some
$t_0 > 0$, such that
\bgl
\sup_{\weyl'\in\Weyl'}
  \betrag{\skalprod{\psi}{\pi\bigl((\weyl_t - \EINS)(\weyl'(f))\bigr) \psi}_\hilb}
 & \geq & \varepsilon
\egl
for all $0 \neq \betrag t < t_0$.
\item
We say \df{$\Weyl'$ splits $\Weyl$ at $\psi$} iff there
is a continuous $\mu$-generating system $\ons$, 
such that $\Weyl'$ splits $\Weyl$ at $\psi$ 
for every $f\in\ons$ with $f \neq \EINS$ and 
$\skalprod{\psi}{\pi(f)\psi}_\hilb \neq 0$.
\eunum
\edf
\noindent
In other words, $\weyl_t$ is not uniformly weakly continuous on the 
$\Weyl'$-span of $f$. Moreover, note that the splitting property
actually refers to the choice of $\epg$. Since, in general,
we will have fixed $\epg$, we drop this notion here. 

\bprop
\label{prop:split->eqmeas}
Assume that $\pi \ident \pi'\einschr{\alg(\Weyl,\mu)}$ is regular 
(w.r.t.\ $\epg$).

If $\Weyl'$ splits $\Weyl$ at $\EINS_{\nu_0}$,
then $\mu_{\nu_0}$ equals $\mu$. 
\eprop
\bpf
Choose a continuous $\mu$-generating system $\ons$, 
such that
$\Weyl'$ splits $\Weyl$ at $\EINS_{\nu_0}$ for every non-constant $f \in \ons$
with $\skalprod{\EINS}{f}_{\nu_0} \ident \skalprod{\EINS_{\nu_0}}{\pi(f) \EINS_{\nu_0}}_\hilb \neq 0$.
Assume, there is such an $f$ with 
$\skalprod{\EINS}{f}_{\nu_0}
 \neq 0$.
Choose a one-parameter $\epg$-subgroup $\weyl_t$ in $\Weyl$,
some sufficiently small $\varepsilon > 0$ and
some $t_0 > 0$, such that

\bgl 
\sup_{\weyl'\in\Weyl'} 
 \betrag{\skalprod{\EINS_{\nu_0}}{\pi\bigl((\weyl_t - \EINS)(\weyl'(f))\bigr) \EINS_{\nu_0}}_\hilb}
 & \geq & \varepsilon
\egl
for all non-zero $\betrag t < t_0$.
Hence, using Lemma \ref{lem:absch(w-1_diffeo_f)},
\bgl
 2 \: \supnorm{f} \: 
               \norm{\pi(\weyl_t)\EINS_{\nu_0} - \EINS_{\nu_0}}_\hilb
 & \geq &
           \sup_{\weyl'\in\Weyl'}
             \betrag{\skalprod{\EINS_{\nu_0}}{\pi\bigl((\weyl_t - \EINS)(\weyl'(f))\bigr) \EINS_{\nu_0}}_\hilb} 
 \breitrel \geq  \varepsilon
\egl
for all non-zero $\betrag t < t_0$.
This, however, is a contradiction to our assumption that
$\pi$ is regular, i.e.,
$t \auf \pi(\weyl_t)$ is 
weakly continuous. Hence, 
$\skalprod{\EINS}{f}_{\nu_0} = 0$ 
for all $f$ in $\ons$. By Lemma \ref{lem:eq(mu-gen_syst)}, we have
$\mu = \mu_{\nu_0}$.
\qed
\epf

%------------------------------------------------------------------------%
%            Abschnitt: Preliminaries                                    %
%------------------------------------------------------------------------%
\subsection{$\mpsg$-Regularity}
\bdf
Let $A$ be any set.

\bunum
\item
A set $\mpsg$ is called \df{set of $A$-functions} iff
its elements are $A$-valued functions (i.e., there is no restriction
for the domains of these functions).
\item
A set $\mpsg$ of $A$-functions is called \df{topological (sequential)} iff
the domain of each $\mps\in\mpsg$ is a topological (sequential topological) 
space.
\eunum
\edf

\bdf
Let $\Weylx$ be some subset of a $C^\ast$-algebra $\alg$,
and let $\pi$ be a representation of $\alg$ on some Hilbert space $\hilb$.
Moreover, let $\mpsg$ be a set of topological $\Weylx$-functions.

Then $\pi$ is called \df{$\mpsg$-regular} iff the
mapping 
\zgl{\skalprod{\psi_1}{\pi(\mps(\:\cdot\:)) \psi_2}_\hilb : \dom\:\mps \nach \C}
is continuous for all $\psi_1,\psi_2\in\hilb$ and each $\mps \in \mpsg$.
\edf

\brem
The ordinary regularity uses 
$\dom\:\mps = \R$, where $\mps : t \auf \weyl_t$
runs over all one-parameter $\epg$-subgroups.
\erem

Let us return to the case that $\pi$ is a representation of $\alg(\Weyl,\mu)$
on $\hilb$.

\bprop
\label{prop:dense_lambda-reg}
Let $\pi$ be $\mpsg$-regular for some set $\mpsg$ of 
$\Weyl$-functions.
Fix for each $\mps\in\mpsg$
some subset $Y_{\mps}$ in $\dom\:\mps$,
such that
$\mps(Y_{\mps})$ consists of $\pi_\nu$-units only and 
$\bigcup_{\mps\in\mpsg} \mps(\quer{Y_{\mps}})$ generates $\Weyl$.

Then every $\weyl\in\Weyl$ is a $\pi_\nu$-unit. 
\eprop
\bpf
For all $\mps\in\mpsg$ and all $y\in Y_{\mps}$, we have
\zgl{
\skalprod{\EINS_{\nu}}{\pi(\mps(y)) \EINS_{\nu}}_\hilb =
 \skalprod{\EINS_{\nu}}{\EINS_{\nu}}_\hilb = 1.
} 
Consequently, by $\mpsg$-regularity, we even have 
$\skalprod{\EINS_{\nu}}{\pi(\mps(y)) \EINS_{\nu}}_\hilb = 1$ 
for all $y\in \quer{Y_{\mps}}$, hence   
$\pi(\mps(y)) \EINS_{\nu} = \EINS_\nu$, i.e.,
$\mps(\quer{Y_{\mps}})$ contains $\pi_\nu$-units only.
Since these sets generate full $\Weyl$ and since, obviously, 
products and inverses of $\pi_\nu$-units are $\pi_\nu$-units again, all
elements of $\Weyl$ are $\pi_\nu$-units.
\qed
\epf

%------------------------------------------------------------------------%
%            Abschnitt: Preliminaries                                    %
%------------------------------------------------------------------------%
\section{Quantum Geometric Background}
\label{sect:qgeo_backgr}
%------------------------------------------------------------------------%
%            Abschnitt: Preliminaries                                    %
%------------------------------------------------------------------------%
\subsection{Quantum Geometric Hilbert Space}
In the remaining sections we will apply the general framework of 
Section \ref{sect:gen_set} to quantum geometry. 
First, however, let us briefly 
recall in this subsection the basic facts and notations needed in the following.
General expositions can be found 
in \cite{a48,a30,a28} for the analytic framework. 
The smooth case is dealt with in \cite{d3,d17,e46}. 
The facts on hyphs and the conventions are due to \cite{paper3,diss,paper2+4}.

Let $\LG$ be some arbitrary connected compact Lie group 
and $M$ be some manifold. We let $M$ be equipped with an
arbitrary, but fixed differential structure. Later, we will restrict ourselves
to analytic (or, if so desired, semianalytic) manifolds. 
A path is a piecewise differentiable map from $[0,1]$ to $M$, whereas
differentiability is always understood in the 
chosen smoothness class. 
Moreover, we may restrict ourselves to use piecewise embedded paths only.
A path is trivial iff its image is a single point.
Two paths $\gamma_1$ and $\gamma_2$ are composable iff the end point 
$\gamma_1(1)$ of the first one coincides with the starting point $\gamma_2(0)$
of the second one. If they are composable, their product is given 
by 
\bgl
(\gamma_1 \gamma_2)(t) 
       \breitrel{:=} \begin{cases}
           \gamma_1(2t)   & \text{ for $t\in[0,\einhalb]$} \\
           \gamma_2(2t-1) & \text{ for $t\in[\einhalb,1]$}
           \end{cases}.
\egl\noindent
An edge $e$ is a path having no self-intersections, i.e., 
$e(t_1) = e(t_2)$ implies that $\betrag{t_1-t_2}$ either equals $0$ or $1$.
Two paths $\gamma_1$ and $\gamma_2$ 
coincide up to the parametrization iff there is some 
orientation preserving piecewise diffeomorphism 
$\phi: [0,1] \nach [0,1]$, such that $\gamma_1 = \gamma_2 \circ \phi$.
A path is called finite iff it equals up to the parametrization
a finite product of edges and trivial paths.
In what follows, every path will be assumed to be finite.
Next, two paths are equivalent iff there is a finite sequence of paths,
such that two subsequent paths coincide up to the parametrization or
up to insertion or deletion of retracings $\delta\delta^{-1}$. 
Finally, we denote the set of all paths by $\Pfgen$, that of all equivalence 
classes of paths by $\Pf$. The multiplication of paths 
naturally turns $\Pf$ into a groupoid. Usually
(but not 
in Subsections \ref{subsect:pathdecomp} and \ref{subsect:surf_flux}),
paths are understood to be equivalence classes of paths.

Initial and final segments of paths are naturally defined.
We will write $\gamma_1\BB\gamma_2$ iff there is some path $\gamma$ being
(possibly up to the parametrization) an initial path of both $\gamma_1$
and $\gamma_2$.
A hyph $\hyph$ is some finite collection 
$(\gamma_1,\ldots,\gamma_n)$ of edges each having a ``free'' point.
This means, for at least one direction none of the segments of $\gamma_i$ 
starting in that point in this direction, is a full segment
of some of the $\gamma_j$ with $j<i$.
Graphs and webs are special hyphs.
The subgroupoid generated (freely)
by the paths in a hyph $\hyph$ will be denoted by $\KG\hyph$.
Hyphs are ordered in the natural way. In particular, $\hyph' \leq \hyph''$ 
implies $\KG{\hyph'} \teilmenge \KG{\hyph''}$. 

The set $\Ab$ of generalized connections $\qa$ is now defined by
\bglklein
\Ab & := & \varprojlim_\hyph \Ab_\hyph \breitrel\iso \Hom(\Pf,\LG),
\eglklein
\noindent
with $\Ab_\gc := \Hom(\Pf_\gc,\LG) \teilmenge \LG^{\elanz\gc}$ 
given the topology which is induced by that of $\LG$,
for all finite tuples $\gc$ of paths. 
Moreover, we define the (always continuous) map
$\pi_\gc : \Ab \nach \LG^{\elanz\gc}$ 
by $\pi_\gc(\qa) := \qa(\gc) \ident h_\qa(\gc)$.
Note, that $\pi_\gc$ is surjective, if $\gc$ is a hyph.
Finally, for compact $\LG$,
the Ashtekar-Lewandowski measure $\mu_0$ is the unique
regular Borel measure on $\Ab$ whose push-forward $(\pi_\hyph)_\ast\mu_0$ 
to $\Ab_\hyph \iso \LG^{\elanz\hyph}$ 
coincides with the Haar measure there for every hyph $\hyph$. 
It is used to span the auxiliary Hilbert space $\haux := L_2(\Ab,\mu_0)$
of quantum geometry with scalar product $\skalprod\cdot\cdot$.

If we
included (generalized) gauge transforms into our considerations and 
studied the analytic category only, we could
use the spin-network states to get a basis of 
$\haux{}_{\mathrm{,inv}} = L_2(\AbGb,\mu_0)$ with $\Gb$ being the group
of generalized gauge transforms. Here, however,
we want to include gauge-variant functions as well and, moreover, do
not want to restrict the smoothness class at the beginning. Therefore,
we will consider now generating systems for $\haux$. 
For this, first of all, let us fix a representative in each equivalence 
class of irreducible representations of 
$\LG$, which we will refer to below. When considering matrix indices
for matrices on some Euclidean space $V$, we assume
that the underlying vectors are normalized. This means that for
all $A\in\End\:V$ we have $\betrag{A^i_j} \leq \norm{A}$, where $\norm\cdot$
denotes the standard operator norm.

\zusatzzeile
\bdf
\bunum
\item
For each non-trivial irreducible representation of $\LG$,
we define $\matfkt^\darst$ to be the set
\bglklein
 \matfkt^\darst & := & \bigcup_{m,n} \{\sqrt{\dim\darst} \:\: \darst^m_n\},
\eglklein
of normalized matrix functions,
where 
$m$, $n$ run over the set of matrix indices for $\darst$.
\item
We define $\matfkt$ to be the set
\bglklein
 \matfkt & := & \bigcup_{\darst} \matfkt^\darst
  \breitrel= \bigcup_{\darst,m,n} \{\sqrt{\dim\darst} \:\: \darst^m_n\},
\eglklein
of normalized matrix functions,
where $\darst$ runs over the set of all (equivalence classes of)
non-trivial irreducible representations of $\LG$.
\item
For every hyph $\hyph$ with edges $\gamma_1,\ldots,\gamma_I$ we define
the set $\matfkt_\hyph$ of \df{gauge-variant spin network states (gSN) of $\hyph$} by
\bglklein
 \matfkt_\hyph & := & \bigtensor_{i} \matfkt \circ \pi_{\gamma_i}.
\eglklein
If $\hyph$ is the empty hyph, we have $\matfkt_\hyph := \{1\}$.
The set of all gauge-variant spin network states will be denoted by
$\matfktgsn$.
\item
More compactly, we 
set $(T_{\darst,\gamma})^m_n := \sqrt{\dim\darst} \: \darst^m_n \circ \pi_\gamma$
and 
\bglklein
(T_{\vekt\darst,\gc})^{\vekt m}_{\vekt n} 
 \breitrel{:=} \sqrt{\dim\vekt\darst} \: \vekt\darst^m_n \circ \pi_\gc
 \breitrel\ident \bigtensor_k \sqrt{\dim\darst_k} \: (\darst_k)^{m_k}_{n_k} \circ \pi_{\gamma_k}.
\eglklein
\eunum
\edf
Observe that we get the same gauge-variant spin network state again if we
simultaneously revert the orientations of an arbitrary number of edges
and dualize the corresponding representations. This trivial overcompleteness
will be ignored in the following, i.e., we will always 
identify graphs and hyphs
differing in the ordering or the orientation of the edges only.

Let us now recall 
\blem
For every hyph $\hyph$, the set $\matfkt_\hyph$ of gauge-variant spin networks 
on $\hyph$ is an orthonormal set in $L_2(\Ab,\mu_0)$.
\elem
Note, that (even after admitting only one edge orientation per hyph) 
$\bigcup_\hyph \matfkt_\hyph$ is a generating system for, 
but {\em not}\/ an orthonormal set in $L_2(\Ab,\mu_0)$. 
This would still be the case, if we were in the (semi)analytic category and
use graphs only (see below).
In particular, we have

\blem
\label{lem:decomp_sns}
\bunum
\item
We have $\matfkt_{\hyph'} \teilmenge \spann\:\matfkt_\hyph$
for all $\hyph \geq \hyph'$.
\item
We have $\matfkt^\darst_\gamma \teilmenge \spann\:\matfkt^\darst_\hyph$
for all $\hyph = \{\gamma_1,\ldots,\gamma_n\}\geq \gamma$ with 
$\prod_i \gamma_i = \gamma$.

Here, 
$\matfkt^\darst_\hyph := \bigtensor_{i} \matfkt^\darst \circ \pi_{\gamma_i}$.
\eunum
\elem

\blem
\label{lem:sns=cont-erzs}
$\matfktgsn$ is a continuous $\mu$-generating system in $L_2(\Ab,\mu_0)$. 
\elem
\zusatzzeile[0.8]

Nevertheless, we will be looking for orthogonal decompositions of
$L_2(\Ab,\mu_0)$. For that purpose, we will have to single out orthogonal
subsets of gauge-variant spin network functions:
Until the end of this subsection we will now consider piecewise
analytic paths only.

In contrast to the standard, i.e.\ gauge-invariant spin network states,
the gauge-variant ones do not form an orthonormal basis for $L_2(\Ab,\mu_0)$
even after dropping some subset of them. The problem are the states arising
in the decomposition of an edge into a product of subedges, i.e.\
having two-valent vertices. In the gauge-invariant case they can be 
dropped since, by invariance, they reproduce the original state. Here,
however, in the gauge-variant case, we get a sum like
\bglklein
(T_{\gamma_1\gamma_2,\darst})^m_n & = & \inv{\sqrt{\dim\darst}} \:
 \sum_r \: (T_{\gamma_1,\darst})^m_r \tensor (T_{\gamma_2,\darst})^r_n
\eglklein\noindent
where the $(\dim\darst)$ gauge-variant spin network states 
together with that at the 
left-hand side span a $(\dim\darst)$-dimensional subspace of $L_2(\Ab,\mu_0)$.
We might simply drop the one at the left-hand side, but this would lead to
consistency 
troubles since we could want to decompose those at the right-hand side again.
A possible solution for this dilemma is given by the extended
spin network states as defined by Ashtekar and Lewandowski
in \cite{a13}. We do not want to introduce that notion here,
but only study the ``most dangerous'' cases in our framework -- namely,
those gSN with ``matching'' indices%
\footnote{Recall that a gSN is said to have 
``matching'' indices at a two-valent vertex $m$
iff the lower index, assigned to the incoming edge at $m$, 
and the upper index for the outgoing one are equal.
Note that we possibly have to invert orientations before, in order to have
an incoming and an outgoing edge at a two-valent vertex.}
at each two-valent vertex.
In the decomposition of the $\gamma_1\gamma_2$-state
above, this concerns the vector at $\gamma_1(1) = \gamma_2(0)$.
\blem
\label{lem:orthrel(sns)}
Let two gauge-variant spin networks states
$T := (T_{\vekt\darst,\gc})^{\vekt m}_{\vekt n}$ and 
$T' := (T_{\vekt\darst',\gc'})^{\vekt m'}_{\vekt n'}$ 
with graphs $\gc$ and $\gc'$ be given.

Then $T$ and $T'$ are orthogonal in $L_2(\Ab,\mu_0)$ if
\bunum
\item
$\im\gc \neq \im\gc'$;
\item
there is a point $m \in \inter\gc \cap \inter\gc'$,
such that the representations for the edges in $\gc$ and $\gc'$
running through $m$ do not coincide;
\item
there is some $m\in M$ being a two-valent vertex
with non-matching
indices for one and being interior for the other graph; or
\item
there is some $m\in M$ being a two-valent vertex for both graphs,
whereas both ``incoming'' or both ``outgoing'' indices 
are different.
\eunum
\elem
Note that matrix indices are regarded as different if they belong to
different representations.
\bpf
The first two cases are obvious.
The third one is clear observing our example above. Namely, 
decompose one of the graphs, say $\gc'$, by inserting $m$ as a vertex.
In the decomposition of $T'$
into a sum of gauge-variant spin network states of the enlarged graph,
the indices of every addend are matching. By the orthogonality
properties of matrix functions w.r.t.\ the Haar measure, 
we get the assertion.
The last case is now clear as well.
\qed
\epf
\bdf
Let $\gamma$ be an edge and $\darst$ be a 
non-trivial irreducible representation of $\LG$ and
let $T := (T_{\vekt\darst,\gc})^{\vekt m}_{\vekt n}$ be a gauge-variant
spin network state.
\bnum{2}
\item
If $\gamma$ is non-closed, then $T$ 
is called \df{$(\gamma,\darst)$-based} iff
\bunum
\item
$\gamma = \gamma_1 \circ \cdots \circ \gamma_{\elanz\gc}$;
\item
$\darst_k = \darst$ for all $k$; and
\item
all indices at two-valent vertices are matching, 
i.e., $m_{k+1} = n_k$ for all $k$.
\eunum
\item
If $\gamma$ is closed, then $T$ 
is called \df{$(\gamma,\darst)$-based} iff
\bunum
\item
$\gamma_1 \circ \cdots \circ \gamma_{\elanz\gc}$
equals $\gamma$ or equals 
$\gamma\einschr{[\tau,1]} \circ \gamma\einschr{[0,\tau]}$
for some $\tau\in(0,1)$;
\item
$\darst_k = \darst$ for all $k$; and
\item
all indices at two-valent vertices are matching, 
i.e., $m_{k+1} = n_k$ for all $k$ and $m_1 = n_{\elanz\gc}$.
\eunum
\enum
The set of all $(\gamma,\darst)$-based gauge-variant spin network
states will be denoted by $\bsns{\gamma,\darst}$.
Moreover, we set 
$\bsns\gamma := 
  \{\EINS\} \cup \bigcup_{\darst} \bsns{\gamma,\darst}$,
where the union runs over all non-trivial irreducible representations
of $\LG$. It contains precisely the \df{$\gamma$-based} gauge-variant
spin network states.
\edf
Note again that $T$ is $(\gamma,\darst)$-based if for {\em some}\/
orientation and {\em some}\/ ordering of $\gc$, the conditions above are
met.

\blem
\label{lem:decomp(bsns)}
$\bsns{\gamma,\darst}$ is orthogonal to its complement in the set of all
gauge-variant spin network states,
for every edge $\gamma$ and every irreducible representation $\darst$ of $\LG$.
\elem
\bpf
Let $T = (T_{\vekt\darst,\gc})^{\vekt m}_{\vekt n}$ 
be a gSN not contained in $\bsns{\gamma,\darst}$. 
If $\im\gc \neq \im\gamma$, the situation is clear. 
The same is true for $\darst_k \neq \darst$ for some $k$.
Let now $\im\gc = \im\gamma$ and $\darst_k = \darst$ for all $k$.
Then,
possibly after modifying ordering or orientations, we have 
$\gamma = \gamma_1 \cdots \gamma_n$. Moreover, every vertex of $\gc$
is at most two-valent. Thus, the proof follows from
Lemma \ref{lem:orthrel(sns)}.
\qed
\epf

\bcorr
\label{corr:decomp(hilb)-bsns}
For every edge $\gamma$, the 
Hilbert space $L_2(\Ab,\mu_0)$ is the closure of
\bglklein
\bigl(\bigdirsum_\darst \spann \: \bsns{\gamma,\darst} \bigr)
& \dirsum & \C \: \EINS \breitrel\dirsum
  \spann \: \bigl(\matfktgsn \setminus \bsns\gamma\bigr).
\eglklein
\ecorr

%------------------------------------------------------------------------%
%            Abschnitt: Preliminaries                                    %
%------------------------------------------------------------------------%
\subsection{Decomposition of Paths}
\label{subsect:pathdecomp}
In the following we will study the intersection behaviour between
paths and (generalized) surfaces. For this, we first consider how
paths can be decomposed.
Most of the relevant definitions and assertions are given in \cite{paper20}.
We will quote where appropriate and will simplify
some assumptions and, therefore, proofs. Note that in this subsection we will 
often distinguish between $\Pf$ and $\Pfgen$; paths here are genuine maps
from $[0,1]$ to $M$, not equivalence classes.

\subsubsection{Completeness}
\label{subsubsect:complete}

\bdf
Let $\gamma$ be some path. 

Then a finite 
sequence $\gc := (\gamma_1,\ldots,\gamma_n)$ in $\Pfgen$ is called
\df{decomposition} of $\gamma$ iff $\gamma_1 \cdots \gamma_n$ equals
$\gamma$ up to the parametrization.
\edf
This definition is well defined, since $\gamma_1 (\gamma_2\gamma_3)$
equals $(\gamma_1\gamma_2) \gamma_3$ up to the parametrization.
Moreover, observe that every reparametrization of $\gamma$ gives
a decomposition of $\gamma$.

If confusion is unlikely, we identify $\gamma_1\cdots\gamma_n$,
$\{\gamma_1,\ldots,\gamma_n\}$, and $(\gamma_1,\ldots,\gamma_n)$.

\bdf
Let $\gc := \gamma_1\cdots\gamma_I$ and 
$\gd := \delta_1\cdots\delta_J$
be decompositions of some path $\gamma$.

Then $\gc$ is a \df{refinement} of
$\gd$ iff there are
$0 = I_0 < I_1 < \ldots < I_J = I$, such that 
$\gamma_{I_{j-1}+1} \cdots \gamma_{I_j}$ is a decomposition of 
$\delta_j$ for all $j = 1, \ldots, J$.
We write%
\footnote{By a little misuse of notation we denote both graphs and decompositions
by $\gc$, $\gd$, etc., and denote 
both the relation on the set of hyphs (or graphs) and
that of refinement by $\geq$. Confusion should be unlikely.}
$\gc \geq \gd$ iff $\gc$ is a refinement of
$\gd$.
\edf
It can easily be shown \cite{paper20} that 
the set of all decompositions of a path $\gamma$ is directed w.r.t.\ $\geq$.

\bdf
\bunum
\item
A subset $\adm$ of $\Pfgen$ is called \df{hereditary} iff 
for each $\gamma\in\adm$
\bnum2
\item
the inverse of $\gamma$ is in $\adm$ again, and
\item
every decomposition of $\gamma$ consists of paths in $\adm$.
\enum
\item
A subset $\adm$ of $\Pfgen$ is called
\df{complete} iff it is hereditary 
and every path in $\Pfgen$ has a decomposition into paths 
in $\adm$.
\eunum
\edf
A decomposition consisting of paths in $\adm$ only, will be called
$\adm$-decomposition.

\blem
Let $\adm \teilmenge \Pfgen$ be complete.

Then for every hyph $\hyph$ there is a hyph $\hyph' \geq \hyph$ 
with $\hyph' \teilmenge \adm$.
\elem
\bpf
First decompose each $\gamma\in\hyph$ into paths in $\adm$.
Collect all these paths in a set $\gc'\geq\hyph$.
Since $\gc'$ may be not a hyph again, refine, if necessary, the paths in $\gc'$
further to get a hyph $\hyph'\geq\gc'\geq\hyph$ \cite{paper3}. By completeness,
$\hyph'$ contains only paths in $\adm$.
\qed
\epf

\blem
The set of all edges and trivial paths in $\Pfgen$ is complete.
\elem

%------------------------------------------------------------------------%
%            Abschnitt: Preliminaries                                    %
%------------------------------------------------------------------------%
\subsubsection{Main Construction}
\label{subsubsect:construction}

%------------------------------------------------------------------------%
%            Konstruktion: neue Zusammenh"ange                           %
%------------------------------------------------------------------------%
\bdf
\label{def:germ}
Let $\adm$ be some hereditary subset of $\Pfgen$. 

Then a map $\qgerm : \adm \nach \LG$ is called
\df{$\adm$-germ} iff for all $\gamma\in\adm$
\bnum2
\item
\label{punkt:inv(qgerm)}
$\qgerm(\gamma^{-1}) = \qgerm(\gamma)^{-1}$, and
\item
\label{punkt:homom(qgerm)}
$\qgerm(\gamma) = \qgerm(\gamma_1) \qgerm(\gamma_2)$
for all decompositions $\gamma_1\gamma_2$ of $\gamma$.
\enum
The set of all $\adm$-germs from $\adm$ to $\LG$
is denoted by $\Germ(\adm,\LG)$.
\edf
Observe that $\qgerm(\gamma)$ and $\qgerm(\delta)$ coincide if 
$\gamma$ and $\delta$ coincide up to the parametrization. In fact, since
every decomposition $\gamma_1 \gamma_2$ of $\gamma$ is also some
for $\delta$, we may apply property \ref{punkt:homom(qgerm)} above.

Note that we will shortly speak about germs instead of $\adm$-germs,
provided the domain $\adm$ is clear from the context.

\bprop
\label{prop:allgkonstr(zush)}
Let $\adm$ be some complete subset of $\Pfgen$,
and 
let $\qgerm: \adm \nach \LG$ be a germ.

Then we have:
\bunum
\item
There is a unique germ $\neuzh\qgerm : \Pfgen \nach \LG$
extending $\qgerm$.

\item
The map $\neuzh\qgerm$ is given by
\bglklein
\neuzh\qgerm(\gamma) & = & \prod_{i=1}^I \qgerm(\gamma_i)
\eglklein
for each $\gamma\in\Pfgen$, where 
$\gamma_1\cdots\gamma_n$ is any%
\footnote{Recall that, by completeness of $\adm$, 
such a decomposition exists always.}
$\adm$-decomposition of $\gamma$.
\item
The map $\neuzh\qgerm$ is constant on equivalence classes in $\Pfgen$.
\item
The induced map
$[\neuzh\qgerm] : \Pf \nach \LG$ is a homomorphism, 
i.e., it is an element of $\Ab$.
\eunum
\eprop

\neueseite

\bpf
Let us first define the desired map $\neuzh\qgerm$ as given in the 
proposition above and now check its properties.
\bnum{3}
\item
$\neuzh\qgerm$ does not depend on the choice of the 
$\adm$-decomposition.

Let $\gc$ and $\gd$ be two $\adm$-decompositions of $\gamma$.
Since, by assumption, 
every path in $\adm$ has $\adm$-decompositions only, and
since the set of decompositions of a path is directed w.r.t.\ $\geq$,
we may assume $\gc \geq \gd$.
But, in this case the well-definedness follows directly 
from the definitions and germ property \ref{punkt:homom(qgerm)} 
of $\qgerm$.
\item
$\neuzh\qgerm$ is constant on equivalence classes in $\Pfgen$.

Let $\gamma$ and $\delta$ in $\Pfgen$ be equivalent. By definition,
it is sufficient to check the following two cases:
\bunum
\item
$\gamma$ and $\delta$ coincide up to the parametrization.

This case is trivial, since every $\adm$-decomposition of $\gamma$
is also one of $\delta$. Hence, $\qgerm(\gamma) = \qgerm(\delta)$.
\item
There is some $\varepsilon$ in $\Pfgen$ and some decomposition 
$\gamma_1 \gamma_2$ of $\gamma$, such that 
$\delta$ equals the product 
of $\gamma_1$, $\varepsilon$, $\varepsilon^{-1}$ and $\gamma_2$.

Now, in this case, 
choose some $\adm$-decompositions $\varepsilon_1 \cdots \varepsilon_K$
of $\varepsilon$ and $\gamma_{s1}\cdots\gamma_{sI_s}$ of $\gamma_s$ with
$s=1,2$. Then 
$\gamma_{11}\cdots\gamma_{1I_1} \: \gamma_{21}\cdots\gamma_{2I_2}$
is a $\adm$-decomposition of $\gamma$ and
$\gamma_{11}\cdots\gamma_{1I_1} \: \varepsilon_1 \cdots \varepsilon_K \:
 \varepsilon_K^{-1} \cdots \varepsilon_1^{-1} \: \gamma_{21}\cdots\gamma_{2I_2}$
one of $\delta$.
Hence, we have
\bglklein
\neuzh\qgerm(\delta) 
  & = & \qgerm(\gamma_{11}) \cdots \qgerm(\gamma_{1I_1}) \: 
         \qgerm(\varepsilon_1) \cdots \qgerm(\varepsilon_K) \:
\\&&\hspace*{3em}
         \qgerm(\varepsilon_K^{-1}) \cdots \qgerm(\varepsilon_1^{-1}) \: 
         \qgerm(\gamma_{21}) \cdots \qgerm(\gamma_{2I_2}) 
        \erl{Definition of $\neuzh\qgerm$} \\
  & = & \qgerm(\gamma_{11}) \cdots \qgerm(\gamma_{1I_1}) \: 
         \qgerm(\gamma_{21}) \cdots \qgerm(\gamma_{2I_2}) 
        \erl{Property \ref{punkt:inv(qgerm)} of $\qgerm$} \\
  & = & \neuzh\qgerm(\gamma).
        \erl{Definition of $\neuzh\qgerm$}
\eglklein
\eunum
\item
$\neuzh\qgerm$ is a germ extending $\qgerm$, 
and $[\neuzh\qgerm]$ is a homomorphism.

This is proven as the statements above.
\item
$\neuzh\qgerm$ is the only germ extending $\qgerm$.

If $\neuzh\qgerm'$ is some other germ extending $\qgerm$ different
from $\neuzh\qgerm$, then there is some $\gamma\in\Pfgen$ with 
$\neuzh\qgerm'(\gamma) \neq \neuzh\qgerm(\gamma)$. Now, choose
a $\adm$-decomposition $\gamma_1 \cdots \gamma_I$ of $\gamma$. By
the properties of a germ, there is some $i$ with
$\neuzh\qgerm'(\gamma_i) \neq \neuzh\qgerm(\gamma_i)$. However, since both
$\neuzh\qgerm'$ and $\neuzh\qgerm$ extend $\qgerm$, both sides
are equal to $\qgerm(\gamma_i)$. Contradiction.
\qed
\enum
\epf

\bprop
\label{prop:allgqfa_cont}
Let $\adm$ be some complete subset of $\Pfgen$.
Let $X$ be some topological space, and let $\qwert : X \nach \Germ(\adm,\LG)$
be some map.
Finally, assume that the map $\bigl(\qwert(\cdot)\bigr)(\gamma) : X \nach \LG$
is continuous for all $\gamma\in\adm$.

Then 
\fktdefabgesetzt{\qfa_\qwert}{X}{\Ab}{x}{[\neuzh{\qwert(x)}]}
is continuous, 
where $\:\neuzh\cdot\:$ is given as in Proposition \ref{prop:allgkonstr(zush)}.
\eprop
\bpf
It is sufficient \cite{paper2+4} to prove that 
$\pi_{\gamma} \circ \qfa_\qwert : X \nach \LG$
is continuous for all edges $\gamma$. Since the multiplication in $\LG$
is continuous and $\adm$ is complete, 
we even may restrict ourselves to the cases 
of $\gamma \in \adm$.
Here, however, the assertion
follows immediately from
\bgl
(\pi_{\gamma} \circ \qfa_\qwert) (x) 
   & \ident & \pi_{\gamma} ([\neuzh{\qwert(x)}]) 
   \breitrel= [\neuzh{\qwert(x)}]([\gamma]) 
   \breitrel= \neuzh{\qwert(x)}(\gamma) 
   \breitrel\ident \qwert(x)(\gamma),
\egl
i.e., $\pi_{\gamma} \circ \qfa_\qwert = \bigl(\qwert(\cdot)\bigr)(\gamma)$ 
for all $\gamma\in\adm$.
\qed
\epf
\blem
\label{lem:coinc_krit}
Two generalized connections coincide iff they coincide for all
(equivalence classes of) paths of a complete subset of $\Pfgen$.
\elem

\neueseite

%------------------------------------------------------------------------%
%            Abschnitt: Preliminaries                                    %
%------------------------------------------------------------------------%
\subsubsection{Application to Weyl-Type Operators}
\label{sect:main}

\bdf
Let $\adm$ be some hereditary subset of $\Pfgen$.

Then a map $\rrr : \adm \nach \LG$ is called \df{admissible} iff
\bunum
\item
$\rrr(\delta_1) = \rrr(\delta_2)$
for all $\delta_1,\delta_2\in\adm$
with $\delta_1 \BB \delta_2$,
and
\item
$\rrr(\gamma_1^{-1}) = \rrr(\gamma_2)$
for all $\gamma\in\adm$ and all decompositions $\gamma_1\gamma_2$ of $\gamma$.
\eunum
\edf
Most relevant for the well-definedness of the Weyl operators to be introduced
below, will be
\bthm
\label{thm:cont+masz(qfa)}
Let $\adm$ be a complete subset of $\Pfgen$
and $\rrr : \adm \nach \LG$ an admissible map.

Then there is a unique map $\qfa : \Ab \nach \Ab$,
such that, for all $\gamma \in \adm$,
\bgl
h_{\qfa(\qa)}([\gamma])  & = & 
\rrr(\gamma)^{-1} \: h_\qa([\gamma]) \: \rrr(\gamma^{-1}).
\egl
Moreover, $\qfa$ is a homeomorphism preserving
the Ashtekar-Lewandowski measure $\mu_0$. Hence, 
the pull-back  
$\qfa^\ast : C(\Ab) \nach C(\Ab)$ is an isometry and the induced operator
on $\bound(L_2(\Ab,\mu_0))$ is well defined and unitary.
\ethm
An more general version is proven in \cite{paper20}. We replay the
corresponding proof.
\bpf
\bunum
\item
Define $\qwert : \Ab \nach \Maps(\adm,\LG)$ by%
\footnote{From now on, we will drop the square brackets 
in all $h_\qa([\ldots])$.}
\bgl
\bigl(\qwert(\qa)\bigr)(\gamma)  & = & 
\rrr(\gamma)^{-1} \: h_\qa (\gamma) \: \rrr(\gamma^{-1}).
\egl
\item
First we show that $\qwert(\qa)$ is indeed in $\Germ(\adm,\LG)$
for all $\qa\in\Ab$. 

In fact,
for all $\gamma\in\adm$ and all decompositions $\gamma_1\gamma_2$ of $\gamma$,
we have
\bgl
\bigl(\qwert(\qa)\bigr)(\gamma^{-1})
 & = & \rrr(\gamma^{-1})^{-1} \: h_\qa (\gamma^{-1}) \: \rrr(\gamma) \\
 & = & \bigl(\rrr(\gamma)^{-1} \: h_\qa (\gamma) \: \rrr(\gamma^{-1})\bigr)^{-1} 
 \breitrel= \bigl(\qwert(\qa)(\gamma)\bigr)^{-1}
\egl
and
\bgl
\bigl(\qwert(\qa)\bigr)(\gamma)
 & = & \rrr(\gamma)^{-1} \: h_\qa (\gamma) \: \rrr(\gamma^{-1}) \\
 & = & \rrr(\gamma_1\gamma_2)^{-1} \: h_\qa (\gamma_1) \:
       h_\qa (\gamma_2) \: \rrr(\gamma_2^{-1}\gamma_1^{-1}) \\
 & = & \rrr(\gamma_1)^{-1} \: h_\qa (\gamma_1) \:
       \rrr(\gamma_1^{-1}) \: \rrr(\gamma_2)^{-1} \:
       h_\qa (\gamma_2) \: \rrr(\gamma_2^{-1}) \\
 & = & \bigl(\qwert(\qa)\bigr)(\gamma_1) \: \bigl(\qwert(\qa)\bigr)(\gamma_2).
\egl
Here, we used the admissibility of $\rrr$ with 
$\gamma_1 \BB \gamma_1 \gamma_2$
and $\gamma_2^{-1} \gamma_1^{-1} \BB \gamma_2^{-1}$.
\item
Next, observe that, for every fixed $\gamma\in\adm$,
\bgl
\bigl(\qwert(\qa)\bigr)(\gamma)
 & = & \rrr(\gamma)^{-1} \: h_{\qa}(\gamma) \: \rrr(\gamma^{-1})
 \breitrel\ident \rrr(\gamma)^{-1} \: \pi_{\gamma} (\qa) \: \rrr(\gamma^{-1})
\egl
depends continuously on $\qa$, by definition of the projective-limit topology
on $\Ab$.
\item
Now, by Proposition \ref{prop:allgqfa_cont},
$\qfa := [\neuzh{\qwert(\cdot)}] : \Ab \nach \Ab$ is continuous, whereas
for $\gamma\in\adm$
\bgl
h_{\qfa(\qa)}(\gamma) 
 \breitrel\ident \bigl(\qfa(\qa)\bigr)([\gamma]) 
 & = & [\neuzh{\qwert(\qa)}] ([\gamma]) 
 \breitrel= \rrr(\gamma)^{-1} \: h_\qa (\gamma) \: \rrr(\gamma^{-1}).
\egl
The uniqueness of $\qfa$ follows from the completeness of $\adm$.
\item
To prove that $\qfa$ is a homeomorphism, we explicitly describe the
inverse of $\qfa$. 
Define $\rrr' : \adm \nach \LG$
by $\rrr'(\gamma) := \rrr(\gamma)^{-1}$.
It is easy to check that $\rrr'$ is admissible. 
As already proven above, there is a unique continuous map
$\qfa' : \Ab \nach \Ab$ with 
\bgl
h_{\qfa'(\qa)}(\gamma)  & = & 
\rrr'(\gamma)^{-1} \: h_\qa(\gamma) \: \rrr'(\gamma^{-1})
\egl
for all $\gamma\in\adm$. Altogether, this gives
\bgl
h_{\qfa'(\qfa(\qa))}(\gamma)  
 & = & \rrr'(\gamma)^{-1} \: 
          h_{\qfa(\qa)} (\gamma) \: 
          \rrr'(\gamma^{-1}) \\
 & = & \rrr'(\gamma)^{-1} \: 
          \rrr(\gamma)^{-1} \: 
          h_\qa(\gamma) \:
          \rrr(\gamma^{-1}) \:
          \rrr'(\gamma^{-1}) \\
 & = & h_\qa(\gamma) 
\egl
for all $\gamma\in\adm$.
The completeness of $\adm$ and Lemma \ref{lem:coinc_krit} prove
$\qfa' \circ \qfa = \ido_\Ab$. Analogously, one shows 
$\qfa \circ \qfa' = \ido_\Ab$.
\item
$\qfa$ even preserves the Ashtekar-Lewandowski measure.

In fact, let $\hyph$ be an arbitrary, but fixed hyph. 
By completeness, there is some hyph $\hyph'\geq\hyph$ with 
$Y'$ edges and $\hyph'\teilmenge \adm$.
By construction, we have 
\bgl
\pi_{\hyph'} \circ \qfa  & = &  
   (\qfa_{\gamma_1} \kreuz \cdots \kreuz \qfa_{\gamma_{\Hyph'}}) \circ \pi_{\hyph'}
\egl
with 
$\qfa_\gamma (g) := 
  \rrr(\gamma)^{-1} \: g \: \rrr(\gamma^{-1})$ for $\gamma\in\adm$.
In other words, each $\qfa_\gamma$ consists of a left and a right translation, 
whence the Haar measure on $\LG$ is $\qfa_\gamma$-invariant.
Since $\pi_\hyph^{\hyph'} \circ \pi_{\hyph'} = \pi_\hyph$ with continuous
$\pi_\hyph^{\hyph'} : \Ab_{\hyph'} \nach \Ab_\hyph$ and
since $(\pi_{\hyph'})_\ast \mu_0$ is the $Y'$-fold product of the 
Haar measure on $\LG$, we get
\bgl
(\pi_\hyph)_\ast (\qfa_\ast \mu_0) 
 & = & (\pi_\hyph^{\hyph'})_\ast (\pi_{\hyph'} \circ \qfa)_\ast \mu_0  \\
 & = & (\pi_\hyph^{\hyph'})_\ast (\qfa_{\gamma_1} \kreuz \cdots \kreuz \qfa_{\gamma_{\Hyph'}})_\ast
       (\pi_{\hyph'})_\ast \mu_0 \\
 & = & (\pi_\hyph^{\hyph'})_\ast (\qfa_{\gamma_1} \kreuz \cdots \kreuz \qfa_{\gamma_{\Hyph'}})_\ast
       \mu_\Haar^{\Hyph'} \\
 & = & (\pi_\hyph^{\hyph'})_\ast \mu_\Haar^{\Hyph'} \\
 & = & (\pi_\hyph^{\hyph'})_\ast (\pi_{\hyph'})_\ast \mu_0 \\
 & = & (\pi_\hyph)_\ast \mu_0.
\egl
Since finite regular Borel measures on $\Ab$ 
coincide iff their push-forwards w.r.t.\ 
all $\pi_\hyph$ coincide, we get the assertion.
\qed
\eunum
\epf

We get immediately
\bcorr
\label{corr:cont(qfa)voll}
Let $\adm$ be some complete subset of $\Pfgen$.
Moreover, let $Y$ be some topological space
and let $\rrr : \adm \kreuz Y \nach \LG$
be some map, such that 
\bunum
\item
$\rrr(\cdot,y) : \adm \nach \LG$ is admissible for all $y\in Y$, and
\item
$\rrr(\gamma,\cdot) : Y \nach \LG$ is continuous for all $\gamma\in\adm$.
\eunum
Then there is a unique map $\qfa : \Ab \kreuz Y \nach \Ab$ with 
\bgl
h_{\qfa(\qa,y)}(\gamma) 
  & = & \rrr(\gamma,y)^{-1} \: h_{\qa}(\gamma) \: 
                 \rrr(\gamma^{-1},y) 
\egl
for all $\gamma \in \adm$. Moreover, $\qfa$ is continuous.
\ecorr

%------------------------------------------------------------------------%
%            Abschnitt: Preliminaries                                    %
%------------------------------------------------------------------------%
\subsection{Surfaces and Fluxes}
\label{subsect:surf_flux}
Originally (see, e.g., \cite{d64}), the action of flux operators on 
cylindrical functions has been
given by self-adjoint differential operators. 
Since these operators are unbounded, 
one has to study their domains very carefully. To avoid this problem,
one usually considers them as generators of unitary, i.e.\  bounded
operators. Now, the flux operators turn into some sort of translation
operators. In this section, we are going to shift this action
to a still deeper level. We will see that it can be regarded as the
pull-back of some continuous action of translations on $\Ab$ itself.

%------------------------------------------------------------------------%
%            Abschnitt: Preliminaries                                    %
%------------------------------------------------------------------------%
\subsubsection{Quasi-Surfaces}
Before we can define this action we study how paths are decomposed by
surfaces. 
\bdf
Let $S$ be a subset of $M$.
\bunum
\item
A path $\gamma\in\Pfgen$ is called \df{$S$-external} iff $\:(\inter\gamma) \cap S = \leeremenge$.
\item
A path $\gamma\in\Pfgen$ is called \df{$S$-internal} iff $\:\inter\gamma \teilmenge S$. 
\eunum
\edf
Observe that the end points of an $S$-external path may be contained in $S$. It
is only required for the ``interior part'' of the path, i.e., for all $\gamma(t)$ with
$0 < t < 1$ to be outside of $S$.
If $S$ is clear from the context, we simply speak about external 
and internal edges. 
\bdf
Let $S$ be some subset of $M$.

Then $\adm_S$ denotes
the set of all paths that
are $S$-external or $S$-internal.
\edf

\bdf
Let $S$ be a subset of $M$ and $\gamma\in\Pfgen$ be an edge. 

Then a decomposition $\gc$ of $\gamma$ is called \df{$S$-admissible}
iff $\gc \teilmenge \adm_S$.
\edf
In other words, 
$\gc = (\gamma_1,\ldots,\gamma_I)$ is 
$S$-admissible 
iff $\gamma$ equals $\gamma_1 \cdots \gamma_I$ up to the parametrization and
each $\gamma_i$ is $S$-internal or $S$-external.

\blem
\label{lem:compl_crit}
Let $S$ be a subset of $M$.

Then $\adm_S$ is complete, if every edge has an $S$-admissible
decomposition.
\elem
\bpf
Heredity is clear. The completeness follows since
any (finite) path can be decomposed into a product of edges and trivial paths,
hence, by assumption, into a product of $S$-external or $S$-internal paths.
\qed
\epf

\bdf
A subset $S$ of $M$ is called \df{quasi-surface}
iff every edge $\gamma\in\Pfgen$ has an $S$-admissible decomposition.
\edf
Examples for quasi-surfaces, 
in case we are in the (semi)analytic category for the paths,
are embedded analytic submanifolds that are even semianalytic.%
\footnote{This, however, is no longer true if we drop 
the semianalyticity 
(for its definition see Subsection \ref{subsect:stratif_anal_iso}). In fact,
consider $\R^2$ and a smooth path $\gamma$ in the closed half-plane
$y\leq0$, such that $\gamma$ connects $(-1,0)$ and $(+1,0)$ and intersects
the straight line $\delta$ between these two points infinitely often without
sharing a full segment. (See similar constructions, e.g., 
in \cite{d17,paper15}.). Now define $S$ to be the upper one of the 
two open sets in $\R^2$ bounded by $\gamma$, by $x=-1$ and by $x=+1$.
Of course, $S$ is an embedded analytic manifold, although 
it is not semianalytic in $\R^2$.
Nevertheless,
$\delta$ leaves $S$ and returns into it infinitely often. Therefore,
there is no $S$-admissible decomposition of $\delta$, whence $S$
is not a quasi-surface.} 
Note that these submanifolds may have any dimension. Therefore, any collection
of points having no accumulation point is a quasi-surface. This even remains true
in the category of piecewise smooth paths. 

On the other hand, there are indeed non-semianalytic
submanifolds that are quasi-surfaces. Consider, e.g., the smooth function $f$ on $\R$ with
$f(x) := \e^{-1/x^2}$
for $x \neq 0$ and $f(0) := 0$.
Of course, it is analytic everywhere except for $x = 0$. But, its graph $S$ does not form a 
semianalytic submanifold in, for simplicity, $\R^2$. Nevertheless, 
it is a quasi-surface. In fact, let $\gamma$ be a piecewise analytic path in $\R^2$. 
If it does not run through
the origin, the statement is trivial. Assume now that $\gamma$ runs through the origin. 
Decomposing $\gamma$ appropriately, if necessary, we may restrict ourselves to the case
of an analytic $\gamma$ starting at the origin without returning there at any other parameter time. 
Assume next that $\gamma$ has infinitely many intersection points with $S$, and let 
the origin $0$ be an accumulation point for $\inter\gamma \cap S$. W.l.o.g.,%
\footnote{Otherwise, restrict the domain of $\gamma$, such that the $x$-component of $\dot\gamma$ 
is non-zero everywhere. If this is not possible, the $x$-component of $\dot\gamma$ vanishes
at $t= 0$. But, then $\gamma$ is $S$-external anyway, at least locally.}
we may consider, finally, $\gamma$ to be the graph of an analytic function on $\R$, again denoted
by $\gamma$. Use now the fact that two $C^\infty$ functions $f_1, f_2$ have identical Taylor
coefficients at $0$ if $0$ is an accumulation point of $f_1 = f_2$, to derive that
$\gamma$ has only zero Taylor coefficients, just because $f$ does. 
Now, analyticity implies that $\gamma$ is
a straight edge along the $x$-axis never intersecting $S$ again.
Using this contradiction, the statement is now trivial.

If we would like to take even more quasi-surfaces into account, we
may reduce the set of paths under consideration.
This might be relevant, e.g., in the case of piecewise linear paths, although there
usually also the set of manifolds is restricted to that of piecewise linear submanifolds
a priori.

The punctures leading to an $S$-admissible decomposition will be relevant
for the definition of Weyl operators. In particular, these operators
depend on the transversality properties between the path and the (oriented)
hypersurface. Therefore, we need to introduce a general notion for the
properties an orientation should encode.

\bdf
Let $S$ be a quasi-surface of $M$.

\bunum
\item
A function
$\sigma_S: \Pfgen \nach \Z$ is called 
\bunum
\item
\df{outgoing intersection function for $S$} iff we have 
\bnum2
\item
$\sigma_S(\gamma) = 0$ if $\gamma(0) \nichtin S$ and
\item
$\sigma_S(\gamma) = \sigma_S(\gamma')$ 
\enum
for all $\gamma,\gamma'\in\Pfgen$ with $\gamma\BB\gamma'$;
\item
\df{incoming intersection function for $S$} iff we have 
\bnum2
\item
$\sigma_S(\gamma) = 0$ if $\gamma(1) \nichtin S$ and
\item
$\sigma_S(\gamma) = \sigma_S(\gamma')$ 
\enum
for all $\gamma,\gamma'\in\Pfgen$ with $\gamma\EE\gamma'$.
\eunum
\item
An outgoing intersection function $\sigma^\ausl_S$ 
and an incoming intersection function $\sigma^\einl_S$
are called \df{compatible}
iff 
$\sigma^\ausl_S(\gamma) + \sigma^\einl_S(\gamma^{-1}) = 0$
for all $\gamma\in\Pfgen$.
\eunum
\edf
For brevity, we will denote a compatible pair 
$(\sigma^\ausl_S,\sigma^\einl_S)$ 
of an outgoing and an incoming intersection function by $\sigma_S$
and call it intersection function for $S$. Even more,
we use $\sigma_S$ and $\sigma_S^\ausl$ synonymously.
Sometimes, we write
$\sigma(S,\gamma)$ instead of $\sigma_S(\gamma)$ to emphasize that
the intersection function may depend on quasi-surface and path as well.
\bdf
Let $S$ be a quasi-surface of $M$, and let $\sigma_S : \Pfgen \nach \Z$ 
be some intersection function for $S$.

Then the intersection function $-\sigma_S$ is called
\df{inverse} to $\sigma_S$.
\edf
\bdf
Let $S$ be a quasi-surface with intersection function $\sigma_S$,
and let $\gamma\in\Pfgen$ be some path. Assume, moreover, that 
there are only finitely many $\tau_i\in[0,1]$ with $\gamma(\tau_i) \in S$.

We say that the \df{orientation of $S$ coincides with} the \df{direction of $\gamma$}
iff 
$\sigma^\ausl_S(\gamma\einschr{[\tau_i,1]}) = 1$ for all $\tau_i \neq 1$
and 
$\sigma^\einl_S(\gamma\einschr{[0,\tau_i]}) = 1$ for all $\tau_i \neq 0$.
\edf

In our applications, we will, e.g., define $\sigma_S(\gamma)$ 
for an $S$-external path $\gamma$ to be $\pm1$ (depending
on the direction of $\gamma$),
if its initial path intersects $S$ transversally,
and equal to $0$, otherwise:

\bdf
Let $S$ be an oriented (embedded)
hypersurface in $M$ being a quasi-surface of $M$. Then we have:
\bnum2
\item
The \df{natural} intersection function $\sigma_S : \Pfgen \nach \Z$ is defined
by:
\bunum
\item
$\sigma_S(\gamma) = 0$ if
$\gamma(0) \nichtin S$ or $\dot\gamma(0)$ is tangent to $S$;
\item
$\sigma_S(\gamma) = \pm1$ if
$\gamma(0) \in S$ and $\dot\gamma(0)$ is not tangent to $S$ and 
some initial path of $\gamma$ lies (except $\gamma(0)$)
above (below) $S$.
\eunum
\item
The \df{topological} intersection function $\sigma^\tif_S: \Pfgen \nach \Z$ is 
defined as follows:
\bunum
\item
$\sigma^\tif_S(\gamma) = 0$ if
$\gamma(0) \nichtin S$ or some initial path of $\gamma$ is contained in $S$;
\item
$\sigma^\tif_S(\gamma) = \pm1$ if
$\gamma(0) \in S$ and no initial path of $\gamma$ is contained in $S$
and some initial path of $\gamma$ lies (except $\gamma(0)$)
above (below) $S$.
\eunum
\enum
\edf
Here, ``above'' and ``below'' refer to the orientation of $S$. Moreover,
initial paths w.r.t.\ a trivial interval are not taken into consideration.
It is easy to check that this definition is well defined. Moreover, 
obviously, for every orientable $S$ there are precisely two natural
(and two topological) intersection functions corresponding to the
two choices of orientations. They coincide up to the sign.

If $S$ is a submanifold of codimension larger than $1$, there
is no longer just a pair of natural orientations. 
Nevertheless, in view of the applications
we aim at, we may define ``natural'' orientations:
\bdf
\label{def:ind_nat_orient}
Let $S$ be some embedded submanifold of $M$ being a quasi-surface
of $M$ and having codimension $2$ or higher. 

Then an intersection function $\sigma_S : \Pfgen \nach \Z$ 
is called \df{natural (topological)} iff there is some oriented 
embedded hypersurface $S'$ in $M$ being a quasi-surface and
having $\sigma_S$ as its natural (topological) intersection function.
\edf
One sees immediately, that the number of natural intersection functions
of such quasi-surfaces with higher codimension may be rather large. 
For instance, let $S$
be (a bounded part of) a line in $\R^3$. Then we may take all the
full circles in $\R^3$ having $S$ as its diameter. Of course, there is
a continuum of such circles each having another pair of 
natural or topological intersection functions.

\bdf
\bunum
\item
A quasi-surface $S'$ is called \df{quasi-subsurface} 
of some quasi-surface $S$ iff $S'$ is contained in $S$.
\item
Let $S'$ be a quasi-subsurface of a quasi-surface $S$ having
intersection function $\sigma_S$.
Then an intersection function $\sigma_{S'}$ 
is called \df{induced by $\sigma_{S}$}
iff $\sigma_S (\gamma) = \sigma_{S'} (\gamma)$ for all
$\gamma$ with $\gamma(0) \in S'$.
\eunum
\edf
Definition \ref{def:ind_nat_orient} gives an example
for the induction of intersection functions.

\blem
The complement of a quasi-surface is a quasi-surface.
\elem
\bpf
An $S$-admissible decomposition of an edge is also 
($M \setminus S$)-admissible.
\qed
\epf

\blem
If $S_1$ and $S_2$ are quasi-surfaces,
then $S_1 \cup S_2$ and $S_1 \cap S_2$ are quasi-surfaces.
\elem
\bpf
If $\gamma$ is some edge, decompose each path of some $S_1$-admissible 
decomposition w.r.t.\ $S_2$. It is easy to check
that this leads to an $S$-admissible
decomposition of $\gamma$
with $S$ being $S_1 \cup S_2$ or $S_1 \cap S_2$.
\qed
\epf

\bcorr
\label{corr:joint_inters_fct}
Let $S_1$ and $S_2$ be quasi-surfaces with intersection functions 
$\sigma_{S_1}$ and $\sigma_{S_2}$, respectively. 
Then $\sigma_{S_1} + \sigma_{S_2}$ is an intersection function
for $S := S_1 \cup S_2$. 
If, additionally, $\sigma_{S_1}$ and $\sigma_{S_2}$ coincide for all
paths starting at $S_1 \cap S_2$, then
the function $\sigma_{S_1 S_2}$ defined by 
\bgl
\sigma_{S_1 S_2}(\gamma) & := & 
 \begin{cases}      
      \sigma_{S_1}(\gamma) & \text{ if $\gamma(0) \in S_1$,} \\ 
                         0 & \text{ if $\gamma(0) \nichtin S_1 \cup S_2$,} \\
      \sigma_{S_2}(\gamma) & \text{ if $\gamma(0) \in S_2$,} 
 \end{cases}
\egl
is an intersection function for $S$. 
It is called
\df{joint} intersection function.
\ecorr
Obviously, the joint intersection function equals $\sigma_{S_1} + \sigma_{S_2}$
if $S_1$ and $S_2$ are disjoint.

Sometimes, it is convenient to use some sort of standard decomposition
of edges. Indeed, there is a minimal decomposition.
\bdf
Let $S$ be a subset of $M$ and $\gamma\in\Pfgen$ be an edge. 

An $S$-admissible decomposition $\gc$ of $\gamma$ 
is called \df{minimal} iff $\gc' \geq \gc$
for any other $S$-admissible decomposition $\gc'$ of $\gamma$.
\edf
In other words, 
$\gc = (\gamma_1,\ldots,\gamma_I)$ is
minimal iff every other
$S$-admissible decomposition $\gc' = (\gamma'_1,\ldots,\gamma'_J)$ 
is a refinement of $\gc$, 
i.e., there are $0 = j_0 < j_1 < \ldots < j_I = J$, such that
$\gamma_i$ equals $\gamma'_{j_{i-1}+1} \cdots \gamma'_{j_i}$ 
up to the parametrization.
\blem
\label{lem:unique_S-admiss_decomp}
If an edge $\gamma$ has any $S$-admissible decomposition, it has also a minimal
$S$-admissible decomposition. Moreover, this minimal decomposition 
is unique up to the parametrization
of its components.
\elem
\bpf
Let $\gd$ be an $S$-admissible decomposition of $\gamma$. 
Since $\gamma$ equals $\delta_1 \cdots \delta_K$ up to the parametrization,
the parameter domain $[0,1]$ of $\gamma$ may be decomposed into nontrivial
closed intervals $R_k = [t_{k-1},t_k]\teilmenge [0,1]$, 
such that each $\gamma\einschr{R_k}$ corresponds to $\delta_k$. 
Cancel now in $T' := \{t_0,t_1,\ldots,t_K\}$ each $t_k \neq 0,1$
with 
$\inter \gamma\einschr{[t_{k-1},t_{k+1}]} \cap S = \leeremenge$
or $\inter \gamma\einschr{[t_{k-1},t_{k+1}]} \teilmenge S$. 
The remaining set $T = \{\tau_0,\ldots,\tau_I\} \teilmenge T'$ 
naturally defines another 
$S$-admissible decomposition $\gc = (\gamma_1,\ldots,\gamma_I)$ of $\gamma$
and a corresponding decomposition of $[0,1]$ into intervals $P_i$. 

Let now $\gc' = (\gamma'_1,\ldots,\gamma'_J)$ be any $S$-admissible
decomposition of $\gamma$. Then each $\gamma'_i$ corresponds to some interval 
$Q_j \teilmenge[0,1]$ with $\gamma\einschr{Q_j} = \gamma'_j$.
Assume that $Q_j$ overlaps two different intervals $P_i$ and $P_{i+1}$,
i.e.\ $\gamma(\tau_i) \in \inter \gamma_j'$. 
\bunum
\item
Let $\gamma(\tau_i) \in S$. 
Then $\inter \gamma_j' = \inter \gamma\einschr{Q_j} \teilmenge S$,
hence $\inter\gamma\einschr{P_i}\teilmenge S$ 
and $\inter\gamma\einschr{P_{i+1}}\teilmenge S$,
by admissibility.
Consequently, 
$\inter\gamma\einschr{P_i\cup P_{i+1}} = 
  \inter\gamma\einschr{P_i} \cup \{\gamma(\tau_i)\} \cup
  \inter\gamma\einschr{P_{i+1}}\teilmenge S$.
This implies $\tau_i \nichtin T$, in contradiction 
to the minimality of $\gc$.
\item
Let $\gamma(\tau_i) \nichtin S$. 
Then, analogously, we get a contradiction.
\eunum
Consequently, $Q_j$ can overlap nontrivially only either $P_i$
or $P_{i+1}$.
\qed
\epf

\bdf
Let $S$ be a quasi-surface with intersection function $\sigma_S$,
let $\gamma$ be an edge and let $\gc = \{\gamma_i\}_{i=0}^n$ 
be its minimal $S$-admissible decomposition.

Then a point $x\in M$ is called 
\bunum
\item
\df{$\gamma$-puncture} in $S$ iff
there is an $i\in[1,n]$ with 
\zgl{
\gamma_{i-1} (1) = x = \gamma_i(0) \breitrel{\breitrel{\text{ and }}}
\sigma^+_S(\gamma_{i-1}) \sigma^-_S(\gamma_i) > 0;}
\item
\df{$\gamma$-half-puncture} in $S$ iff
there is an $i\in[0,n]$ with 
\zgl{
x = \gamma_i(0)
\breitrel{\breitrel{\text{ and }}}
\sigma^-_S(\gamma_i) \neq 0\phantom.}
or 
\zgl{
x = \gamma_i(1)
\breitrel{\breitrel{\text{ and }}}
\sigma^+_S(\gamma_i) \neq 0.}
\eunum

We say that $\gamma$ intersects $S$ \df{completely transversally}
iff there are no $S$-internal edges in the minimal $S$-admissible decomposition
of $\gamma$ and each $\gamma$-half-puncture is also a $\gamma$-puncture.
\edf
Roughly speaking, $x$ is a $\gamma$-puncture iff $\gamma$ intersects
$S$ (w.r.t.\ $\sigma_S$) transversally at $x$.

%------------------------------------------------------------------------%
%            Abschnitt: Preliminaries                                    %
%------------------------------------------------------------------------%
\subsubsection{Quasi-Flux Action}
In this subsection, $S$ is some quasi-surface and 
$\sigma_S$ some intersection function for $S$. 

\bprop 
\label{prop:qfa}
There
is a unique map 
$\qfa^{S,\sigma_S} : \Ab \kreuz \Maps(M,\LG) \nach \Ab$,
such that

\bgl
\!\!\!\!h_{\qfa^{S,\sigma_S}(\qa,d)} (\gamma)  & \!\!\! = \!\!\! & 
 \begin{cases}
    d(\gamma(0))^{\sigma^\ausl_S(\gamma)} \: 
    h_\qa(\gamma) \: 
    d(\gamma(1))^{\sigma^\einl_S(\gamma)}
    & \text{for $S$-external $\gamma$} \\
    \phantom{d(\gamma(0))^{\sigma^\ausl_S(\gamma)}} \:
    h_\qa(\gamma) \: 
    & \text{for $S$-internal $\gamma$} 
  \end{cases}.\!\!\!
\egl
If $\Maps(M,\LG) \iso \LG^M$ is given the product topology, then
$\qfa$ is continuous. 
Moreover, the map 
\bgl \qfa^{S,\sigma_S}_d & : & \Ab \nach \Ab, \egl given
by $\qfa^{S,\sigma_S}_d(\qa) := \qfa^{S,\sigma_S} (\qa,d)$, 
is a homeomorphism and preserves
the Ashtekar-Lewandowski measure for each $d\in\Maps(M,\LG)$.
Finally, the inverse of $\qfa^{S,\sigma_S}_d$ is given 
by $\qfa^{S,\sigma_S}_{d^{-1}}$.

\eprop
\bpf
\bunum
\item
$\qfa^{S,\sigma_S}$ exists uniquely and 
is continuous for the product topology on $\Maps(M,\LG)$.

First note that $\adm_S$ is complete by Lemma \ref{lem:compl_crit}.
Let now $Y := \Maps(M,\LG)$ and define
\bgl
\rrr(\gamma,d) & := & 
 \begin{cases}
    d(\gamma(0))^{-\sigma^\ausl_S(\gamma)}    
    & \text{if $\gamma$ is $S$-external } \\
    e_\LG \: 
    & \text{if $\gamma$ is $S$-internal } 
  \end{cases}.
\egl
The only nontrivial property of $\rrr$ 
in Corollary \ref{corr:cont(qfa)voll} to be checked is 
$\rrr(\gamma_1^{-1},d) = \rrr(\gamma_2,d)$ for 
decompositions $\gamma_1 \gamma_2$ of $S$-external $\gamma$.
Observe, however, that here 
$\gamma_1^{-1} (0) \ident \gamma_1(1) \ident \gamma_2(0)$ is not contained
in $S$, hence 
$\rrr(\gamma_1^{-1},d) = e_\LG = \rrr(\gamma_2,d)$.
The claim now follows from 
$\sigma^\ausl_S(\gamma) + \sigma^\einl_S(\gamma^{-1}) = 0$
and Corollary \ref{corr:cont(qfa)voll}.
\item
$\qfa^{S,\sigma_S}_d$ is a homeomorphism and 
leaves $\mu_0$ invariant.

This now follows from Theorem \ref{thm:cont+masz(qfa)}.
\qed
\eunum
\epf

%------------------------------------------------------------------------%
%            Konstruktion: neue Zusammenh"ange                           %
%------------------------------------------------------------------------%

\bdf
$\qfa^{S,\sigma_S} : \Ab \kreuz \Maps(M,\LG) \nach \Ab$ is called 
\df{quasi-flux action}.
\edf

\brem
Note that $\qfa^{S,\sigma_S}$ is, in general, {\em not}\/ a
{\em group}\/ action of $\Maps(M,\LG)$. 
\erem
But, we have
\blem
Let $S_1$ and $S_2$ be two quasi-surfaces,
and let $d_1, d_2 : M \nach \LG$ 
be functions commuting on $S_1 \cap S_2$. 
Let $d : M \nach \LG$ be any function with 
\bgl
d & := &  \begin{cases} d_1 & \text{on $S_1 \setminus S_2$} \\ 
                    d_1 d_2 & \text{on $S_1 \cap S_2$} \\ 
                        d_2 & \text{on $S_2 \setminus S_1$} 
          \end{cases}.
\egl
If $\sigma_{S_1}$ and $\sigma_{S_2}$ 
coincide for all paths starting in $S_1 \cap S_2$ 
and vanish both for $S_1$- and
$S_2$-internal paths, 
then
\bgl
\qfa^{S_1,\sigma_{S_1}}_{d_1}  \circ \qfa^{S_2,\sigma_{S_2}}_{d_2}
 & = & \qfa^{S_1 \cup S_2,\sigma_{S_1 S_2}}_{d}
 \breitrel= \qfa^{S_2,\sigma_{S_2}}_{d_2}  \circ \qfa^{S_1,\sigma_{S_1}}_{d_1}.
\egl
\elem
\bpf
By direct calculation.
\qed
\epf

\bcorr
Let $d_1,d_2 : M \nach \LG$ be two functions.

If $d_1$ and $d_2$ commute pointwise, we have 
$\qfa^{S,\sigma_S}_{d_1} \circ \qfa^{S,\sigma_S}_{d_2} 
  = \qfa^{S,\sigma_S}_{d_1 d_2}$.
\ecorr
\bpf
Straightforward.
\qed
\epf

%------------------------------------------------------------------------%
%            Abschnitt: Preliminaries                                    %
%------------------------------------------------------------------------%
\subsection{Weyl Operators}

Recall that every continuous map $\psi : X \nach X$ on a
topological space $X$ defines a continuous pull-back map 
$\psi^\ast : C(X) \nach C(X)$. This map is an isometry if $\psi$
is surjective. If $X$ is even a compact Hausdorff space, $\psi$ is 
surjective and
$\mu$ a (finite) regular Borel measure on $X$ with $\psi_\ast \mu = \mu$,
then $\psi^\ast$ is a unitary operator on $L_2(X,\mu)$.
This motivates 
\bdf
The operators
\zgl{\weyl^{S,\sigma_S}_d := (\qfa^{S,\sigma_S}_d)^\ast}
with $S$ being a quasi-surface, $\sigma_S$ an intersection function and
$d:M \nach \LG$ being any function
are called \df{Weyl operators}.
\edf
Note that each Weyl operator is both a map on $C(\Ab)$ and
$L_2(\Ab,\mu_0)$. In fact,
Proposition \ref{prop:qfa} gives
\neueseite
\bprop
\bunum
\item
Every Weyl operator is an isometry on $C(\Ab)$.
\item
Every Weyl operator is a unitary operator on $L_2(\Ab,\mu_0)$.
\eunum
\eprop
Note, however, that measures, in general,
lead to Weyl operators that are ill defined on the
$L_2$-functions:
For instance, let us work in the analytic category, fix some 
hypersurface $S$ and some intersection function $\sigma_S$.
Assume now that, $g$ running over $\LG$, we have all Weyl operators 
at our disposal that are given by
\bgl
\weyl^{S,\sigma_S}_{g} \breitrel{:=} \weyl^{S,\sigma_S}_{d_g},
\egl\noindent 
where $d_g$ is the constant function 
on $M$ with value $g\in\LG$.
To make all these Weyl operators well defined as
operators on $L_2(\Ab,\mu)$ for some $\mu$,
we have at least to demand that, for each $S$-external edge $\gamma$ 
(having only one end attached to $S$), 
the support of the push-forward measure $(\pi_\gamma)_\ast \mu$ 
equals $\LG$. Of course,
there are many measures without this property.

Let us now collect some additional properties of Weyl operators, again
following directly from the properties of $\qfa$ and
the definition of Weyl operators by pull-backs.
\blem
Let $S$ be a quasi-surface and let $d, d_1, d_2 : M \nach \LG$ 
be some functions. 

Then we have (dropping always the upper indices $S,\sigma_S$ in 
$\weyl^{S,\sigma_S}_d$):
\bnum2
\item 
$\weyl_d(f_1 f_2) = \weyl_d(f_1) \weyl_d(f_2)$ for all
functions $f_1, f_2$ on $\Ab$.
\item
$\weyl_{d_1} \weyl_{d_2} = \weyl_{d_1 d_2}$, if $d_1 d_2 = d_2 d_1$.
\enum
\elem

\bcorr
For all quasi-surfaces $S$, all intersection functions $\sigma_S$ and
all functions 
$d : M \nach \LG$, we have
\zgl{
\weyl_d^{S,-\sigma_S}
  \breitrel= \weyl_{d^{-1}}^{S,\sigma_S}
  \breitrel= (\weyl_d^{S,\sigma_S})^{-1}
  \breitrel\ident (\weyl_d^{S,\sigma_S})^\ast
. }
\ecorr
The preceding corollary implies that the inversion of the orientation of 
a quasi-surface leads to the adjoint Weyl operator. The uniqueness proof
in Section \ref{sect:repr} will heavily use this fact.

\bcorr
Let $\hyph = \{\gamma_1,\ldots,\gamma_n\}$ be a hyph. 
Then 
\bgl
w^{S,\sigma_S}_d (T_1 \tensor \cdots \tensor T_n) 
  & = &  w^{S,\sigma_S}_d (T_1) \tensor \cdots \tensor w^{S,\sigma_S}_d (T_n)
\egl
for all $T_i \in \matfkt_{\gamma_i}$ and all 
functions $d : M \nach \LG$. 
\ecorr

Corollary \ref{corr:joint_inters_fct} implies
\blem
\label{lem:disjqusurf->commut}
Let $S_1$ and $S_2$ be disjoint quasi-surfaces with intersection
functions $\sigma_{S_1}$ and $\sigma_{S_2}$, respectively.
Let, moreover, $d_1, d_2 : M \nach \LG$ 
be some functions.

Then we have 
\bgl
\weyl^{S_1,\sigma_{S_1}}_{d_1}  \circ \weyl^{S_2,\sigma_{S_2}}_{d_2}
 \breitrel= \weyl^{S_2,\sigma_{S_2}}_{d_2}  \circ \weyl^{S_1,\sigma_{S_1}}_{d_1}.
\egl
\elem

\blem
\label{lem:weyl(matfkt)_teil_spann(matfkt)}
Let $\hyph$ be a hyph and $\weyl$ be a Weyl operator for some quasi-surface $S$.

Then there is a hyph
$\hyph' \geq \hyph$ with $\weyl(\matfkt_\hyph) \teilmenge \spann\:\matfkt_{\hyph'}$.
If, moreover, $\hyph$ contains $S$-external and $S$-internal edges only,
then $\weyl(\matfkt_\hyph) \teilmenge \spann\:\matfkt_{\hyph}$.
\elem
\bpf
Choose a hyph $\hyph' \geq \hyph$ containing $S$-external
and $S$-internal edges only. One checks immediately, that 
$\weyl(\matfkt_{\hyph'}) \teilmenge \spann\:\matfkt_{\hyph'}$.
Using $\matfkt_{\hyph} \teilmenge \spann\:\matfkt_{\hyph'}$ (Lemma \ref{lem:decomp_sns}),
we get the assertion.
\qed
\epf

\neueseite

%------------------------------------------------------------------------%
%            Abschnitt: Preliminaries                                    %
%------------------------------------------------------------------------%
\subsection{Regularity}
\bprop
\label{prop:reg(pi_0)}
Fix some quasi-surface $S$ and some intersection function for $\sigma_S$.
Next,
let $\mpsg_0$ be a set of sequential $\Maps(M,\LG)$-functions,
such that%
\footnote{Here, $\pr_x : \Maps(M,\LG) \nach \LG$ assigns to each
function from $M$ to $\LG$ its value in $x$.}
\zgl{\pr_x \circ \mps_0 : \dom\:\mps_0 \nach \LG} is 
sequentially continuous for every $x\in M$ and each $\mps_0\in\mpsg_0$. 
Finally, assign to each $\mps_0$ some 
\fktdefabgesetzt{\mps}{\dom\:\mps_0}{\Weyl,}{y}{\weyl^{S,\sigma_S}_{\mps_0(y)}}
$\Weyl$ being the set of Weyl operators,
and collect all such $\mps$ into $\mpsg$.

Then 
\zgl{\mps(\:\cdot\:) \psi : \dom\:\mps_0 \nach \hilb_\aux \ident L_2(\Ab,\mu_0)}
is continuous for all $\psi\in \hilb_\aux$ 
and each $\mps \in \mpsg$.

\eprop

\bpf
Fix some $\mps\in\mpsg$ with corresponding $\mps_0\in\mpsg_0$
and recall that sequential continuity equals 
continuity, if the domain is sequential. To avoid cumbersome notation,
we write shortly $\weyl_y$ instead of $\weyl^{S,\sigma_S}_{\mps_0(y)}$.
\bunum
\item
Of course, $\weyl_y(\EINS) = \EINS$ for all $y$.
\item
Let $\gamma$ be an edge and $T\in\matfkt_\gamma$ some
gauge-variant spin network state
over $\gamma$.
\bunum
\item
If $\gamma$ is internal, then $\weyl_y(T) = T$ for all $y$, hence 
$y \auf \weyl_y(T)$ is continuous.
\item
If $\gamma$ is external, then with $T = \sqrt{\dim\darst} \: \darst^k_l$
and after a straightforward calculation, we have 
\bgl[1ex]
 &   & \norm{\weyl_y(T) - \weyl_{y'}(T)}_{\hilb_\aux}^2 \s
 & = & 2 - 2 \: \re 
         \darst^k_k\bigl([\mps_0(y')](\gamma(0))^{\sigma^\ausl_S(\gamma)} \: 
	                   [\mps_0(y)](\gamma(0))^{-\sigma^\ausl_S(\gamma)}\bigr) \:\: \cdot \\
 &   & \phantom{2 - 2 \: \re } \cdot \:\:
         \darst^l_l\bigl([\mps_0(y)](\gamma(1))^{-\sigma^\einl_S(\gamma)} \: 
	                   [\mps_0(y')](\gamma(1))^{\sigma^\einl_S(\gamma)}\bigr) \\
 & \ident & 2 - 2 \: \re 
         \darst^k_k\bigl([\pr_{\gamma(0)}\circ\mps_0](y')^{\sigma^\ausl_S(\gamma)} \: 
	                   [\pr_{\gamma(0)}\circ\mps_0](y)^{-\sigma^\ausl_S(\gamma)}\bigr) \:\: \cdot \\
 &   & \phantom{2 - 2 \: \re } \cdot \:\:
         \darst^l_l\bigl([\pr_{\gamma(1)}\circ\mps_0](y)^{-\sigma^\einl_S(\gamma)} \: 
	                   [\pr_{\gamma(1)}\circ\mps_0](y')^{\sigma^\einl_S(\gamma)}\bigr).
\egl
(There is no summation over $k$ and $l$.)
Since, by assumption each $\pr_x \circ \mps_0$ is a continuous mapping
from $\dom\:\mps_0$ to $\LG$, we get 
$\norm{\weyl_y(T) - \weyl_{y'} (T)}_{\hilb_\aux} \gegen 0$ for $y \gegen y'$,
implying the desired continuity of $y \auf \weyl_y(T)$.

\eunum
\item
Let $\hyph$ contain external and internal edges only.
Let, moreover, 
$T = T_1 \tensor \cdots \tensor T_\Hyph$
be in $\matfkt_\hyph$.
Then we have
\bglklein
 &   & \norm{\weyl_y(T) - \weyl_{y'}(T)}_{\hilb_\aux}^2 \\
 & = & 2 - 2\:\re \skalprod{\weyl_{y}T}{\weyl_{y'}T}_{\hilb_\aux} \\
 & = & 2 - 2\:\re \skalprod{\weyl_{y}T_1 \tensor \cdots \tensor \weyl_{y}T_\Hyph}%
                          {\weyl_{y'}T_1 \tensor \cdots \tensor \weyl_{y'}T_\Hyph }_{\hilb_\aux} \\
 & = & 2 - 2\:\re \prod_i \skalprod{\weyl_{y}T_i}{\weyl_{y'} T_i}_{\hilb_\aux} \\
 & \gegen & 2 - 2\:\re\prod_i \skalprod{\weyl_{y'}T_i}{\weyl_{y'}T_i}_{\hilb_\aux} \\
 &   & 
                \erl{$\weyl_{y}T_i \gegen \weyl_{y'}T_i$ by the preceding step} \\
 & = & 0
\eglklein
for $y \gegen y'$.
The factorization of the scalar products was possible, because
$\weyl_y$ leaves the span of (non-trivial) matrix functions over $\gamma_i$ invariant
and because such spans are orthogonal w.r.t.\ $\mu_0$ for paths in a hyph.
\item
Let now $T\in\matfktgsn$ be an arbitrary gauge-variant spin network function,
i.e., there is a hyph $\hyph$ with $T \in \matfkt_\hyph$.
Then there is some hyph $\hyph' \geq \hyph$ containing external and internal
edges only. Since $\matfkt_{\hyph} \teilmenge \spann\: \matfkt_{\hyph'}$ 
by Lemma \ref{lem:decomp_sns}, $\weyl_{y}(T) \gegen \weyl_{y'}T$ 
for $y \gegen y'$.
\item\enlargethispage{0.4\baselineskip}
Now, Lemma \ref{lem:strongcont_crit} gives
the proof: The span of 
$\matfktgsn = \bigcup_\hyph \matfkt_\hyph$
is dense in $L_2(\Ab,\mu_0)$, and $\norm{\weyl_y} = 1$ for all $y$ by unitarity.
\qed
\eunum
\epf
A typical example is given by the continuous (or differentiable) functions
w.r.t.\ the supremum norm:
\bdf
Let $S$ be some quasi-surface and $\sigma_S$ some intersection
function for $S$. 

Now let $\mpsg^{p,S,\sigma_S}$ for $p\in\N\cup\{\infty,\omega\}$ 
contain precisely all mappings 
\fktdefabgesetzt{\weyl^{S,\sigma_S}_{\cdot}}{C^p(M,\LG)}{\Weyl,}{d}{\weyl^{S,\sigma_S}_{d}}
where $C^p(M,\LG)$ is equipped with the supremum norm on $S$.
\edf

We now may transfer this result to one-parameter subgroups.
Using the one-parameter subgroups on $\LG$ induced by the 
elements of the Lie algebra $\Lieg$, we have 
\bcorr
\label{corr:reg(pi_0)}
Let $\gotd : M \nach \Lieg$ be a (not necessarily continuous) function, and
define 
\fktdefabgesetzt{E_\gotd}{\R}{\Maps(M,\LG).}{t}{(\e^{t\gotd(x)})_{x\in M}}
Then we have:
\bnum3
\item
$E_\gotd (t_1) E_\gotd (t_2) = E_\gotd (t_1 + t_2)$
for all $t_1,t_2\in\R$.
\item
$\pr_x \circ E_\gotd$ is continuous for every $x\in M$.
\item
The one-parameter subgroup 
\zgl{t \auf \weyl^{S,\sigma_S}_{E_\gotd(t)}}
is strongly continuous w.r.t.\ to $L_2(\Ab,\mu_0)$
for each quasi-surface $S$ with intersection function $\sigma_S$.
\enum
\ecorr

\bpf
The first two assertions are trivial. To see the strong continuity,
apply Proposition \ref{prop:reg(pi_0)}
to the case $\mpsg_0 := \{E_\gotd : \R \nach \Maps(M,\LG)\}$.
\qed
\epf

%------------------------------------------------------------------------%
%            Abschnitt: Preliminaries                                    %
%------------------------------------------------------------------------%
\subsection{Graphomorphisms}
One of the particular features of quantum geometry is its 
invariance w.r.t.\ diffeomorphisms of $M$. More precisely, diffeomorphisms
act naturally on the paths inducing a $\mu_0$-invariant 
action on $\Ab$ and, consequently, a unitary action
on $\haux$. It remains the question, what kind of diffeomorphisms are to be
admitted: analytic, piecewise analytic, smooth or something else?
Anyway, we will postpone this discussion to Section \ref{sect:weyl_alg} and
consider here only some sort of minimal requirements. For this,
let us again fix some smoothness class for the manifold and the paths
in it.
\bdf
A map $\diffeo : M \nach M$ is called \df{graphomorphism} iff $\diffeo$ 
is bijective and
induces a groupoid isomorphism on $\Pf$. \cite{paper20}
\edf
Here, $\diffeo(\gamma) := \diffeo \circ \gamma$. 
Graphomorphisms have a convenient characterization \cite{paper20}:
\blem
\label{lem:krit(graphom)}
A bijection $\diffeo$ on $M$ is a graphomorphism iff $\diffeo$ 
and 
$\diffeo^{-1}$
map edges to $\Pfgen$.
\elem

The action of graphomorphisms on $\Pf$ can be lifted to an action
on $\Ab$. In fact, each graphomorphism $\diffeo$ defines
via
\bgl\
[\diffeo (\qa)](\gamma) & := & h_\qa(\diffeo^{-1}\circ\gamma) \breitrel{\breitrel{\text{for all $\gamma\in\Pf$}}}
\egl\noindent
a map from $\Ab$ to $\Maps(\Pf,\LG)$,
again denoted by $\diffeo$. We have

\bprop
Every graphomorphism $\diffeo$ maps $\Ab$ homeomorphically to $\Ab$.
\eprop
\bpf
Of course, $\diffeo$ maps $\Ab$ to $\Ab$. 
Moreover, $\pi_\gamma \circ \diffeo = \pi_{\diffeo^{-1}\circ\gamma}$
is continuous for all $\gamma\in\Pf$. Hence, $\diffeo$ is continuous.
The proof now follows, since 
$\diffeo \circ {\diffeo^{-1}}$ is the identity on $\Ab$.
\qed
\epf

\bprop
The Ashtekar-Lewandowski measure $\mu_0$ 
is $\diffeo$-invariant for all graphomorphisms $\diffeo$. 
\eprop
\bpf
This follows, because for all hyphs $\hyph$ 
\zgl{
(\pi_\hyph)_\ast (\diffeo_\ast \mu_0)
 \breitrel= (\pi_{\diffeo^{-1} \circ \hyph})_\ast \mu_0 
 \breitrel= \mu_\Haar^{\elanz(\diffeo^{-1}\circ\hyph)}
 \breitrel= \mu_\Haar^{\elanz\hyph}
 \breitrel= (\pi_\hyph)_\ast \mu_0.
}
\qed
\epf

\bdf
For each graphomorphism $\diffeo$ define 
$\dwirk\diffeo$ to be the pull-back of $\diffeo^{-1}$.
\edf

\bprop
For every graphomorphism $\diffeo$, 
\bunum
\item
$\dwirk\diffeo$ is an isometry on $C(\Ab)$;
\item
$\dwirk\diffeo$ is a unitary operator on $L_2(\Ab,\mu_0)$.
\eunum
\eprop
The map $\diffeo \auf \dwirk\diffeo$ is even 
a representation of the group of graphomorphisms on $L_2(\Ab,\mu_0)$, because
$\dwirk{\diffeo_1\circ\diffeo_2} = \dwirk{\diffeo_1}\circ\dwirk{\diffeo_2}$
and 
$\dwirk{\diffeo^{-1}} = \dwirk{\diffeo}^{-1}$.%
\footnote{Note that we did not care about the corresponding
covariance property for the Weyl operators.
In fact, there $\weyl$ is given by the pull-back of $\qfa$,
not of $\qfa^{-1}$. Since, however, the
$\qfa$-transforms do not form a group, that does not matter.}

Graphomorphisms do not only act on graphs, but also on quasi-surfaces,
intersection and other functions. 
\bdf
Let $\diffeo$ be a graphomorphism. Then we set:
\bunum
\item
$\diffeo(S) := \diffeo \circ S$ for every quasi-surface $S$;
\item
$\diffeo(d) := d \circ \diffeo^{-1}$ for every function $d : M \nach \LG$;
\item
$[\diffeo(\sigma)](S,\gamma) := \sigma(\diffeo^{-1}(S),\diffeo^{-1}(\gamma))$
for every intersection function $\sigma$.
\eunum
\edf
We, therefore, will have to guarantee that admissible 
homeomorphisms do not only 
preserve the set of paths under consideration, 
but also that of quasi-surfaces, and have to avoid ill-defined intersection
functions -- in particular, if we aim at an ``intrinsic'' assignment of
intersection functions to quasi-surfaces.
All that will be provided by using stratified analytic isomorphisms
as to be discussed below. 

Directly from the definitions, we get finally 
\bprop
Let $\diffeo : M \nach M$ be a graphomorphism, $S$ a quasi-surface,
$\sigma$ an intersection function and $d : M \nach \LG$ 
a function.
Then we have

\bgl
\weyl^{\diffeo(S),\diffeo(\sigma)}_{\diffeo(d)} & = & 
\dwirk\diffeo (\weyl^{S,\sigma}_d) \breitrel\ident
\dwirk\diffeo \circ \weyl^{S,\sigma}_d \circ \dwirk\diffeo^{-1}.
\egl 

\eprop

%------------------------------------------------------------------------%
%            Abschnitt: Preliminaries                                    %
%------------------------------------------------------------------------%
\subsection{Generalized Gauge Transforms}
\label{sect:gaugetrfs}
Any gauge theory incorporates gauge invariance. Therefore, we close this section
with a few remarks on gauge transformations and, more general, bundle automorphisms.
\bdf
The elements of $\Gb := \Maps(M,\LG)$ are called \df{generalized gauge transforms}.%
\footnote{Starting from Section \ref{sect:weyl_alg}, we will usually drop the word 
``generalized'' for simplicity.}
$\Gb$ is given the product topology inherited from the canonical isomorphism
$\Maps(M,\LG) \iso \LG^M$ and its group structure is given by pointwise multiplication.
\edf
\bprop
$\Gb$ is a topological group and acts continuously on $\Ab$ via
\bgl
h_{\qa\circ\qg} (\gamma) 
 & := & \qg(\gamma(0))^{-1} \: h_\qa(\gamma) \: \qg(\gamma(1))
\breitrel{\breitrel{\text{for all $\gamma\in\Pf$}}}.
\egl\noindent
\eprop

\bprop
The Ashtekar-Lewandowski measure $\mu_0$ 
is invariant w.r.t.\ all generalized gauge trans\-forms.
\eprop

\bdf
For each generalized gauge transform $\qg$ define 
$\gwirk\qg$ on functions on $\Ab$ by 
\bgl
(\gwirk\qg \: f)(\qa) & := & f(\qa \circ \qg).
\egl
\edf
Observe that 
$\gwirk{\qg_1\circ\qg_2} = \gwirk{\qg_1}\circ\gwirk{\qg_2}$
and 
$\gwirk{\qg^{-1}} = \gwirk{\qg}^{-1}$.
\bprop
\bunum
\item
$\qg \auf \gwirk\qg$ is a representation of $\Gb$ on $C(\Ab)$ by isometries.
\item
$\qg \auf \gwirk\qg$ is a representation of $\Gb$ on $L_2(\Ab,\mu_0)$ by unitaries.
\eunum
\eprop

Generalized gauge transforms 
do also act on the $\LG$-valued functions labelling the quasi-surfaces. 
\bdf
Let $\qg$ be a generalized gauge transform. Then we set:
\bunum
\item
$\qg(d) := \qg \cdot d \cdot \qg^{-1}$ for every function $d : M \nach \LG$.
\eunum
\edf

Again, directly from the definitions, we get 
\bprop
\label{prop:weyl-gauge-covariance}
Let $\qg : M \nach M$ be a generalized gauge transform, $S$ a quasi-surface,
$\sigma$ an intersection function and $d : M \nach \LG$ a function.
Then we have

\bgl
\weyl^{S,\sigma}_{\qg(d)} & = & 
\gwirk\qg (\weyl^{S,\sigma}_d) \breitrel\ident
\gwirk\qg \circ \weyl^{S,\sigma}_d \circ \gwirk\qg^{-1}.
\egl 

\eprop

%------------------------------------------------------------------------%
%            Abschnitt: Preliminaries                                    %
%------------------------------------------------------------------------%
\subsection{Bundle Automorphisms}
Up to now, we have widely ignored the bundle structure of the gauge theory.
Without a real need, we tacitly assumed to deal with a trivialized bundle,
as we focused on the manifold $M$ and the structure group $\LG$ only. 
Of course, it made the notations simpler and can, moreover, be justified a posteriori:
$\Ab$ contains the $C^p$ connections
of {\em any}\/ $\LG$-principal bundle over $M$,
independently from the bundle we started from.
Similarly, $\Gb$ contains
all $C^p$ gauge transforms in any such bundle. 
But, conceptually, it is much 
more desirable to include the full bundle structure.
Then we would also like to include the full group of bundle automorphisms.
Note, here, that given any bundle automorphism $\bdlauto : P \nach P$
of the $\LG$-bundle $P$ over $M$, we may extract from it a
diffeomorphism $\diffeo_\bdlauto : M \nach M$ via
\bgl
\diffeo_\bdlauto \circ \pr_M & = & \pr_M \circ \bdlauto
\egl\noindent
where $\pr_M$ denotes the canonical projection $\pr_M : P \nach M$.
Moreover, the (smooth) gauge transforms correspond to vertical automorphisms; these are
the bundle automorphisms with $\diffeo_\bdlauto = \id_M$.

Nevertheless, the full information on any (possibly stratified) $C^p$ bundle automorphism can be 
encoded in a (again, possibly stratified) $C^p$ diffeomorphism and a generalized
gauge transform (even of any other bundle). 
The only danger arising from taking all the generalized gauge 
transforms of Subsection \ref{sect:gaugetrfs} is to take too many gauge transforms. 
However, observe that, at least for the piecewise analytic category, the set of gauge
orbits $\ag$ is densely embedded into $\AbGb$ and no two piecewise analytic connections fall
into the same equivalence class by moding out the group of gauge transforms \cite{paper10}.

Finally, as it will turn out, the diffeomorphisms and the gauge transforms will play 
different r\^oles in the following proofs. Therefore, to make the basic ideas clearer
and to sometimes allow for relaxed assumptions in the assertions,
we will refrain from considering the fully automorphism invariant treatment of the
Weyl algebra. Thus, w.l.o.g., we may pragmatically consider the bundle-automorphism invariance
given by implementing both diffeomorphism and gauge invariance.
The translation into the fully invariant language has to be left to the interested reader.

%------------------------------------------------------------------------%
%            Abschnitt: Preliminaries                                    %
%------------------------------------------------------------------------%
\section{Weyl Algebra of Quantum Geometry}
\label{sect:weyl_alg}
%------------------------------------------------------------------------%
%                                                                        %
%------------------------------------------------------------------------%
\subsection{Structure Data}
\label{subsect:struct_data}
In what follows, we are going to apply the above definitions and results
to quantum geometry.
Usually, this means to use piecewise analytic paths $\gamma$ and 
oriented hypersurfaces $S$ in $M$,
whereas the intersection functions encode whether $\gamma$ intersects
$S$ transversally or not and how its direction is related to the orientation
of $S$. Moreover, (piecewise) analytic diffeomorphisms
act on these objects. However, is it obvious that we should
consider precisely these ingredients?

Before we discuss this question, let us collect these assumptions
to avoid cumbersome notation.
\bdf
The \df{structure data} of the theory under consideration contain:
\bunum
\item
a manifold $M$;
\item
a Lie group $\LG$;
\item
a smoothness class used for the definition of the set $\Pf$ of paths in $M$,
\item
a subset $\qsf$ of the set of quasi-surfaces in $M$,
\item
for each $S \in \qsf$ a subset $\isf(S)$ of the set of intersection functions for $S$,
\item
for each $S \in \qsf$ a subset $\dsf(S)$ of the set of functions from $M$ to $\LG$,
\item
a subset $\gaugetrfs$ of the set $\Gb$ of gauge transforms acting
covariantly on $\dsf$;
\item
a subset $\Diffeo$ of the set of graphomorphisms on $M$ that leave 
$\qsf$ invariant and act covariantly on $\isf$ and $\dsf$;
\eunum
\edf

Indeed, at the first glance, there seems to be an enormous freedom in choosing
structure data of a theory. 
However, there are several antagonists in the game. For instance,
if we would enlarge $\Pf$, we might have to reduce $\qsf$, simply because
we have to guarantee that there are at most finitely many (genuine) 
intersections of paths and quasi-surfaces. In fact, this practically excludes
the choice of the smooth category for the paths: There are even analytic
submanifolds having an infinite number of isolated transversal intersections
with smooth paths. Therefore, we are -- from the mathematical, technical
point of view -- quite forced to admit at most (piecewise) analytic paths. This
however reduces the number of graphomorphisms in $\diffeo$. Namely, 
they have to map analytic paths to (piecewise) analytic ones. This 
would lead directly into conflicts, if general smooth diffeomorphisms were
allowed. They have to be ``analyticity preserving'' -- at least for
one-dimensional submanifolds. There are indeed classes of homeomorphisms
having this property: At first, of course, analytic diffeomorphisms
fulfill this requirement. However, this will not be sufficient for two reasons:
On the one hand, analyticity usually implies high non-locality -- a feature
not desired in gravity for physical reasons. On the other hand, in the sequel,
the proofs will, in general, crucially depend on the locality 
for technical reasons as we will see later. Thus, some sort
of piecewise analytic diffeomorphisms are to be admitted. In a natural way,
this leads to stratified diffeomorphisms, because they map
semianalytic sets (disjoint unions of analytic submanifolds forming
stratifications)
into semianalytic sets. 

Next, we have to take care of the intersection
functions. Given some oriented submanifold, say, a hypersurface,
we would like to use this
orientation to define such a function. However, this might lead to 
problems again: Using piecewise analytic diffeomorphisms, it may happen
that a surface (including its orientation)
is kept invariant, but an originally transversally intersecting
path may now be mapped to a tangential one.%
\footnote{Let $M$ be $\R^2$ and divide $M$ by the two lines $x=\pm1$ 
into three open parts and the two lines. 
Now define 
$\diffeo$ on the open strip between these two lines
by $\diffeo(x,y) := (x,y+\sqrt{1-x^2})$ and let $\diffeo$
be the identity otherwise. Of course, $\diffeo$ is continuous everywhere
and an analytic diffeomorphism on each of these five parts. 
Nevertheless, the path $\gamma$ with $\gamma(t) = (t,0)$ is transversal w.r.t.\
$x=1$, but $\diffeo(\gamma)$ is tangent to it.}
This would contradict the concept that 
the intersection function encodes the transversality properties of a surface 
and its orientation,
i.e., is assigned naturally and uniquely to an oriented surface.
Of course, in contrast to the previous arguments, this rather is 
a conceptual demand and not a technical one. 
Moreover, it can be overcome using a slightly more special kind
of piecewise analytic diffeomorphisms, as we will see later.

Third, the selection of
functions is to be discussed. Since we have argued that mostly
analytic (or piecewise analytic) objects are to be used, we could 
restrict ourselves again to (piecewise) analytic functions (at least 
for the restrictions to the respective surface). However, although this is
possible, we may consider more general classes. In particular, 
after decomposing a surface into several submanifolds, we may 
admit functions that are analytic only on these submanifolds, but do not 
satisfy any continuity condition at their ``boundaries''. In fact, assume,
e.g., that we are given a $2$-surface $S$ and divide it by a line $S_0$
into two pieces $S_1$ and $S_2$ plus $S_0$
(like the interior of a circle is divided by a diameter). 
We now want to label $S$ on each $S_i$ by some analytic function $d_i$.
We may take 
the Weyl operator $\weyl_0$ for
$S$ and $d_0$, then $\weyl_{j0}$ for $S_j$ with $(d_0)^{-1}$, and, finally,
$\weyl_j$ for $S_j$ and $d_j$ ($j = 1, 2$). 
Now, $\weyl \circ \weyl_{10} \circ \weyl_{20} \circ \weyl_1 \circ \weyl_2$ 
is the Weyl operator for $S$ with a function whose restriction on each $S_i$
is $d_i$. We should remark, that this way one may even define 
submanifolds with codimension $2$ or larger to be (quasi-)surfaces. 
This, however, brings back the problem that the intersection function
is not necessarily given directly by the orientation of the submanifold itself:
the transversality between paths and such lower-dimensional submanifolds 
would, in general, be destroyed already by analytic diffeomorphisms.
Thus, one should restrict oneself to hypersurfaces (or at least semianalytic
sets of pure codimension $1$) and control lower-dimensional surfaces
by including labellings of hypersurfaces with functions $d$ 
that are nontrivial only on these ``sub''-surfaces.
Or, equivalently, one may give lower-dimensional surfaces orientations that are 
induced by hypersurfaces containing them. We will exploit this idea.
Anyway, after all, it does not seem necessary to impose
very strong smoothness restrictions on $\dsf(S)$ 
from the conceptual point of view.
Nevertheless, as we will see, there will be some technical difficulties
that lead to restrictions.

To summarize, in what follows we will always assume to work with
``nice'' structure data having the following minimal properties:
\bdf
The structure data are called \df{nice} iff
\bunum
\item
$M$ is an at least two-dimensional analytic manifold;
\item
$\LG$ is a nontrivial, connected compact Lie group;
\item
$\Pf$ consists of all piecewise analytic paths in $M$;
\item
$\qsf$ contains at most the stratified analytic sets in $M$;
\item
$\isf(S)$ contains at least the natural%
\footnote{In contrast to Definition \ref{def:ind_nat_orient}, we consider
an intersection function on $S$ with $\codim_M S \geq 2$ 
to be natural iff it is induced
by an embedded hypersurface $S'$ that is contained {\em in $\qsf$}, not
just in $M$. 
Moreover, 
one can directly extend the definition of natural intersection
functions to stratified sets, e.g., using triangulations. 
However, since, at the end, we are interested mostly
in the orientation of genuine submanifolds (possibly with boundary) only,
we do not consider this issue in this paper in detail. Thus,
at the moment, the statement ``$\isf(S)$ contains at least the natural
intersection functions of $S$'' only refers to such submanifolds $S$.}
intersection functions of $S$;
\item
$\dsf(S)$ contains at least the constant functions on $M$;
\item
$\gaugetrfs$ contains at least the trivial gauge transform;
\item
$\Diffeo$ contains at most the stratified analytic diffeomorphisms in $M$.
\eunum
\edf
The requirements regarding regularity will be discussed
in Subsection \ref{subsect:weyl_alg_reg}.
The precise definitions of stratified objects will be given 
in Section \ref{sect:strat_diffeo}.
Note, that whether we consider
closed manifolds only or include open ones, is not decided here.
The remaining ``fine-tuning'' will be made if needed.

%------------------------------------------------------------------------%
%            Abschnitt: Preliminaries                                    %
%------------------------------------------------------------------------%

\subsection{Weyl Algebra}
Assume we are working with some arbitrary, but fixed ``consistent''
structure data. We define 
\bglklein
\Weyl & := & 
   \erz{ \: \bigcup_{S \in \qsf} \bigcup_{\sigma_S \in \isf(S)} \bigcup_{d\in\dsf(S)} 
           \{\weyl^{S,\sigma_S}_d\} \: }
\eglklein\noindent
and set 
\bglklein
\Weyl' & := & 
   \erz{ \: \bigcup_{\diffeo \in \Diffeo} \{\dwirk\diffeo\} \: }.
\eglklein\noindent
and
\bglklein
\Weyl'' & := & 
   \erz{ \: \bigcup_{\qg \in \gaugetrfs} \{\gwirk\qg\} \: }.
\eglklein\noindent

%------------------------------------------------------------------------%
%            Abschnitt: Weyl Algebra                                     %
%------------------------------------------------------------------------%

\bdf
The $C^\ast$-subalgebra 
$\alg := \alg(\Weyl,\mu_0)$ of $\bound(L_2(\Ab,\mu_0))$,
generated by $C(\Ab)$ and $\Weyl$, 
is called 
\df{Weyl algebra of quantum geometry}. 
\edf

\bdf
\bunum
\item
$\algdiff := \alg(\Weyl\cup\Weyl',\mu_0)$ denotes the 
$C^\ast$-subalgebra of $\bound(L_2(\Ab,\mu_0))$ generated by $\alg$ and
$\Weyl'$.
\item
$\algauto := \alg(\Weyl\cup\Weyl'\cup\Weyl'',\mu_0)$ denotes the 
$C^\ast$-subalgebra of $\bound(L_2(\Ab,\mu_0))$ generated by $\alg$, 
$\Weyl'$ and $\Weyl''$.
\eunum
\edf
\bdf
Let $\pi'$ be a representation of $\algdiff$ on some Hilbert space $\hilb$.
\bunum
\item
$\psi\in\hilb$ is called \df{diffeomorphism invariant} (w.r.t.\ $\pi'$)
iff $\pi'(\dwirk\diffeo) \psi = \psi$ for all $\diffeo\in\Diffeo$.
\item
$\pi'$ is called \df{diffeomorphism invariant} iff it has a
diffeomorphism invariant vector.
\eunum
Often we write ``$\Diffeo$-invariant'' instead of ``diffeomorphism invariant''.
\edf
Analogously, we speak about $\Diffeo$-natural representations
meaning $\Weyl'$-natural representations.

\bdf
Let $\pi''$ be a representation of $\algauto$ on some Hilbert space $\hilb$.
\bunum
\item
$\psi\in\hilb$ is called \df{automorphism invariant} (w.r.t.\ $\pi''$)
iff $\pi''\einschr{\algdiff}$ is diffeomorphism invariant and 
$\pi''(\gwirk\qg) \psi = \psi$ for all $\qg\in\gaugetrfs$.
\item
$\pi''$ is called \df{automorphism invariant} iff it has an
automorphism invariant vector.
\eunum
Usually we write ``$\autos$-invariant'' 
instead of ``automorphism invariant''.
\edf

\bdf
$\pi_0$ denotes the fundamental (i.e.\ identical) 
representation of $\alg$ on $L_2(\Ab,\mu_0)$
(and, analogously, that of $\algdiff$ and $\algauto$, 
respectively).
\edf
Since $\EINS \in L_2(\Ab,\mu_0)$ is already cyclic for 
$C(X) \teilmenge \alg$, and $\dwirk\diffeo(\EINS)$ equals $\EINS$
for all $\diffeo\in\Diffeo$ as well as $\gwirk\qg(\EINS)$ does for all
$\qg\in\gaugetrfs$, we have
\bprop
$\EINS$ is a cyclic, diffeomorphism and automorphism invariant
vector for $\pi_0$.
\eprop
The irreducibility of $\pi_0$ will be proven separately 
in Section \ref{sect:irred}.

%------------------------------------------------------------------------%
%            Abschnitt: Preliminaries                                    %
%------------------------------------------------------------------------%

\subsection{Regularity}
\label{subsect:weyl_alg_reg}
One of our goals in this paper is a uniqueness proof for certain 
representations of $\alg$. However, we will only be able to
do this for certain regularity conditions. It is now reasonable 
to presuppose as little of them as possible. In other words, 
$\epg$ which encodes the one-parameter subgroups to be 
mapped to weakly continuous ones, should be chosen as small as possible.
As we will see, it will be sufficient to include
that all 
$t \auf \weyl_t = \weyl^{S,\sigma_S}_{d(t)}$ with 
$d(t) := \e^{t\gotd} \in \dsf(S)$ for constant 
$\gotd : M \nach \Lieg$. Of course, more regularity, hence larger $\epg$, 
will not reduce uniqueness, but may even lead to the case that there is
no such regular representation at all. Therefore, we are faced with some
maximality conditions as well.
First of all, we may at most allow for those one-parameter subgroups
that map to the Weyl operators given by the structure data.
Typically such restrictions are induced by the functions $d$ at our
disposal. 
For instance, 
let $\LG$, $M$ and $S$ be not simply connected,
allow $\dsf(S)$ to contain continuous functions only,
and let $d : M \nach \LG$ have nontrivial mapping degree. 
Then, in general, it is not possible to deform $d$ in $\dsf(S)$ continuously 
into the trivial function on $\LG$. This shows that it need not
be possible to connect any Weyl operator to the identity within the
limits of the structure data. 
Of course, using non-continuous $d$, it {\em is}\/ always
possible: Choose at every point $x$ in $M$ some 
$\gotd(x) \in \Lieg$ with $\e^{\gotd(x)} = d(x)$ and define
$\weyl_t := \weyl^{S,\sigma_S}_{E_\gotd(t)}$ for all $t$.
But, moreover, even if we might find for each $t$ some allowed $d(t)$ 
with $d(t_1 + t_2) = d(t_1) d(t_2)$, the corresponding 
maps $t \auf (d(t))(x)$ need not be continuous at all. 
The reason behind is that the functional equation $f(x+y) = f(x) + f(y)$
has non-continuous, ``cloudy'' solutions. Then the corresponding 
one-parameter subgroups of Weyl operators are no longer weakly
continuous, as one immediately checks.
Therefore, we should 
restrict ourselves indeed to the functions generated by the 
Lie algebra functions. We summarize these considerations in
\bdf
Let $\qsf$ contain some quasi-surfaces in $M$ and, for each $S \in \qsf$,
let $\isf(S)$ contain intersection functions for $S$
and $\dsf(S)$ contain functions from $M$ to $\LG$.

A set $\epg$ of one-parameter subgroups in the set of Weyl operators
is called \df{full-consistent with $\qsf$, $\{\isf(S)\}$ and $\{\dsf(S)\}$} 
iff for every element $t \auf \weyl_t$ in $\epg$ 
there is some 
function $\gotd : M \nach \Lieg$ and some quasi-surface $S \in \qsf$ 
with intersection function $\sigma_S \in \isf(S)$, such that
$d(t) := \e^{t\gotd} \in \dsf(S)$ and 
$\weyl_t = \weyl^{S,\sigma_S}_{d(t)}$ for all $t$.

$\epg$ is called 
\df{consistent with $\qsf$, $\{\isf(S)\}$ and $\{\dsf(S)\}$} 
iff $\epg$ equals $\erz{\epg_0}$ for some $\epg_0$ being
full-consistent with $\qsf$, $\{\isf(S)\}$ and $\{\dsf(S)\}$.
\edf

After all, we enlarge the structure data above by
some subset $\epg$ of the set of one-parameter subgroups in $\Weyl$.
\bdf
The enlarged structure data are called \df{nice} iff the structure data are nice and
\bunum
\item
$\epg$ contains at most the one-parameter subgroups of Weyl operators 
consistent
with $\qsf$, $\{\isf(S)\}$, $\{\dsf(S)\}$ and at least those consistent
with $\qsf$, $\{\isf(S)\}$ and the constant functions.
\eunum
\edf
Using 
Corollary \ref{corr:reg(pi_0)} and
Proposition \ref{prop:reg(pi_0)}, we have
for nice enlarged structure data
\bprop
\bnum2
\item
$\pi_0$ is regular w.r.t.\ $\epg$.
\item
$\pi_0$ is $\mpsg$-regular with $\mpsg$ given in Proposition \ref{prop:reg(pi_0)}.
\enum
\eprop
In particular, $\pi_0$ is $\mpsg^{p,S,\sigma_S}$-regular 
for all $p\in\N\cup\{\infty,\omega\}$, $S\in\qsf$ and $\sigma_S \in \isf(S)$.

%------------------------------------------------------------------------%
%            Abschnitt: Preliminaries                                    %
%------------------------------------------------------------------------%
\section{Irreducibility}
\label{sect:irred}
In this section we are going to prove the irreducibility of $\alg$
for nice structure data.\ \cite{paper21}
Additionally, we assume 
that $\qsf$ contains at least the closed, oriented
hypersurfaces of $M$.
Since we do not need diffeomorphisms, there will be
no restrictions for $\Diffeo$. 
Note that given the irreducibility of the Weyl algebra of quantum geometry 
for these
structure data, we get it immediately for all larger structure data.
In fact, since the Weyl algebra cannot shrink if the structure data
get larger, the commutant of the Weyl algebra cannot get larger in this
case. Since, however, we will see it is already trivial for the 
assumptions above, the enlarged Weyl algebra is again irreducible.

%------------------------------------------------------------------------%
%            Abschnitt: Preliminaries                                    %
%------------------------------------------------------------------------%
\subsection{Nice Intersections}
In this subsection, properties of intersections between 
graphs and surfaces, together with their implications for
certain scalar products are studied. 
\bdf
Let $\gamma$ be an edge
and let $\gc$ be a (possibly trivial) graph.

A surface $S$ is called \df{$(\gamma,\gc)$-nice} iff
\bnum2
\item
$S$ is naturally oriented;
\item
$S$ and (the image of) $\gc$ are disjoint; and
\item
$\gamma$ intersects $S$ in precisely one interior point $x$ of $\gamma$
transversally,
such that the orientation of $S$ coincides with the direction of $\gamma$.
\enum
In this case, $x$ is called \df{puncture} of $S$ and $(\gamma,\gc)$.
\edf

\blem
\label{lem:ex(nice_surf)}
Let $\gamma$ be an edge and $\gc$ be a (possibly empty) graph, such that
$\gamma$ and (the edges in) $\gc$ intersect at most at their end points.

Then for every interior point $x$ of $\gamma$, there is a
$(\gamma,\gc)$-nice hypersurface $S$ 
with corresponding puncture $x$.
\elem
Note that it does not matter whether we restrict ourselves to the case
of closed surfaces or to that of open ones.
\bpf
If we admit open surfaces $S$, then the assertion is trivial, since
we may always find some neighbourhood of $x$ disjoint to $\gc$, 
where $\gamma$ is a straight line. Take for $S$ some 
sufficiently small hyperplane ``orthogonal'' to $\gamma$ and that contains $x$. 

Let us, therefore, consider the case of closed surfaces. 
Roughly speaking, the problem here is
that if $\gamma$ ``enters'' $S$ at some point, it has to ``leave''
it somewhere else. Thus, we have to ensure that at only one point this
intersection is transversal. For that purpose, we consider  
some (real) analytic curve $c$ in $\R^2$ that has an inflection point,
such that the corresponding tangent $t$ intersects $c$ in precisely
one other point $y$ transversally. 
Such curves exist -- take, e.g., an appropriate Cassini curve \cite{Z1}. 
As in the case of open surfaces, consider now some neighbourhood of $x$
isomorphic to $\R^n \obermenge \R^2$ and disjoint with $\gc$, such that
$x$ is mapped to $y$ and such that (the image of) $\gamma$ 
coincides with $t$ in some sufficiently large neighbourhood of $y$.
Let now $S$ be the rotational surface given by $c$ and, e.g., the
$x^1$-axis in $\R^2 \teilmenge \R^n$. By construction, $S$ has the
required properties. 
(If the direction of $\gamma$ and the orientation of $S$ at the puncture
do not coincide, simply mirror $S$ at the hyperplane ``orthogonal''
to $\gamma$.)
\qed
\epf

\blem
\label{lem:skalprod-spur}
Let $\gamma$ be an edge and let $\gc$ be some (possibly trivial) graph,
such that $\gamma$ and the edges in $\gc$ intersect each other at most at their end points.
Moreover, let $S$, $S_1$ and $S_2$ be $(\gamma,\gc)$-nice surfaces,
such that the corresponding punctures are
different.
Finally, let $T$ be a
gauge-variant spin network function of the form 
$T = (T_{\darst,\gamma})^m_n \tensor T'$
with $T' \in \matfkt_\gc$.

Then we have 
\bgl
       \skalprod{\weyl^{S_1}_{g_1} (T)}{\weyl^{S_2}_{g_2} (T)}
 & = & \frac{\quer{\charakt_\darst(g_1^2)} \: \charakt_\darst(g_2^2)}
            {(\dim\darst)^2}
\egl
for all $g_1, g_2 \in\LG$. Moreover, if $\darst$ is abelian%
\footnote{Recall that a representation is called \df{abelian} (or \df{linear})
iff its character $\charakt_\darst:\LG \nach \C$ is multiplicative, 
i.e.\ $\charakt_\darst(g_1) \charakt_\darst(g_2) = \charakt_\darst(g_1 g_2)$
for all $g_1,g_2\in\LG$. An irreducible abelian representation of a 
connected compact group is necessarily one-dimensional, i.e.\
$\darst(g) = \charakt_\darst(g) \EINS$ with $\betrag{\charakt_\darst(g)} = 1$
for all $g\in\LG$. Moreover, every compact connected 
$\LG$ equals $(\LG_\he \kreuz \LG_\ab)/\LN$
for some semisimple $\LG_\he$, some torus $\LG_\ab$ and some discrete
$\LN$ being central in $\LG_\he \kreuz \LG_\ab$.
Hence, 
for every irreducible representation $\darst$ of $\LG$ 
there are irreducible representations $\darst_\he$ and $\darst_\ab$ of
$\LG_\he$ and $\LG_\ab$, respectively, such that 
$\darst \circ \pi = \darst_\he \tensor \darst_\ab$ 
with $\pi : \LG_\he \kreuz \LG_\ab \nach \LG$ being the canonical projection.
Then $\darst$ is abelian iff $\darst_\he$ is trivial.}, 
we have 
\bgl
  \weyl^{S}_{g} (T) & = & \darst(g^2) \: T 
          \breitrel= \charakt_\darst(g^2) \: T
\egl
for all $g\in\LG$.
\elem
Here, $\weyl^S_g$ is a shorter notation for $\weyl^{S,\sigma_S}_{d_g}$ 
with $\sigma_S$ given by the natural orientation of $S$
and with $d_g$ being the function on $M$ constantly equal $g\in\LG$.

\bpf
First of all, note that 
$\weyl^{S}_{g} (T) = \weyl^{S}_{g}((T_{\darst,\gamma})^m_n) \tensor T'$ 
for all $g\in\LG$ and for all $(\gamma,\gc)$-nice $S$.
Assume now, that $t_1 < t_2$, where $\gamma(t_j)$ is the intersection point of
$S_j$ and $\gamma$. Decompose $\gamma$ into the three segments
$\gamma_1$, $\gamma_0$ and $\gamma_2$ according to the parameter intervals
$[0,t_1]$, $[t_1,t_2]$ and $[t_2,1]$, respectively. 
Then we have 
\bgl
            (T_{\darst,\gamma})^m_n 
   & = &    (T_{\darst,\gamma_1 \gamma_0 \gamma_2})^m_n 
 \breitrel= \inv{\dim\darst} \:
            (T_{\darst,\gamma_1})^m_p \tensor
            (T_{\darst,\gamma_0})^p_q \tensor
            (T_{\darst,\gamma_2})^q_n.
\egl
Consequently,
\bgl
            \weyl^{S_1}_{g_1}((T_{\darst,\gamma})^m_n)
   & = &    \inv{\dim\darst} \:
            (T_{\darst,\gamma_1})^m_{r_1} 
            \darst(g_1)^{r_1}_p \tensor
            \darst(g_1)^{p}_{s_1}
            (T_{\darst,\gamma_0})^{s_1}_q \tensor
            (T_{\darst,\gamma_2})^q_n \\
   & = &    \inv{\dim\darst} \:
            \darst(g_1^2)^{r_1}_{s_1} \:
            (T_{\darst,\gamma_1})^m_{r_1} \tensor
            (T_{\darst,\gamma_0})^{s_1}_q \tensor
            (T_{\darst,\gamma_2})^q_n 
\egl
and, analogously,
\bgl
            \weyl^{S_2}_{g_2}((T_{\darst,\gamma})^m_n)
   & = &    \inv{\dim\darst} \:
            \darst(g_2^2)^{r_2}_{s_2} \:
            (T_{\darst,\gamma_1})^m_p \tensor
            (T_{\darst,\gamma_0})^p_{r_2} \tensor
            (T_{\darst,\gamma_2})^{s_2}_n.
\egl
Since $\gamma_1$, $\gamma_0$, $\gamma_2$ and $\gc$ are independent,
we get
\bgl[1ex]
   &   &    \skalprod{\weyl^{S_1}_{g_1} T}{\weyl^{S_2}_{g_2} T} \s
   & = &    \skalprod{\weyl^{S_1}_{g_1}((T_{\darst,\gamma})^m_n)}
                     {\weyl^{S_2}_{g_2}((T_{\darst,\gamma})^m_n)} \:\cdot\:
            \skalprod{T'}{T'} \s 
   & = &    \inv{(\dim\darst)^2} \:\:
            \quer{\darst(g_1^2)^{r_1}_{s_1}} \:
                  \darst(g_2^2)^{r_2}_{s_2}  \: \cdot \\
   &   &    \phantom{\inv{(\dim)}} \cdot \:
            \skalprod{(T_{\darst,\gamma_1})^m_{r_1}}
                     {(T_{\darst,\gamma_1})^m_p} \:
            \skalprod{(T_{\darst,\gamma_0})^{s_1}_{q}}
                     {(T_{\darst,\gamma_0})^p_{r_2}} \:
            \skalprod{(T_{\darst,\gamma_2})^{q}_{n}}
                     {(T_{\darst,\gamma_2})^{s_2}_n} \s
   & = &    \inv{(\dim\darst)^2} \:\:
            \quer{\darst(g_1^2)^{r_1}_{s_1}} \:
                  \darst(g_2^2)^{r_2}_{s_2}  \:
            \delta_{r_1 p} \: 
            \delta^{s_1 p} \delta_{q r_2} \:
            \delta^{q s_2} \\
   & = &    \inv{(\dim\darst)^2} \:\:
            \quer{\tr\darst(g_1^2)} \:\: \tr\darst(g_2^2).
\egl
If $t_1 > t_2$, the calculation is completely analogous.

The assertion $\weyl^{S}_{g} (T) = \darst(g^2) \: T $
for abelian $\darst$ follows directly from the definition of 
$\weyl^{S}_{g}$. Recall that every abelian representation is one-dimensional
and maps $\LG$ to $U(1)\:\EINS$.
\qed
\epf

%------------------------------------------------------------------------%
%            Abschnitt: Preliminaries                                    %
%------------------------------------------------------------------------%
\subsection{Irreducibility Proof}
\bthm
The Weyl algebra $\alg$ of quantum geometry is irreducible 
on $L_2(\Ab,\mu_0)$.
\ethm
Before proving the theorem, we set
$\linf := L_\infty(\Xb,\mu_0)$ and $\lzwo := L_2(\Xb,\mu_0)$.
\bpf
We are now going to prove the irreducibility of $\alg$ by verifying
that the commutant of $\alg$ consists of scalars only \cite{BratRob1}.

Since $C(\Xb) \teilmenge \alg$, we have
$\alg' \teilmenge C(\Xb)' = \linf$ for the commutants \cite{EMS124}.
Next, one checks immediately,  
that $\weyl(f) \weyl(\psi) = \weyl(f\psi)$ for all $\weyl\in\Weyl$, 
$f\in \linf$ and $\psi \in \lzwo$. In other words,
$\weyl(f) \circ \weyl = \weyl \circ f$ in $\bound(\lzwo)$.

Let now $f\in \alg' \teilmenge \linf$. Then
we have $f \circ \weyl = \weyl \circ f = \weyl(f) \circ \weyl$
for all $\weyl\in\Weyl$,
hence $\weyl(f) = f$ in $\linf \teilmenge \lzwo$ 
by invertibility of $\weyl$.
Consider additionally some non-trivial 
gauge-variant spin network function $T$. 
It can be written as $T = (T_{\darst,\gamma})^m_n \tensor T'$ with
nontrivial $\darst$,
where $T' \in \matfkt_\gc$ is a 
(possibly trivial) spin network function, such that
$\gamma$ and the edges in $\gc$ intersect at most at their end points.
By $\weyl(f) = f$ and $\weyl^\ast\in\Weyl$ for all $\weyl\in\Weyl$, we have
$\skalprod{T}{f} = \skalprod{T}{\weyl^\ast(f)} = \skalprod{\weyl(T)}{f}$
and, therefore, 
\bgl
            \skalprod{\weyl(T)}{f} 
 \breitrel= \skalprod{T}{f} 
      & = & \skalprod{\weyl'(T)}{f} 
\egl
for all $\weyl,\weyl'\in\Weyl$.

\bnum2
\item
Let $\darst$ be abelian. 

Choose some $(\gamma,\gc)$-nice surface $S$ by Lemma \ref{lem:ex(nice_surf)}.
Then we have 
$\weyl^S_{g} (T) = \darst(g^2) \: T$ 
for all $g\in\LG$, by Lemma \ref{lem:skalprod-spur}.
Consequently,
\bgl
            \skalprod{T}{f} 
      & = & \skalprod{\weyl^S_g(T)}{f} 
 \breitrel= \quer{\darst(g^2)} \: \skalprod{T}{f}.
\egl
Since $\darst$ is nontrivial, there is some $g\in\LG$ with 
$\darst(g^2) \neq 1$. 
Hence, $\skalprod{T}{f} = 0$.
\item
Let $\darst$ be nonabelian.

Since $\LG$ is compact and connected, there is a square root
for each element of $\LG$. 
Moreover, by \cite{m3}, each nonabelian irreducible character has a zero. 
Hence, there is a $g\in\LG$ with $\charakt_\darst(g^2) = 0$.

Choose now, by Lemma \ref{lem:ex(nice_surf)}, 
infinitely many $(\gamma,\gc)$-nice surfaces $S_i$,
whose punctures with $\gamma$ are mutually different.
Then, by Lemma \ref{lem:skalprod-spur},
we have 
\bgl[1ex]
            \skalprod{\weyl^{S_i}_{g} (T)}{\weyl^{S_j}_{g} (T)}
   & = &    \frac{\quer{\charakt_\darst(g^2)} \: \charakt_\darst(g^2)}{(\dim\darst)^2}
 \breitrel= 0
\egl
for $i\neq j$, due to the choice of $g$. 
Since $\weyl^{S_i}_g$ is unitary, $\{\weyl^{S_i}_{g} (T)\}$ is an orthonormal
system. 
Using 
\bgl
\skalprod{\weyl^{S_i}_{g}(T)}{f} & = & \skalprod{\weyl^{S_j}_{g}(T)}{f}
\egl
for all $i,j$, this implies $\skalprod{\weyl^{S_i}_{g}(T)}{f} = 0$ and thus
$\skalprod{T}{f} = 0$.
\enum
Altogether, we have proven 
$\skalprod{T}{f} = 0$ for all nontrivial 
gauge-variant spin network functions $T$. Therefore, $f \in \C \: \EINS$,
hence $\alg'= \C \: \EINS$. 
\qed
\epf

\bcorr
$\algdiff$ and $\algauto$ are irreducible.
\ecorr

%------------------------------------------------------------------------%
%                                                                        %
%------------------------------------------------------------------------%
\section{Stratified Diffeomorphisms}
\label{sect:strat_diffeo}
As we have mentioned in Section \ref{sect:weyl_alg} and we will see in the 
proofs, analytic graphomorphisms will not always be sufficient for 
studying representations of $\alg$. 
A natural extension is stratified analytic isomorphisms.
The theory of stratifications we will use here is motivated by \cite{m4}. 
The first definition will be quoted almost literally, 
however, that of stratified maps is slightly sharpened. 
Although we will later apply the whole framework to 
the analytic category, we assume at this point only
that we have fixed some smoothness category $C^p$ with $p\in\N$ or $p = \infty$
or $p=\omega$.

Let $M$ and $N$ be $C^p$ manifolds.
\bdf
Let $A$ be some subset in $M$.
\bunum
\item
A \df{stratification} $\strat M$ of $M$ is a locally finite, disjoint
decomposition of $M$ into connected embedded $C^p$ submanifolds $M_i$ of $M$ 
(the so-called \df{strata}), such that
\bgl
M_i \cap \del M_j \neq \leeremenge
 & \impliz & 
M_i \teilmenge \del M_j \breitrel{\text{ and }} \dim M_i < \dim M_j
\egl
for all $M_i, M_j \in \strat M$.
\item
A stratification $\strat M$ of $M$ is called stratification of 
$A$ in $M$ iff $A$ is the union of certain elements in $\strat M$.
\item
$A$ is a \df{stratified set} (w.r.t.\ $M$) iff there 
is a stratification of $A$ in $M$.
\eunum
\edf

\bdf
Let $\strat M_1$ and $\strat M_2$ be two stratifications of
some subset of $A$.

Then $\strat M_1$ is called \df{finer} than $\strat M_2$
iff each stratum in $\strat M_2$ is a union of strata in $\strat M_1$.
\edf

\bdf
A map $f : M \nach N$ is called 
\bunum
\item
\df{stratified map} iff 
$f$ is continuous and 
there are stratifications $\strat M$ and $\strat N$ of $M$ and $N$,
respectively, such that for every $M_i \in \strat M$ 
there is an open $U_i \teilmenge M$ 
and a $C^p$ differentiable 
map $f_i : M_i \teilmenge U_i \nach N$ with
\zgl{
\quer{M_i} \teilmenge U_i,
\breitrel{}
f_i\einschr{M_i} = f\einschr{M_i},
\breitrel{}
f_i(M_i) \in \strat N, 
\breitrel{}
\rank f\einschr{M_i} = \dim f(M_i);
}
\item
\df{stratified monomorphism} iff, additionally, $f\einschr{M_i}$ 
is injective;
\item
\df{stratified isomorphism}%
\footnote{Sometimes we will use 
``stratified diffeomorphism'' synonymously.} 
iff, additionally, $f$ is a homeomorphism and
each $f_i : U_i \nach f_i(U_i)$ is a $C^p$ diffeomorphism.
\eunum
If we drop the above conditions that $U_i$ is open and that
$\quer M_i$ is contained in $U_i$, we speak about
\df{weakly} stratified maps.

\edf

\bdf

A stratified map $f : M \nach M$ 
is called \df{localized} iff $f$ is the identity outside
some compact subset of $M$.
\edf

\bdf

Two stratified sets $S_1$ and $S_2$ in $M$
are called \df{(weakly) strata equivalent} iff there is a product  
of localized (weakly) stratified isomorphisms mapping $S_1$ onto $S_2$.
They are called \df{oriented-strata equivalent} iff there is such a 
product mapping additionally the orientation of $S_1$ to that
of $S_2$.
\edf

%------------------------------------------------------------------------%
%                                                                        %
%------------------------------------------------------------------------%
\subsection{Localized Stratified Diffeomorphisms in Linear Spaces}

In the sections below, we will have to study the local transformation behaviour 
of geometric objects in manifolds. To get prepared for this,
we will now investigate first the corresponding problems in linear spaces.
In particular, we will be able to 
rotate, scale and translate these objects locally, i.e.\
by transformations that are the identity outside some 
bounded region. This guarantees that 
we may lift the corresponding operations to manifolds.

We recall that a $q$-simplex $S$ in $\R^k$ with $q \leq k$ is the
closed convex hull of $q+1$ points in general position. The
corresponding interior of $S$ is called open simplex. Moreover, 
the (open) faces of $S$ are the (open) simplices spanned by subsets of these 
$q+1$ points. Additionally, we denote by $B_r^q(x)$,
or shortly $B_r(x)$,
some closed $q$-dimensional ball in $\R^k$
with radius $r$ around $x$. If $x$ is the origin,
we simply write $B_r$. We remark that, 
in this subsection,
nice orientations of some simplex or ball $S$ 
will always mean an orientation induced
by that of some hyperplane (i.e.\ not by some more general hypersurface
as for natural orientations)
containing $S$. This implies, e.g., that the nice orientation
of a $q$-simplex $S$ is always induced by some $(k-1)$-simplex having
$S$ as one of its faces.

Finally, let us remark that in most of the statements of this subsection 
we will use $0$ as a base point. It should be clear that all
these statements hold analogously if $0$ is replaced by any point in $\R^k$.
%------------------------------------------------------------------------%
%                                                                        %
%------------------------------------------------------------------------%
\subsubsection{Strata Equivalence of Star-Shaped Regions}
\blem
\label{lem:nice-diffeo}
Let $k$ be a positive integer and let $U$ be an open
subset of $\R^k$ not containing $0$.
Next, let $a,b,p : U \nach \R$ be $C^p$-functions, such that
both $a$, $p$ and $p a + b$ are positive on $U$. Moreover, for every 
$\lambda > 0$, let
\bglklein
p(\lambda x) & = & \lambda p(x), \\
a(\lambda x) & = & \phantom\lambda a(x), \\
b(\lambda x) & = & \phantom\lambda b(x), 
\eglklein
whenever both $\lambda x$ and $x$ are contained in $U$.
Finally define $\qwer,\qwer_\invers: U \nach \R$ by
\bgl
\qwer \breitrel{:=} a + \frac b p 
& \breitrel{\text{ and }} & 
\qwer_{\invers} \breitrel{:=} \inv a \Bigl(1 - \frac b p\Bigr).
\egl

Then $\dach\qwer : U \nach \R^k$ defined by
\bgl
\dach\qwer(x) & := & \qwer(x)\:x
\egl
is a $C^p$ diffeomorphism between $U$ and $\dach\qwer(U)$
and maps (subintervals of) each half-ray
$\R_+ x$ into (subintervals of) the same half-ray. Moreover, 
its inverse is given by
\bgl
\dach\qwer^{-1}(x) & = & \qwer_{\invers}(x)\:x.
\egl
\elem
\bpf
$\dach\qwer$ is indeed $C^p$, since $p$ never vanishes. 
Since $\dach\qwer$ at a single $x$ is just a positive scalar multiplication,
it maps (subintervals of) each half-ray
$\R_+ x$ into (subintervals of) the same half-ray.
Moreover, $\qwer$ is injective and the image of $\qwer$ is an open subset of 
$\R^k$. 
Finally, one checks immediately, that $\dach\qwer^{-1}$ is $C^p$ and that
it is the inverse of $\qwer$ by $pa+b > 0$.
\qed
\epf

\blem
\label{lem:diffeo-2surfs}
Let $k$ be a positive integer. Let $S_0$ and $S_1$ be
the boundaries of two bounded open regions $R_0$ and $R_1$ in $\R^k$ 
both containing $0$. 
Assume, moreover, that each $R_i$ is star-shaped,
that the corresponding Minkowski functional $p_i$ for $R_i$ is $C^p$
and that each $S_i$ is an embedded $C^p$ submanifold of $\R^k$. 

Then, for all real $\lambda_\pm$ and $\lambda_{0,\pm}$ with
\bgl
0 \breitrel< \lambda_- \breitrel< 
\inf_{\R^k \setminus \{0\}} \: \frac{p_1}{p_0}
\breitrel\leq \lambda_{0,-}
 \breitrel{\text{ and }} 
\lambda_{0,+} \breitrel\leq 
\sup_{\R^k \setminus \{0\}} \: \frac{p_1}{p_0} 
\breitrel< \lambda_+,
\egl
there are $C^p$ mappings $\dach\qwer_+$ and $\dach\qwer_-$ 
with the following properties:
\bnum3
\item
$\dach\qwer_\pm$ is a $C^p$ diffeomorphism 
from some open neighbourhood of $V_\pm$ 
onto some neighbourhood of $W_\pm$. Here,
\bgl 
 V_- & = & \{x \in \R^k \mid \lambda_- \leq p_1(x) \:\text{ and }\: p_0(x) \leq 1\}, \\
 V_+ & = & \{x \in \R^k \mid 1 \leq p_0(x) \:\text{ and }\: p_1(x) \leq \lambda_+ \}, \\
 W_- & = & \{x \in \R^k \mid \lambda_- \leq p_1(x) \leq \lambda_{0,-}\}, \\
 W_+ & = & \{x \in \R^k \mid \lambda_{0,+} \leq p_1(x) \leq \lambda_+\}
\egl
are compact sets with nonempty interior.
\item
$\dach\qwer_\pm$ maps $S_0$ to $\lambda_{0,\pm} S_1$;
\item
$\dach\qwer_+$ and $\dach\qwer_-$ coincide on $S_0$ 
if $\lambda_{0,-} = \lambda_{0,+}$;
\item
$\dach\qwer_\pm$ is the identity on $\lambda_\pm S_1$; 
\item
$\dach\qwer_\pm$ maps subintervals of half-rays to subintervals of the
same half-ray.
\item
The restrictions of $\dach\qwer_\pm$ 
to (an appropriate open subset of) 
any linear subspace of $\R^k$ are diffeomorphisms into 
that linear subspace.
\enum
\elem

\bcorr
\label{corr:diffeo-2surfs}
Given the assumptions of Lemma \ref{lem:diffeo-2surfs},
there is a stratified $C^p$ diffeomorphism $\diffeo$ mapping
$S_0$ to $\lambda_0 S_1$ and $R_0$ to $\lambda_0 R_1$ for some
$\lambda_- \leq \lambda_0 \leq \lambda_+$, such that 
$\diffeo$ is the identity inside $\lambda_- R_1$ and 
outside $\lambda_+ R_1$. Moreover, $\diffeo$ can be chosen, such that
it preserves half-rays and its restrictions to linear subspaces of
$\R^k$ are stratified $C^p$ diffeomorphisms again.
\ecorr
\bpf
Simply define $\diffeo$ to equal $\dach\qwer_\pm$ on $V_\pm$ 
and to be the identity otherwise. Since these mappings coincide
on the corresponding overlaps $\lambda_- S_1$, $S_0$ and $\lambda_+ S_1$,
we get the assertion.
\qed
\epf
Note that $\lambda_\pm$ does only depend on the {\em relative}\/
shape of $S_0$ and $S_1$. In particular, $\lambda_\pm$ need
not be changed if both $S_0$ and $S_1$ are scaled by the same factor.

\bpf[Lemma \ref{lem:diffeo-2surfs}]
Denote $R_\pm := \lambda_\pm R_1$ and, correspondingly, 
$S_\pm := \del R_\pm \ident \lambda_\pm S_1$.
By choice of $\lambda_\pm$, we have 
$\quer{R_-} \teilmenge R_0 \teilmenge \quer{R_0} \teilmenge R_+.$
Furthermore, let us define $q := \frac{p_1}{p_0}$ on $V := \R^k \setminus \{0\}$
and let 
\bgl
a_\pm \breitrel{:=} \frac{\lambda_\pm - \lambda_{0,\pm}}{\lambda_\pm - q}
& \breitrel{\text{ and }} & 
b_\pm \breitrel{:=} \lambda_\pm\:\frac{\lambda_{0,\pm} - q}{\lambda_\pm - q}
\egl
define functions $a_\pm,b_\pm : V \nach \R$.
Of course, $a_\pm$ is positive.
Since Minkowski functionals are semilinear%
\footnote{This means $p(\lambda x) = \lambda p(x)$ for all $\lambda > 0$.}, 
we see immediately that $q$, and so $a$ and $b$ as well,
are constant on each half-ray $\R_+ x$.
Observe that 
$a$ and $b$ are well defined by choice of $\lambda_\pm$ and $\lambda_{0,\pm}$.
Finally, we define
$\dach\qwer_\pm(x) := \qwer_\pm(x) \: x $ on $V$
by
\bgl
\qwer_\pm & := & a_\pm + 
\frac{b_\pm}{p_1} 
 \breitrel= \frac{\lambda_\pm\:(p_1+\lambda_{0,\pm} -q) - \lambda_{0,\pm} p_1}{p_1 \: (\lambda_\pm - q)}.
\egl
We have 
\bglklein
(q - \lambda_\pm) (p_1 a_\pm + b_\pm)
 & = & p_1(\lambda_{0,\pm} - \lambda_\pm) + \lambda_\pm(q - \lambda_{0,\pm}) \\
 & = & p_1(\lambda_\pm(\inv{p_0} - 1) + \lambda_{0,\pm}) - \lambda_\pm \lambda_{0,\pm} \\
 & = & (p_1 - \lambda_\pm) \lambda_{0,\pm} + p_1 \lambda_\pm (\inv{p_0} - 1).
\eglklein
Let us check the properties of $\dach\qwer_\pm$:
\bunum
\item
$\dach\qwer_\pm$ is obviously a $C^p$ function mapping 
subintervals of half-rays
to subintervals of the same half-ray.
\item
Let $x\in S_0$, i.e.\ $p_0(x) = 1$, hence $q(x) = p_1(x)$. Then 
$p_1(\qwer_\pm(x)x) = \qwer_\pm(x) p_1(x) = \lambda_{0,\pm}$, 
i.e.\ $\dach\qwer_\pm(x) \in \lambda_{0,\pm} S_1$.
In particular, $\dach\qwer_-(x) = \dach\qwer_+(x)$ if 
$\lambda_{0,-} = \lambda_{0,+}$.
\item
Let $x\in\lambda_\pm S_1$, i.e.\ $p_1(x) = \lambda_\pm$. 
Then $\qwer_\pm(x) = 1$, hence $\dach\qwer(x) = x$.
\item
Let $x\in V_-$, i.e.\ $p_0(x) \leq 1$ and $p_1(x) \geq \lambda_-$.
From the lines above, we see that this implies $(p_1 a_- + b_-)(x) \geq 0$,
the equality holding iff $p_0(x) = 1$ and $p_1(x) = \lambda_-$. This, however,
is impossible, since $q(x)$ would be equal $\lambda_- < \inf_V q$.
Therefore, $p_1 a_- + b_- > 0$ on $V_-$. Since, by construction,
$V_-$ is compact, there is some open neighbourhood of $V_-$ where 
$p_1 a_- + b_- > 0$.
Lemma \ref{lem:nice-diffeo} now shows that $\dach\qwer_\pm$ is a $C^p$ diffeomorphism
on that neighbourhood. By the previous items we see that 
$\dach\qwer_-(V_-) = W_-$.
\item
The corresponding properties of $\dach\qwer_+$ are proven completely
analogously.
\item
By intersecting $R_i$ and $S_i$ with linear subspaces we get 
$C^p$ boundaries with $C^p$ Minkowski functionals again. The remaining
statements are now clear.
\qed
\eunum
\epf

%------------------------------------------------------------------------%
%                                                                        %
%------------------------------------------------------------------------%
\subsubsection{Scaling}
To study geometric objects in charts, it may be necessary to first
shrink them to have enough ``space''. That this is (almost) always possible
using stratified diffeomorphisms, guarantees the following
\blem
\label{lem:locstrat_scaling}
Let $k$ be a positive integer and 
let $R$ be a bounded star-shaped open region in $\R^k$ containing $0$
and having a $C^p$ differentiable Minkowski functional $p$.
Moreover, assume that the boundary $S$ of $R$ is an embedded $C^p$ submanifold
of $\R^k$.

Then for all $\lambda > 0$ and all $\varepsilon > 0$, there 
is a stratified $C^p$ isomorphism $\diffeo$ preserving
half-rays, such that 
$\diffeo = \lambda \id$ on $\quer R$ and
$\diffeo = \ido$ outside $(1+\varepsilon) \max(\lambda,1) R$.
\elem
\bpf
\bunum
\item
Assume first $\lambda \geq 1$.

Choosing 
$\lambda_{0,+} := \sqrt{\lambda}$ and $\lambda_+ := (1+\varepsilon)\sqrt{\lambda}$,
we may apply 
Lemma \ref{lem:diffeo-2surfs} to $R_0 := R$ and $R_1 := \sqrt\lambda \: R$ with
$p = p_0 = \sqrt\lambda \: p_1$.
For this, define 
\bgl
\diffeo(x) & := & \begin{cases}
   \lambda \id  & \text{ on \: $p^{-1}\bigl([0,1]\bigr)$} \\
   \dach\qwer_+ & \text{ on \: $p^{-1}\bigl([1,(1+\varepsilon)\lambda]\bigr)$} \\
   \ido         & \text{ on \: $p^{-1}\bigl([(1+\varepsilon)\lambda,\infty)\bigr)$} 
   \end{cases}.
\egl
Now, let $x\in S$, i.e.\ $p(x) = 1$. Then, by construction,
$p_1(\dach\qwer_+(x)) = \lambda_{0,+} = \sqrt\lambda$, hence 
$p_0(\dach\qwer_+(x)) = \lambda = p_0(\lambda x)$. 
For $x\in (1+\varepsilon)\lambda S$,
we have $p(x) = (1+\varepsilon)\lambda$, hence 
$p_1(x) = (1+\varepsilon)\sqrt\lambda = \lambda_+$.
By definition, we get $\dach\qwer_+(x) = x$. Altogether,
$\dach\qwer_+$ equals $\lambda \id$ on $S = \del R$ and
$\ido$ on $(1+\varepsilon)\lambda S = (1+\varepsilon)\lambda \:\del R$.
Therefore, $\diffeo$ is a stratified diffeomorphism having the desired
properties.
\item
Assume now $\lambda \leq 1$.

Define $\lambda_+ := \sqrt{1+\varepsilon}$ and 
$\lambda_{0,+} := \lambda (\sqrt{1+\varepsilon})^{-1}$,
and apply 
Lemma \ref{lem:diffeo-2surfs} to $R_0 := R$ and 
$R_1 := \sqrt{1+\varepsilon} \: R$ with
$p = p_0 = \sqrt{1+\varepsilon} \: p_1$:
Define 
\bgl
\diffeo(x) & := & \begin{cases}
   \lambda \id  & \text{ on \: $p^{-1}\bigl([0,1]\bigr)$} \\
   \dach\qwer_+ & \text{ on \: $p^{-1}\bigl([1,1+\varepsilon]\bigr)$} \\
   \ido         & \text{ on \: $p^{-1}\bigl([1+\varepsilon,\infty)\bigr)$} 
   \end{cases}.
\egl
For $x\in S$, i.e.\ $p(x) = 1$, we have 
$p_1(\dach\qwer_+(x)) = \lambda_{0,+} = \lambda (\sqrt{1+\varepsilon})^{-1}$. 
Therefore, we get 
$p_0(\dach\qwer_+(x)) = \lambda = p_0(\lambda x)$. 
If, on the other hand, $x\in (1+\varepsilon) S$,
we have $p(x) = 1+\varepsilon$ and 
$p_1(x) = \sqrt{1+\varepsilon} = \lambda_+$,
implying $\dach\qwer_+(x) = x$. Consequently,
$\dach\qwer_+$ equals $\lambda \id$ on $S = \del R$ and
$\ido$ on $(1+\varepsilon) S = (1+\varepsilon) \:\del R$.
Now, $\diffeo$ is a stratified diffeomorphism having the desired
properties.
\qed
\eunum
\epf
%
%------------------------------------------------------------------------%
%                                                                        %
%------------------------------------------------------------------------%
\subsubsection{Rotation}
\blem
\label{lem:locstrat_rot}
Let $k$ be a positive integer and
let $r_1 > r_2 > 0$ be real. Let 
$X \in \gotso(k)$, define $A := \e^X \in SO(k)$
and denote the orthogonal projection from $\R^k$ to $(\ker X)^\senk$ by $P$.

Then there is a stratified diffeomorphism $\diffeo$ 
of $\R^k$, such that
\bunum
\item
$\diffeo$ coincides with $A$ on $B_{r_2}$;
\item
$\diffeo$ is the identity outside of $B_{r_1}$;
\item
$\diffeo$ is norm preserving;
\item
$\diffeo$ is homotopic to the identity;
\item
$P\:\diffeo = P$.
\eunum
\elem
\bpf
We stratify $\R^k$ into 
\zgl{\inter B_{r_2} \cup \del B_{r_2} \cup (\inter B_{r_1} \setminus B_{r_2}) 
  \cup \del B_{r_1} \cup (\R^k\setminus B_{r_1})}
and define three auxiliary $C^p$ functions $a_i : \R \nach \R$ with
\zgl{
a_{12} (r) := 1, \breitrel{} 
a_{234} (r) := \frac{r_1 - r}{r_1 - r_2} \breitrel{ \text{ and } }
a_{45} (r) := 0.}
One now immediately checks that
\zgl{ 
\diffeo(x) \breitrel{:=} 
  \begin{cases}
     \e^{a_{12}(\norm{x}) X} \: x & \text{ if $x \in \inter B_{r_2} \cup \del B_{r_2}$} \\
     \e^{a_{234}(\norm{x}) X} \: x & \text{ if $x \in \del B_{r_2} \cup (\inter B_{r_1} \setminus B_{r_2}) \cup \del B_{r_1}$} \\
     \e^{a_{45}(\norm{x}) X} \: x & \text{ if $x \in \del B_{r_1} \cup (\R^k\setminus B_{r_1})$} \\
  \end{cases} 
}
gives the desired map.%
\footnote{Moreover, note that the three functions used to define $\diffeo$
are defined on full $\R^k$ (possibly, up to the origin).}
For the homotopy property define
$\diffeo_t$ as above with $tX$ instead of $X$. Then $\diffeo_1 = \diffeo$ 
and $\diffeo_0 = \id$.
\qed
\epf
Immediately from the proof, we get with the above assumptions 
\bcorr
\label{corr:loc_rot(path)}
Let $k$ be a nonnegative integer and let $\varepsilon > 0$. 
Moreover, let $\gamma_{\winkel}$ be the straight line in $\R^2$ connecting
$(-\cos\winkel,-\sin\winkel)$ and $(\cos\winkel,\sin\winkel)$.

Then for each $\winkel\in\R$ 
there is a stratified isomorphism $\diffeo_\winkel$
of $\R^2 \dirsum \R^k$, such that 
\bunum
\item
$\diffeo_\winkel$ 
is the identity outside $B_{1+\varepsilon} \teilmenge \R^{k+2}$;
\item
$\diffeo_\winkel$ is norm preserving; 
\item
$\diffeo_\winkel$ is homotopic%
\footnote{The mapping is given by $t \auf \diffeo_{t\winkel}$.} 
to the identity; 
\item
$P \diffeo_\winkel = P$,
where $P$ is the canonical projection from $\R^2\dirsum\R^k$ 
to $\R^k$;
\item
$\diffeo_\winkel$ maps $\gamma_0$ to $\gamma_\winkel$.
\eunum
\ecorr
\bpf
Choose 
$X = \winkel \: \bmat \phantom-0 & 1 \\ -1 & 0 \emat \in \gotso(2) 
\teilmenge \gotso(2) \kreuz \gotso(k) \teilmenge \gotso(2+k)$.
\qed
\epf

%------------------------------------------------------------------------%
%                                                                        %
%------------------------------------------------------------------------%
\subsubsection{Translation}
\blem
\label{lem:locstrat_translat}
Let $k$ be some positive integer.
Let $\gamma$ be some edge in $\R^k$ and $U$ be some
neighbourhood of $\gamma$. Choose $r > 0$, such that
the balls with radius $r$ centered at $\gamma(0)$ and $\gamma(1)$, respectively,
are contained in $U$.

Then there is a finite product of stratified $C^p$ diffeomorphisms of $\R^k$ 
being the identity outside $U$ and the translation by $\gamma(1) - \gamma(0)$
on $B_r(\gamma(0))$.
\elem
\bpf
We only give the idea of the proof. The technical details
are similar to that for the preceding statements. Moreover, in 
Lemma \ref{lem:winddiffeo-bumps} we will give a proof for a more specific type of translation.

Here, we cover $\gamma$ by (non-trivial) balls. By compactness, there is some
$r'$, such that finitely many balls with radius $r'$ centered at 
points of $\gamma$ will cover $\gamma$ and such that the convex hull
of ``neighbouring'' balls is contained in $U$.
The idea now is to first shrink
$B_r(\gamma(0))$ to $B_{r'}(\gamma(0))$, then move parallelly
this ball through the convex hulls of neighbouring balls and finally
blow it up to its original size. All these operations are possible
by the statements above without moving any point outside $U$.
\qed
\epf

%------------------------------------------------------------------------%
%                                                                        %
%------------------------------------------------------------------------%
\subsubsection{Strata Equivalence of Simplices and Balls}

Let us now show that all $q$-simplices are not only isomorphic
as simplices themselves, but can also be mapped into each other 
by localized stratified $C^p$ diffeomorphisms. Moreover, 
they are equivalent to $q$-dimensional balls.

\bprop
\label{prop:strateq_simplball}
Let $q \leq k$ be two positive integers. 

Then all $q$-simplices and all $q$-dimensional balls in $\R^k$ are
strata equivalent.
\eprop
For this, we first show that each $q$-simplex
can be mapped to a $q$-dimensional ball.
\blem
\label{lem:strateq_simplball}
Let $q \leq k$ be two positive integers.
Moreover, let $V := \{v_0,\ldots,v_q\} \teilmenge \R^k$
contain $q+1$ points in general position, such that 
$0$ is contained in the interior of the $q$-simplex $R_V$ spanned
by $V$. Finally, fix some $\varepsilon > 0$ and some $r > 0$, such that 
$R_V$ is contained completely in the interior of $B_r$.

Then there is a stratified $C^p$ diffeomorphism $\diffeo$,
being the identity outside of $B_{(1+\varepsilon)r}$, such that 
$R_V$ is mapped to $B_r \cap \spann_\R V$.
\elem
\bpf
Choose some set $V' = \{v'_0,\ldots,v'_{k-q}\} \teilmenge \R^k$
of $k-q+1$ points in general position,
such that its span is complementary to that of $V$ and such that
the $(k-q)$-simplex spanned by $W$ contains $0$ in its interior
and is contained in $\inter B_r$.
Define for every $0\leq i \leq q$ and $0\leq j \leq k-q$ the
set 
\zgl{V_{ij} := \{0\} \cup (V \setminus \{v_i\}) \cup (V \setminus \{v'_j\})}
now containing $k+1$ points in general position, hence
each defining a $k$-simplex $R_{ij}$. These simplices form a complex, i.e.,
in particular, they share at most lower-dimensional faces.
Let $R_0$ be the union of all these $(k-q+1)(q+1)$ simplices.
Its boundary is the union of the simplices $V^0_{ij}$ spanned
by $V_{ij} \setminus \{0\}$.

Let us now invoke Corollary \ref{corr:diffeo-2surfs}. First of all,
observe that the statement there
can be extended directly to the case that $R_0$ is
formed by a finite number of cones each having tip at $0$
and each defined by $k$-simplices, such that these cones fill $\R^k$
completely and share at most the boundaries with each other. Of course,
the requirements for $S_0$ have to be relaxed accordingly.
We refrained
from explicitly giving this form of Corollary \ref{corr:diffeo-2surfs}
(and Lemma \ref{lem:diffeo-2surfs}, respectively),
since it would have made the proof even more technical without 
introducing new ideas.
One simply has to construct the stratified diffeomorphism in the more
general case for every cone (more precisely, some open appropriate 
neighbourhood of it) and then use that these mappings fit together
at the boundaries. This however, follows from the coincidence of
the Minkowski functionals at these boundaries, the 
construction of the maps in the proofs above and the invariance
of half-rays. 

Coming back to the present proof, define $R_1$ to be $B_r$. Then,
by $R_0 \teilmenge B_r$, the corresponding 
Minkowski functionals fulfill $p_1 \leq p_0$, and we may choose
$\lambda_+ = 1+\varepsilon > 1$. This means that, by 
Corollary \ref{corr:diffeo-2surfs}, there is a stratified 
$C^p$ diffeomorphism $\diffeo$ being the identity outside $\lambda_+ B_r$,
mapping $R_0$ to $R_1$ and $\del R_0$ to $\del R_1$. Now
the assertion follows, since $\diffeo$ preserves linear subspaces.
Therefore, $R_V$ (being the intersection of $R_0$ with 
$\spann_\R V$) is mapped to $B_r \cap \spann_\R V$ being
a $q$-dimensional ball. 
\qed
\epf

\bpf[Proposition \ref{prop:strateq_simplball}]
Let two $q$-simplices be given. Using Lemma \ref{lem:locstrat_translat},
translate both, such that they contain $0$ in their interior. 
Then each of them is strata equivalent to some $q$-dimensional sphere
in $\R^k$, by Lemma \ref{lem:strateq_simplball}. 
Shrinking these balls, if necessary,
we make them of identical radius. Finally, by Lemma \ref{lem:locstrat_rot},
we may find some localized stratified $C^p$ diffeomorphism rotating
one ball into the other. Hence, these two $q$-simplices are strata equivalent
to a (hence, any) $q$-dimensional ball.
\qed
\epf

Now we are going to mirror simplices and balls into each other.
\neueseite

\bprop
\label{prop:inv_orient_simplball}
Let $q < k$ be two non-negative integers. 

Then every $q$-simplex and every $q$-dimensional ball in $\R^k$ 
having a nice orientation, is strata equivalent to
itself having inverse orientation.
\eprop
\bpf
First assume that $q = k-1$ and consider some $q$-dimensional ball $B$
around the origin. Choose $X \in \gotso(k)$, such that 
$X$ is zero on some $(k-2)$-dimensional linear subspace $V$ of $\spann_\R B$
and generates a rotation in the two-dimensional
complement in $\R^k$ spanned by the normal of $V$ in $\spann_\R B$
and the normal of $\spann_\R B$ in $\R^k$. 
In particular, it generates some map $A := \e^{tX} \in SO(k)$,
being minus the identity on this two-dimensional space for some $t$.
Since only one of its ``dimensions'' belongs to $B$, 
the rotation $A$ inverts the orientation of $B$. Now,
Lemma \ref{lem:locstrat_rot} guarantees 
the existence of some stratified diffeomorphism
inverting the orientation of $B$.

To prove the statement for $q = k-1$ and a given $q$-simplex $S$,
we map it to some $q$-dimensional ball $B$, invert its rotation
and take the inverse of the first mapping to get $S$ back. Of course,
the orientation of $S$ has been flipped.

Next, let $q$ be arbitrary and consider a $q$-simplex $S$. 
Since we work with nice orientations
only, 
there is some $(k-1)$-simplex $S'$
in $M$ having $S$ as one of its faces
and inducing its orientation. 
Since we may invert the orientation of $S'$, we also may invert 
that of $S$ by localized stratified diffeomorphisms.

To prove the remaining case of $q$-balls for arbitrary $q$, reuse
the argumentation above for $q = k-1$ and reduce to the case
of $q$-simplices.
\qed
\epf
\bcorr
\label{corr:orientstrateq_simplball}
Let $q < k$ be two non-negative integers. 

Then all $q$-simplices and all $q$-balls in $\R^k$ are
oriented-strata equivalent, provided they have nice orientations.
\ecorr
\bpf
Assume, first of all, that $S$ is a $q$-simplex or a $q$-ball containing
the origin, and
let $S$ be given two nice 
orientations. This means, there are 
linear subspaces $T_1$ and $T_2$ inducing these orientations by their
own nice ones. There is now some $A \in SO(k)$ leaving the $q$-plane
spanned by $S$ invariant and mapping $T_1$ onto $T_2$. Hence 
$A$ maps the one orientation of $S$ to either the other one or
the inverse of it. 
Hence, by Lemma \ref{lem:locstrat_rot}, there is some localized stratified 
isomorphism
mapping $S$ onto itself and transforming the orientations by $A$.
By adding, if necessary, some localized stratified isomorphism inverting
the orientation as given by Proposition \ref{prop:inv_orient_simplball}, 
we get such a transformation mapping the
two orientations of $S$ onto each other.

Let now $S_i$ be a $q$-simplex or a $q$-ball for $i=1,2$. Then we may
map them by localized stratified diffeomorphisms to some $q$-simplex $S$
containing the origin. Since, as one checks immediately, these
mappings can be chosen, such that the corresponding orientations of $S$
are nice%
\footnote{One sees that all necessary transformations
are locally ``affine''.},
there is a localized stratified diffeomorphism mapping one orientation of $S$
to the other, by the arguments above.
\qed
\epf
Without explicitly stating the proof, we have by arguments as 
in the proposition above:
\bcorr
\label{corr:gerade-inv}
For every nicely oriented $1$-dimensional ball $S$ in $\R^k$ with $k\geq 3$,
there are finitely many localized stratified isomorphisms,
whose product is the identity on $S$, but inverts the orientation of $S$.
\ecorr

Finally, we are looking for objects that can be divided into
two parts, such that the original one is, on the one hand,
strata equivalent to both of them and, on the other hand,
is the disjoint union of them. Moreover, the orientation
should be preserved. For example, consider an open $2$-simplex, i.e.\ a 
full open triangle.
Intersecting it by a line through one corner and some point of the
opposite edge, we get two triangles. If we take their
interiors, then they are strata equivalent to the original 
triangle, however not a decomposition of it -- simply the border line
is missing. One the other hand, if we were taking it to just 
one of the subtriangles, then they are no longer strata equivalent.
The solution of this problem is to consider at the beginning an open 
triangle plus one of its open edges. Then, as above, we may divide the
triangle by a line, now through some boundary point of the added edge.
Now it is clear that the triangle plus edge is divided into
twice a triangle plus edge and all three objects are strata equivalent.
The generalization to higher dimensions is straightforward,
but more technical:
\bprop
\label{prop:two-in-one-simplex}
Let $q < k$ be two positive integers. 
Let $S$ be some open $q$-simplex in $\R^k$, and let 
$F$ one of its open $(q-1)$-faces.
Finally, give $R := S \cup F$ the orientation induced
by one of the nice orientations of $\quer S \obermenge R$.

Then there are products $\diffeo_0$ and $\diffeo_1$
of localized stratified isomorphisms, such that 
$R$ is the disjoint union of $\diffeo_0 R$ and $\diffeo_1 R$
and the intersection function of $R$ is the joint intersection
function of $\diffeo_0 R$ and $\diffeo_1 R$.
\eprop
\bpf
First of all, 
choose some open $(k-1)$-simplex $\dach S$, such that 
its orientation is induced by one of the nice orientations
of its closure which, on the other hand,
induces the orientations of $S$ and $R$. Let $V$ be the set of $k$ points
$\{v_0,\ldots,v_{k-1}\}$ in $\R^k$,
such that
\bgl
\dach S & \text{is the interior of the simplex spanned by} & \{v_0,\ldots,v_{k-1}\}, \\
\dach F & \text{is the interior of the simplex spanned by} & \{v_1,\ldots,v_{k-1}\}, \\
S & \text{is the interior of the simplex spanned by} & \{v_0,\ldots,v_q\}, \\
F & \text{is the interior of the simplex spanned by} & \{v_1,\ldots,v_q\}. 
\egl
Define $\dach R := \dach S \cup \dach F \cup S \cup F$.
Now choose some $v$ in the open $1$-face connecting $v_0$ and $v_1$,
and cut $\dach R$ by the plane spanned by $\{v,v_2,\ldots,v_{k-1}\}$
into two parts $\dach R_0$ and $\dach R_1$, whereas the intersection
of this plane with $\dach R$ is added to $\dach R_0$,
and $\dach R_1$ is that ``half'' whose closure
contains $v_1$. We now may decompose each
$\dach R_i$ into $\dach S_i \cup \dach F_i \cup S_i \cup F_i$,
where 
\bgl
\dach S_i & \text{is the interior of the simplex spanned by} & \{v,v_i,v_2,\ldots,v_{k-1}\}, \\
\dach F_i & \text{is the interior of the simplex spanned by} & \{x_i,v_2,\ldots,v_{k-1}\}, \\
S_i & \text{is the interior of the simplex spanned by} & \{v,v_i,v_2,\ldots,v_q\}, \\
F_i & \text{is the interior of the simplex spanned by} & \{x_i,v_2,\ldots,v_q\}
\egl
with $x_0 = v$ and $x_1 = v_1$. Obviously, 
$S \cup F = S_0 \cup F_0 \cup S_1 \cup F_1$.

Let now $\diffeo_i$ be products of localized stratified isomorphisms
that leave $v_j$ with $j \geq 2$ and $v_i$
invariant, map $v_{1-i}$ to $v$ and map
the simplex spanned by $\{v_0,\ldots,v_{k-1}\}$
onto that spanned by $\{\diffeo_i(v_0),\diffeo_i(v_1),v_2,\ldots,v_{k-1}\}$.
It is easy to check that 
$\diffeo_i (S \cup F) = S_i \cup F_i$ and that $\diffeo_i$
may be chosen to have the desired orientation
properties.
\qed
\epf

%------------------------------------------------------------------------%
%                                                                        %
%------------------------------------------------------------------------%
\subsection{Localized Stratified Diffeomorphisms in Manifolds}
We are now going to transfer the results of the previous 
subsection to the case of general $C^p$ manifolds $M$.

\bdf
A subset $S$ in $M$ is called 
\df{(nicely oriented) $q$-simplex in the chart $(U,\karte)$} iff
$S \teilmenge U$ is mapped by $\karte$ to a $q$-simplex 
in $\R^{\dim M}$ (and the orientation of $S$ is induced 
by one of the natural orientations of some hyperplane in $\karte(U)$).
\edf
Analogously, we may define $q$-balls. The definition of faces 
of $q$-simplices should be clear as well. We will speak about
$q$-simplices and $q$-balls in general iff there is a chart of $M$,
which they are $q$-simplices or $q$-balls in. Note that, at least locally,
every simplex or ball $S$ having a natural orientation is 
nicely oriented, i.e., it is induced by some 
hypersurface being (an open set of) a hyperplane in some chart.
In fact, let $N$ be some embedded submanifold in $M$ 
containing $S$ as an embedded submanifold and inducing its orientation.
Then we may find some chart mapping $N$ locally into some hyperplane in the
local chart image of $M$ and mapping $S$ locally into some plane 
in the local image of $N$.

\bprop
\label{prop:locstrat_mfs}
The statements of 
Propositions \ref{prop:strateq_simplball}, \ref{prop:inv_orient_simplball} and \ref{prop:two-in-one-simplex}
as well as of 
Corollaries \ref{corr:orientstrateq_simplball} and \ref{corr:gerade-inv}
remain valid if we replace $\R^k$ by $M$ and 
assume all $q$-simplices and $q$-balls to be in one and the same 
connected chart and, moreover, nicely oriented there. 
\eprop

\bpf
The only point to be shown is the case that the localized stratified 
isomorphism $\diffeo$ needs more space 
in $\R^{\dim M}$ than provided by the chart denoted by $(U,\karte)$.
If this is the case, first shrink any occurring object $S$ (being
a ball or a simplex)
to a sufficiently small size. Indeed, since 
simplices and balls are 
assumed to be closed and the chart is open, $S$ 
-- magnified (in the chart) by $1+\varepsilon$ w.r.t.\ some interior point --
is again in $U$ for small $\varepsilon$. Therefore, the scaling lemma 
(Lemma \ref{lem:locstrat_scaling}) is applicable in order to shrink $S$
by any factor $\lambda \leq 1$.
Now it may be necessary to move $S$ to some other place in $U$ inside this
chart. To do this, we choose some path that $S$ is moved along.
By compactness and continuity reasons, there is a finite number of 
open $k$-balls in $U$ covering this path. We now assume 
that $\lambda$ is chosen small enough that the accordingly shrunk $S$ can be 
transferred between any non-disjoint two of these balls 
by means of Lemma \ref{lem:locstrat_translat}.
This way it can be (parallelly) shifted between any 
two points in the chart.
Using these ingredients of shrinking and shifting, 
it is now easy to generate the desired localized
stratified isomorphisms by means of their counterparts in $\R^{\dim M}$.
\qed
\epf

%------------------------------------------------------------------------%
%            Abschnitt: Introduction                                     %
%------------------------------------------------------------------------%
\subsection{Application to the Analytic Category}
\label{subsect:stratif_anal_iso}
Let us now come back to the analytic case, i.e.\ $p = \omega$.
Recall \cite{m4}
that a subset $A$ of an analytic manifold $M$ is called
semianalytic iff $M$ can be covered by some open sets $U_\iota$, 
such that each $U_\iota \cap A$
is a union of connected components of a set 
$f_1^{-1}(0) \setminus f_2^{-1}(0)$, for $f_1$ and $f_2$ 
belonging to some finite family of real-valued functions analytic in $U_\iota$.
Complements, finite intersections and finite unions of semianalytic sets
are semianalytic again \cite{m6}.
Moreover, it can be shown \cite{m4,m9}
that every semianalytic set admits a semianalytic stratification, i.e.\ a
stratification consisting of semianalytic strata only.%
\footnote{When speaking about semianalytic sets in the following, we will 
tacitly assume that the
corresponding stratifications are semianalytic.}
\blem
Let $\strat M_1$ and $\strat M_2$ be two stratifications of $M$. 

Then there is a stratification $\strat M$ of $M$
being finer than $\strat M_1$ and $\strat M_2$.
\elem
\bpf
For every semianalytic, hence stratifiable set $A$ in $M$ 
and every nonnegative integer
$k$, we choose some semianalytic stratification $\strat N(A)$ of $A$ and 
let $\strat N_k(A)$ contain precisely the $k$-dimensional 
strata in $\strat N(A)$ contained in $A$.
Moreover, let $n$ be the dimension of $M$.

Since the intersection of 
any two semianalytic sets is semianalytic, we may define
\bglklein[0.7ex]
\strat N_{n,k} & := & \bigcup_{M_1\in\strat M_1, M_2\in\strat M_2} \: \strat N_k(M_1\cap M_2), \s
 \strat N'_{n} & := & \bigcup_{k < n} \: \strat N_{n,k}, \s
 \strat N_{n} & := & \strat N_{n,n}. 
\eglklein
This means, $\strat N_{n,k}$ contains 
the $k$-dimensional strata given by all the intersections
of elements in $\strat M_1$ and $\strat M_2$.
Since the boundary of every semianalytic set is semianalytic again \cite{m5}, 
hence stratifiable, we may define successively for decreasing $i$
\bglklein[0.7ex]
 \strat N^\del_{i} & := & 
  \bigcup_{N_1, N_2 \in \strat N_{i+1}} 
  \bigcup_k \: \strat N_k(\del N_1 \cap \del N_2), \s
 \strat N_{i,k} & := & 
  \bigcup_{N \in \strat N^\del_{i}} 
  \bigcup_{N' \in \strat N'_{i+1}} \: \strat N_k(N \cap N'), \s
 \strat N'_{i} & := & \bigcup_{k < i} \: \strat N_{i,k}, \s
 \strat N_{i} & := & \strat N_{i,i}. 
\eglklein
Finally, we set 
\bglklein
\strat M & := & \bigcup_{i=0}^n \strat N_i. 
\eglklein
One immediately checks that $\strat M$ is a stratification. Moreover, by 
construction, it is finer than $\strat M_1$ and $\strat M_2$.
\qed
\epf

\bcorr
Every (weakly) stratified isomorphism is a graphomorphism.
\ecorr
\bpf
Let $\diffeo$ be a weakly stratified isomorphism on $M$ mapping 
$\strat M_1$ to $\strat M_2$. Moreover, let $\gamma$ be some
analytic edge. Since $\im\gamma$ is a semianalytic set, 
there is some stratification $\strat M_3$ of $\im\gamma$.
Choosing some stratification $\strat M$ finer than $\strat M_1$ and
$\strat M_3$, the image of $\gamma$ is a union of strata
in $\strat M$; even a finite one, since $\im\gamma$ is compact.
Refining $\strat M_2$ w.r.t.\ $\diffeo$ and w.r.t.\ the refinement
of $\strat M_1$ to $\strat M$, we see that $\diffeo$ maps
$\im\gamma$ into a finite union of strata. This means that $\diffeo(\gamma)$
is piecewise analytic. The assertion now follows from
Lemma \ref{lem:krit(graphom)}.
\qed
\epf

\bcorr
The (weakly) stratified isomorphisms of $M$ form a subgroup of the group of
homeomorphisms of $M$. 
\ecorr
\bpf
Of course, the inverse of a (weakly) stratified isomorphism is
a (weakly) stratified isomorphism. Therefore, consider two (weakly)
stratified isomorphisms $f_1$ and $f_2$, and choose some
stratifications $\strat M_{11}$, $\strat M_{12}$, $\strat M_{21}$ 
and $\strat M_{22}$ of $M$, 
where $f_i$ maps $\strat M_{i1}$ to $\strat M_{i2}$.
By the previous lemma, we may find some $\strat M$ refining
$\strat M_{12}$ and $\strat M_{21}$. Using the isomorphy property of $f_1^{-1}$
and $f_2$, we refine $\strat M_{11}$ and $\strat M_{22}$ accordingly as well.
With respect to these refined stratifications, $f_1 \circ f_2$ is
a (weakly) stratified isomorphism.
\qed
\epf

\bcorr
The localized (weakly) stratified isomorphisms of $M$ 
form a subgroup of the group of homeomorphisms of $M$. 
\ecorr

The deeper reason behind the investigation of simplices above is the fact
that every manifold can be triangulized, this means, roughly speaking, it
is isomorphic to some union of (open) simplices. Originally known for
nonanalytic manifolds (see, e.g., \cite{Whitehead, Whitney}),
this result has been extended later to semianalytic sets in analytic manifolds 
(see, e.g., \cite{m5}). Here, however, we need a notion 
somewhat stronger than the usual one. In fact, recall that all the results
above on (closed) simplices require that they are contained in some
chart in $M$. Therefore we first quote the definition of a
triangulation from \cite{m5} (dropping, however, some condition)
and then extend this notion to the case we need.
\neueseite
\bdf
Let $\{M_i\}$ be a locally finite collection of semianalytic subsets of $M$.
\bunum
\item
A \df{triangulation} of $\{M_i\}$ is a simplicial complex%
\footnote{A simplicial complex is a locally finite collection $K$
of disjoint open simplices in some finite-dimensional linear space,
such that each face of any simplex belongs to $K$ again. 
Moreover, $\betrag K$ denotes the union of all these simplices.}
$K$ 
together with
a homeomorphism $f : \betrag K \nach M$, such that for every $\sigma\in K$
\bnum3
\item
$f(\sigma)$ is an embedded analytic submanifold of $M$;
\item
$f_\sigma := f\einschr{\sigma} : \sigma \nach f(\sigma)$ 
is an analytic diffeomorphism;
\item
$f(\sigma) \teilmenge M_i$ or $f(\sigma) \teilmenge M \setminus M_i$
for all $M_i$.
\enum
\item
A triangulation $(K,f)$ of $\{M_i\}$  is called \df{wide} iff
for every $\sigma\in K$ there is some open chart 
in $M$ containing the closure of $f(\sigma)$
and mapping it to a simplex in that chart.

If each $M_i$ is given a natural orientation, then we additionally require
$f$ to map this orientation to a nice one on each of these simplices.
\item
$\{M_i\}$ is called \df{(widely) triangulizable}
iff there is a (wide) triangulation of $\{M_i\}$.
\eunum
\edf

\bprop
Every semianalytic set is triangulizable. \cite{m5}
\eprop
One immediately checks that (nicely oriented) $q$-balls and $q$-simplices are 
widely triangulizable.
What remains unsolved is
\bquest
Is every semianalytic set {\em widely}\/ triangulizable?
\equest
Until now, we did not find any proof for either answer in the literature%
\footnote{Nor were we able to find our definition in the literature.}
nor are we able to decide it ourselves. There may be some hints 
for this answer to be affirmative. In fact, 
as proven by Ferrarotti (cf.\ \cite{m7a}),
there is a so-called strong triangulation $(K,f)$ 
of any analytic submanifold $M$ of $\R^n$.
This means that, firstly, 
for any $\sigma\in K$ there is some neighbourhood $U$ of $\sigma$
in the Euclidean space containing $K$, and some analytic
$F_\sigma : U \nach \R^n$ with $F_\sigma = f_\sigma$ and,
secondly, for every vertex $v$ in $K$, the derivative 
$\dd f_v : \quer{{\mathrm{St}}(v,K)} \nach \R^n$ is injective.
But, nevertheless, our case remains open.

%------------------------------------------------------------------------%
%                                                                        %
%------------------------------------------------------------------------%
\subsection{Two Types of Localized Stratified Diffeomorphisms}
\label{subsect:2types(strat_diffeo)}
In this subsection%
\footnote{Note that all results of this subsection remain true in
arbitrary smoothness categories, provided one enlarges
the definition of intersection functions a little bit. Actually, they
are defined only for quasi-surfaces; but only in the analytic
category, hypersurfaces are always quasi-surfaces. The reason for that
was that not {\em every}\/ $\gamma$ can be $S$-admissibly decomposed. 
Here, however, we might study the intersection behaviour of {\em certain}\/ 
paths with $S$. This, of course, is possible in the general case of 
smoothness as well.} 
we will investigate in detail the types of
stratified diffeomorphisms to be used for quantum geometry.

Firstly, we present a more elaborate version
of winding diffeomorphisms introduced originally by Sahlmann \cite{d65}.
The aim is to produce stratified diffeomorphisms that 
wind an edge such that it has a certain number of punctures 
at some given surface. This would be possible even in the analytic
category if only the pure number of punctures would count and 
the precise parameter values of the edge at the punctures
would not matter. In fact, then one can use the 
approximation theorems for smooth mappings by analytic ones \cite{m8}.
If, however, the precise location of the punctures becomes relevant, then 
probably this is no longer sufficient. Therefore,
--~nevertheless reusing the main idea by Sahlmann~--
we present here a more general statement in the 
stratified analytic category. 

Secondly, we study how one can transform a given graph into 
a very large set of independent graphs, but minimally
modifying other geometric objects. There will be two cases 
depending whether a graph is contained in (the closure of) some surface
or not. If not, we may leave the surface invariant pointwise. If,
on the other hand, the graph is (partially) contained in the interior
of the surface, then we may, at least, slightly transform the
surface into itself getting an infinite number of different graphs.
This, of course, is possible, only if this surface provided enough
space, i.e.\ is at least two-dimensional.

%------------------------------------------------------------------------%
%                                                                        %
%------------------------------------------------------------------------%
\subsubsection{Winding Diffeomorphisms}
\label{subsubsect:winddiffeo}
\bprop
\label{prop:wind_diffeo_1surf}
Let $\dim M \geq 3$. 
Let $\gc$ be a graph, let $\gamma\in\gc$ be one of its edges and set
$\gc' := \gc \setminus\{\gamma\}$.
Moreover, let $G$ be a finite subset in $\LG$.

Then there is 
\bunum
\item
some subinterval $I\teilmenge[0,1]$,
\item
some nicely oriented, open, embedded, analytic hypersurface $S$,
disjoint to $\im\gc$, 
such that $S$ and $\del S$ have a finite wide triangulation;
\item
some analytic function $d : S \nach \LG$
and 
\item
some $s\in\Z$, 
\eunum
such that
for any sequences $(g_j) \teilmenge G$ and $(\tau_j) \teilmenge I$
(with $\tau_j < \tau_{j'}$ for $j < j'$) having even length,
there is a stratified (analytic) diffeomorphism $\diffeo$
with the following properties:
\bnum5
\item
$S$ and $\im\diffeo(\gc')$ are disjoint;
\item
$d(\diffeo(\gamma(\tau_j))) = g_j$ for all $j$;
\item
$\diffeo(\gamma)$ intersects $S$ completely transversally;
\item
$\{\diffeo(\gamma(\tau_j))\}_j$ is the set of $\diffeo(\gamma)$-punctures of $S$;
\item
$\sigma^+(S,\diffeo(\gamma)\einschr{[\tau_{j-1},\tau_{j}]}) 
 = (-1)^{j + s} = 
 \sigma^-(S,\diffeo(\gamma)\einschr{[\tau_{j},\tau_{j+1}]})$
for all $j$.
\enum
\eprop

\bpf
First of all, since $\gamma$ is embedded into $M$, there is some neighbourhood
of some subpath $\gamma\einschr I$ of $\gamma$ and some cubic chart, 
such that $\gamma(I)$ is exactly the part of $\im\gc$ in that chart.
Any diffeomorphism constructed below will be constant outside 
this chart. Therefore, to simplify notation, we may restrict ourselves to the
case that $M$ is $\R^n$ coordinatized by $x\in\R$, $y\in\R$ and 
$\vec z \in \R^{n-2}$ (where $z$ is the first coordinate of $\vec z$)
and that $\gamma(I)$ is the intersection of the chart and the $x$-axis.
Next, for the surface $S$ we choose some hypersurface $y = a$ 
parallel to the $x$-axis for some $a > 0$. 
Here, $a$ is selected under the assumption that
$y = 2a + 2\varepsilon_0$ and $y = -2\varepsilon_0$ are still hypersurfaces
in the chart for some $\varepsilon_0 > 0$ 
(of course, only that part of the hypersurfaces whose
$x$- and $\vec z$-values are admitted in the chosen cubic chart).
Finally, choose some analytic 
$d : S \nach \LG$ depending on $z$ only, such that
every element of $G$ occurs somewhere (in a sufficiently close)
neighbourhood of $z=0$.%
\footnote{Such a function indeed exists: 
Choose some $b > 0$, such that the surfaces defined by $y = a$
and $z = i b$ are contained in the chosen chart
for all $i=1,\ldots,\elanz G$. Choose, additionally, for each $i$,
some polynomial $p_i$ with $p_i(l) = \delta_{il}$ and some 
$X_i \in \Lieg$ with $g_i = \e^{X_i}$. Now, define 
$d(x,y,\vec z) := \prod_{i} \e^{p_i(\frac z b) X_i}$.
This function fulfills $d(x,y,ib) = g_i$ for all $i$.}

Let now finitely many (mutually different) points $\tau_j \in I$ be given. 
The $\gamma$-images of these points will turn 
into the intersection points of the
transformed $\gamma$ with $S$.
Fix, additionally, some small $\varepsilon < \varepsilon_0$, 
such that the distance of any two of the marked points in $I$
is greater than $2\varepsilon$ (and that, if necessary, 
each of the $\varepsilon$-neighbourhoods of the $\tau_k$ are both
in $I$ and in the fixed chart). 

Now, in the first step we move $\gamma$ the following way
inside the $x$-$y$-plane: On the one hand,
each segment of $\gamma$
outside the $\varepsilon$-neighbourhoods of the marked points
is again parallel to the $x$-axis, now, however, alternately
with $y = 0$ and $y = 2a$. The $\varepsilon$-neighbourhoods, on the other
hand, are the straight lines connecting these alternatingly lifted 
and unlifted segments. This
way, the center of these neighbourhoods, i.e.\ the marked points themselves
are mapped to half-way between the levels $y=0$ and $y=2a$. In other words,
precisely the marked points are the intersection points of the transformed
$\gamma$ with $S$. Note, in particular, that this transformation of $\gamma$
can be done by a stratified isomorphism that does
only change the $y$-coordinates of any point in $M$, but
neither the $x$- nor the $\vec z$-values (see Lemma \ref{lem:winddiffeo-bumps}).
Moreover, note that 
we have tacitly used that there is an even number of $\tau_k$
to end up at level $y=0$ again after moving the largest $\tau_k$.
This finishes the first step. 

We are now left with
the problem to find the intersection points having the correct values of $d$
in the second step. Nicely, the idea of the first step can be used again. 
To see this, assume that $n = 3$ and look at the scene from above the
$(x,z)$-plane. Since we only changed the $y$-values of $\gamma$,
no change can be seen from this perspective. 
Using our assignment of $d$ to $S$,
we move the $\varepsilon$-neighbourhoods of the marked points of 
(the transformed) $\gamma$. Slightly more generally than in the
first step, however, we let the ``bumps'' that these neighbourhoods 
are mapped to, return to the original line before this neighbourhood ends. 
More precisely, the segments outside these neighbourhoods are not shifted
again, and the ``bumps'' map each $\tau_k$ to the correct ``level''
(i.e.\ $z$-coordinate) in order to get mapped to the point with the 
correct value $g_k$ of $d$. Note, that here we only need to change 
the $z$-coordinates,
but leave, in particular, the $y$-coordinates unchanged. This implies
that the parameter values where $\gamma$
intersects $S$ after having been transformed by both steps,
are precisely those of the $\gamma$ after the first transformation.
If $n > 3$, this step is completely analogous.

\begin{figure}\begin{center}
\epsfig{figure=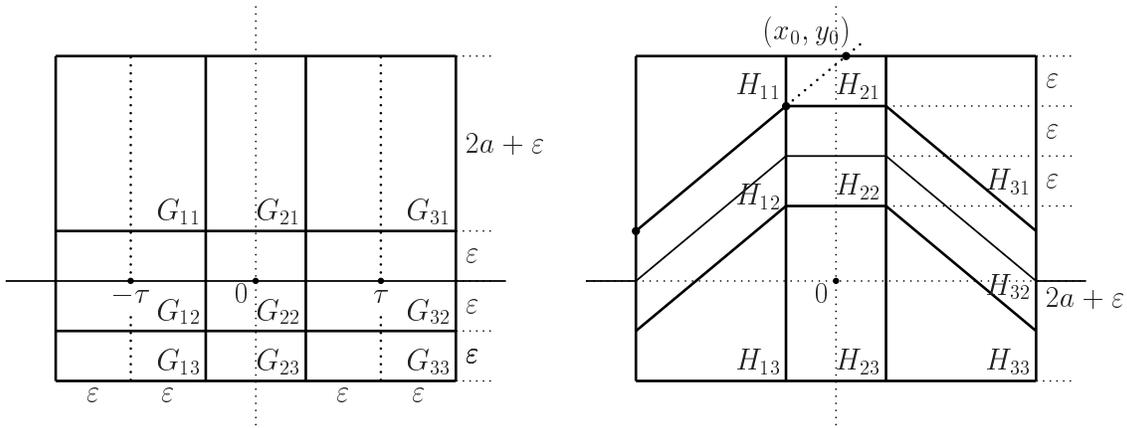,scale=0.95}
\caption{Stratified Diffeomorphism in Lemma \ref{lem:winddiffeo-bumps}}
\label{fig:wind_diffeo}
\end{center}\end{figure}

To summarize: It is clear that the constructed isomorphism
has all the desired properties and that $s$ can be chosen obviously.
Originally, we looked for a stratified
isomorphism mapping $\gamma$, such that its transform intersects
$S$ precisely for the parameter values $\tau_k$ and at points having
the desired values of $d$. By the arguments above, we reduced this
problem to the existence of 
a diffeomorphism in $\R^n$ as indicated in Figure \ref{fig:wind_diffeo}
that, in particular, does not move any point outside the given
square (times some $\varepsilon$-ball in the remaining $n-2$ dimensions
not drawn there). The existence of such a diffeomorphism, however, 
is guaranteed by Lemma \ref{lem:winddiffeo-bumps}. This
furnishes the present proof.
\qed
\epf

The crucial idea in the proof of Proposition \ref{prop:wind_diffeo_1surf}
was to define for each element in $G$ some domain on the surface $S$, 
such that for a given sequence in $G$, 
the transformed graph punctures $S$ 
at the correct points and in the correct ordering, i.e., 
leading to the correct sequence of values for $d$.
We constructed above a single surface with an analytic $d$ on it.
However, we even might use constant $d$, if we admit $S$ to consist
of more than one connected component. In other words, for any finite number
we may find such a number of hypersurfaces $S_i$, such that $\gamma$ may always
be transformed to puncture these different surfaces in an arbitrarily
given ordering. 
More precisely,
choose for $S_i$ some open (cubic) subspace in $S$, and let
the only restriction to $S_i$ be that its $z$-coordinate is in some 
sufficiently small interval $I_i$. We may assume that the closures of 
these intervals are disjoint.
Moreover, each $S_i$ is a hypersurface of $M$. Reusing the
argumentation of the proof of Proposition \ref{prop:wind_diffeo_1surf},
we have shown
\bprop
\label{prop:wind_diffeo}
Let $\dim M \geq 3$. 
Let $\gc$ be a graph, let $\gamma\in\gc$ be one of its edges and set
$\gc' := \gc \setminus\{\gamma\}$. Moreover, let $K$ be a positive integer.

Then there is 
\bunum
\item
some subinterval $I\teilmenge[0,1]$,
\item
some nicely oriented, open, embedded analytic hypersurfaces $S_i$
with $i = 1, \ldots, K$,
such that each $S_i$ and each $\del S_i$ has a finite wide triangulation,
each $\quer{S_i}$ is disjoint to $\im\gc$,
and all $\quer{S_i}$ are mutually disjoint; 
and 
\item
some $s\in\Z$, 
\eunum
such that for any even integer $J > 0$,
any function $l : [1,J] \nach [1,K]$ and any
sequence $(\tau_j) \teilmenge I$
(with $\tau_j > \tau_{j'}$ for $j > j'$) having length $J$,
there is a stratified analytic isomorphism $\diffeo$
with the following properties:
\bnum5
\item
$\bigcup_i \quer{S_i}$ and $\im\diffeo(\gc')$ are disjoint;
\item
$\diffeo(\gamma(\tau_j)) \in S_{l(j)}$ for all $j$;
\item
$\diffeo(\gamma)$ intersects each $S_i$ completely transversally;
\item
$\{\diffeo(\gamma(\tau_j))\}_j$ is the set of $\diffeo(\gamma)$-punctures 
of $\bigcup_i S_i$;
\item
$\sigma^+(S_{l(j)},\diffeo(\gamma)\einschr{[\tau_{j-1},\tau_{j}]}) 
 = (-1)^{j + s} = 
 \sigma^-(S_{l(j)},\diffeo(\gamma)\einschr{[\tau_{j},\tau_{j+1}]})$
for all $j$.
\enum
\eprop

%------------------------------------------------------------------------%
%            Abschnitt: Introduction                                     %
%------------------------------------------------------------------------%
\subsubsection{Generation of Independent Paths}

Transferred to the case of manifolds, Corollary \ref{corr:loc_rot(path)} 
yields
\bprop
\label{prop:generate_indep_locally}
Let $M$ be some $n$-dimensional manifold with $n\geq 2$ and let $S \teilmenge M$.
Assume that $S$ and $\del S$ are connected
embedded submanifolds in $M$ (without boundary)
and that $\quer S$ is an embedded submanifold in $M$
having boundary $\del S$.
Moreover, let $\gc$ be some nontrivial graph in $M$,
such that the image of $\gc$ is neither equal to 
$S$, $\del S$ nor $\quer S$.

Then there is a nontrivial path $\gamma$,
a neighbourhood $U$ of some $m\in\im\gamma$ in $M$
and infinitely many stratified diffeomorphisms $\diffeo_i$ of $M$,
such that
\bunum
\item
$\gamma$ is the only edge in $\gc$ not disjoint to $U$;
\item
$\diffeo_i$ is the identity outside $U$;
\item
$\diffeo_i$ leaves the set $S$ invariant;
\item
$\{\diffeo_i (\gamma)\}_i$ is a hyph%
\footnote{Here, we extended the notion of a hyph naturally to the case
of infinitely many paths.}. 
\eunum
If we additionally assume, that $S$ has one of its natural orientations,
then each $\diffeo_i$ may be chosen such that,
additionally, it leaves the orientation of $S$ invariant.
\eprop
\bpf
\bnum2
\item
$\im\gc$ is not contained in $\quer S$.

Let $\gamma$ be an edge of $\gc$ not contained in $\quer S$.
Choose some interior point $m$ of $\gamma$ outside $\quer S$, and
let $U$ be some open neighbourhood of $m$ disjoint to $S$
and disjoint to all other edges in $\gc$ except for $\gamma$. 
Choose some
chart whose closure is contained in $U$ and 
whose intersection with (the image of)
$\gamma$ is mapped to a straight line with $m$ mapped to the origin.
Corollary \ref{corr:loc_rot(path)} now gives a collection $\diffeo_\winkel$ of 
stratified diffeomorphisms being the identity outside the chart 
that, therefore,
may be extended to stratified diffeomorphisms of $M$ that are the identity 
outside, at least, $U$. Since each $\diffeo_\winkel(\gamma)$ with
$\winkel\in[0,\pi)$ has some interior point
not passed by any other $\diffeo_{\winkel'}(\gamma)$, these paths
are independent. The invariance of $S$ is trivial as well as the
fact that the orientation of $S$ is preserved
and that $\{\diffeo_{\winkel}(\gamma)\}_\alpha$ is a hyph.
\item
$\im\gc$ is contained in $\del S$.

In particular, this implies that $\del S$ is at least one-dimensional.
In fact, otherwise $\del S$ would be a point and $\gc$ trivial.
In the case that $\dim S < n-1$ and that we consider orientations, 
let, moreover, $S' \obermenge S$ 
be some $(n-1)$-dimensional embedded submanifold 
of $M$ inducing the orientation of $S$.
\bnum2
\item
$\dim \del S \geq 2$.

Choose some interior point $m$ of some edge $\gamma$ in $\gc$ and
some open neighbourhood $U$ of $m$
whose closure is disjoint to all 
edges in $\gc$ except for $\gamma$.
By assumption, there is some chart whose closure is contained in $U$,
such that the intersection of the chart 
\bunum
\item 
with $S'$ (if applicable) is some open subset of $\R^{n-1}$,
\item 
with $S$ is some open subset of $\R^{\dim S}$  (if applicable, in $\R^{n-1}$),
\item
with $\del S$ is some open subset of $\R^{\dim S - 1} \teilmenge \R^{\dim S}$,
\item
with $\im\gamma$ is a straight line in $\R \teilmenge \R^{\dim S - 1}$, and 
\item
with $m$ is mapped to the origin.
\eunum
Since $\dim S > \dim\del S \geq 2$,
Corollary \ref{corr:loc_rot(path)} provides us, analogously to the first case, 
with stratified diffeomorphisms having the desired properties.
In particular, observe that, although they are not the identity neither
on $S$ nor on $\del S$,
they leave both $S$ and $\del S$ (and, if applicable, $S'$) invariant.

The orientation of $S$ is obviously preserved for $\dim S = n$.
For $\dim S = n-1$, use the fact that
$t \auf \diffeo_{t\winkel}$ is a homotopy over
diffeomorphisms having the properties above,
whence the natural orientation
of $S$ is preserved by each diffeomorphism.
If $\dim S < n-1$, then that the natural orientations of $S'$ 
are preserved as above, whence the induced orientations on $S$ are so as well. 

\item
$\dim \del S = 1$.

Since $\del S$ is one-dimensional, it is isomorphic to either a line
or a circle. Moreover, $\del\del S = \leeremenge$.
Since the compact set $\im\gc$ does not equal $\del S$,
there is some point $m \in \del(\im\gc) \teilmenge \del S$. Moreover,
there is a (unique) edge, say $\gamma$, 
having $m$ as one of its endpoints. We may assume $\gamma(0) = m$
and choose some open neighbourhood $U$ of $m$
whose closure is disjoint to all 
edges in $\gc$ except for $\gamma$.
Now, we select some chart whose closure is contained in $U$,
such that the intersection of the chart 
\bunum
\item 
with $S'$ is (if applicable) some open subset of $\R^{n-1}$,
\item 
with $S$ is some open subset of $\R^2$ (if applicable, in $\R^{n-1}$),
\item
with $\del S$ is some open subset of $\R \teilmenge \R^2$,
\item
with $\im\gamma$ equals $[0,\tau) \teilmenge \R$, and 
\item
with $m$ is mapped to the origin,
\eunum
and such that $B_\tau \teilmenge U$ for some $\tau > 0$.
By Lemma \ref{lem:locstrat_scaling}, for $\alpha\in[0,\inv3\tau)$,
there are now stratified diffeomorphisms
$\diffeo_\alpha$, taking $-\inv3\tau \in \R \teilmenge \R^n$ as the origin,
such that $(-\tau,+\tau)$ is mapped onto itself, such that
$\diffeo_\alpha\bigl([0,\tau]) = [\alpha,\tau]$ and
such that $\diffeo_\alpha$ is the identity outside $U$.
In particular, each $\diffeo_\alpha$ leaves both $S$, $\del S$ and $S'$ invariant.
Choosing some monotonously decreasing, infinite sequence $\alpha_i \gegen 0$,
we get a hyph $\{\diffeo_{\alpha_i}(\gamma)\}_{i\in\N}$,
since $(\alpha_i,\alpha_{i-1})$ is passed by no $\diffeo_{\alpha_j}(\gamma)$
with $j < i$.

The preservation of orientation by $\diffeo_{\alpha_i}$ is shown analogously to
the case above.
\enum
\item
$\im\gc$ is contained in $\quer S$, but not in $\del S$.

As above, this implies that $S$ is at least one-dimensional.
Again, for $\dim S < n-1$ and if we consider orientations, 
we let $S' \obermenge S$ be some $(n-1)$-dimensional embedded submanifold 
of $M$ inducing the orientation of $S$.
\bnum2
\item
$\dim S \geq 2$.

Choose in $\gc$ some edge $\gamma$ not fully contained in $\del S$.
We now may find some
interior point $m$ of $\gamma$ being in the interior of $S$
and fix some open neighbourhood $U$ of $m$,
whose closure is disjoint to $\del S$ and disjoint to all 
edges in $\gc$ except for $\gamma$.
By assumption, there is some chart whose closure is contained in $U$,
such that the intersection of the chart 
\bunum
\item 
with $S'$ is (if applicable) some open subset of $\R^{n-1}$,
\item 
with $S$ is some open subset of $\R^{\dim S}$  (if applicable, in $\R^{n-1}$),
\item
with $\im\gamma$ is a straight line in $\R \teilmenge \R^{\dim S}$, and 
\item
with $m$ is mapped to the origin.
\eunum
As above, we may find stratified diffeomorphisms of the desired type,
by Corollary \ref{corr:loc_rot(path)}.
\item
$\dim S = 1$.

Since $S$ is one-dimensional, it is isomorphic to either a line
or a circle. Hence $\del S$ consists of at most two points.
Consequently, $\quer S$ is isomorphic either to a circle, a line,
a ray or a closed interval. Since $\im\gc \echteteilmenge \quer S$
is compact, 
there is some point $m \in \del(\im\gc) \cap S$. 
Picking, as above, the (unique) edge $\gamma$
having $m$ as one of its endpoints,
we now may find some open neighbourhood $U$ of $m$
whose closure is disjoint to $\del S$ and to all 
edges in $\gc$ except for $\gamma$.
Again, we select some chart whose closure is contained in $U$,
such that the intersection of the chart 
\bunum
\item 
with $S'$ is (if applicable) some open subset of $\R^{n-1}$,
\item 
with $S$ is some open subset of $\R$ (if applicable, in $\R^{n-1}$),
\item
with $\im\gamma$ equals $[0,\tau) \teilmenge \R$, and 
\item
with $m$ is mapped to the origin,
\eunum
and such that $B_\tau \teilmenge U$ for some $\tau > 0$.
Again, as in the case $\im\gc \teilmenge \del S$ and $\dim\del S = 1$,
we find the desired 
stratified diffeomorphisms by Lemma \ref{lem:locstrat_scaling}.
\qed
\enum
\enum
\epf

\bprop
\label{prop:generate_indep_locally2}
Let $M$ be some $n$-dimensional manifold with $n\geq 2$ and 
let $S \teilmenge M$.
Assume that $S$ and $\del S$ are connected
embedded submanifolds in $M$ (without boundary)
and that $\quer S$ is an embedded submanifold in $M$
having boundary $\del S$.
Moreover, let either $S$ or $\del S$ be an embedded $1$-circle $S^1$.
Finally, let $\gc$ be a graph whose image is this $S^1$
and let $m$ be some vertex of $\gc$.

Then there is a neighbourhood $U$ of $m$ in $M$,
infinitely many different $m_i$ in $S^1 \cap U$ and  
for each $i$ a stratified diffeomorphism $\diffeo_i$ of $M$ 
with the following properties:
\bunum
\item
$\diffeo_i$ is the identity outside $U$;
\item
$\diffeo_i$ leaves the set $S$ invariant;
\item
$\diffeo_i$ maps $m$ to $m_i$;
\item
$\diffeo_i$ is the identity on all edges of $\gc$ not adjacent to $m$.
\eunum
If we additionally assume, that $S$ has one of its natural orientations,
then each $\diffeo_{m'}$ may be chosen such that,
additionally, it leaves the orientation of $S$ invariant.
\eprop
\bpf
This proof is very analogous to that of Proposition \ref{prop:generate_indep_locally2}.
Therefore, we only present its main idea.

First choose some $U$, small enough to intersect $\im\gc$ only at 
its edges adjacent to $m$ and such that $U \cap \quer S$ is a domain of
a straight line (if $\im\gc = S$) or of a half plane (if $\im\gc = \del S$).
Now choose some point near $m$ as the origin for a local
scaling as in Lemma \ref{lem:locstrat_scaling}. 
This way, we may move $m$ to 
every other sufficiently near-by point, leaving $\del S$ or $S$, respectively,
invariant, without moving any point outside $U$. 
\qed
\epf

%------------------------------------------------------------------------%
%            Abschnitt: Introduction                                     %
%------------------------------------------------------------------------%
\section{Representations of the Weyl Algebra}
\label{sect:repr}
Now we are prepared to give a rigorous proof of 
(a stronger version of the) uniqueness
theorem 
claimed by Sahlmann and Thiemann \cite{d60}.
As well, we will proceed in two steps: First we use regularity and
diffeomorphism invariance to show that the first-step decomposition
contains the Ashtekar-Lewandowski measure. This will follow from
the fact that the diffeomorphisms split the Weyl operators,
i.e., the weak convergence of Weyl operators is not uniformly on
states related by diffeomorphisms.
Second, using diffeomorphism invariance again,
we show that each Weyl operator is a scalar at this
component. This enables us to use the
naturality of the action of diffeomorphisms in order
to prove that each Weyl operator is even a unit there.
Cyclicity will give the proof.
At the end, we discuss the technical assumptions made in the proofs.
%------------------------------------------------------------------------%
%            Abschnitt: Introduction                                     %
%------------------------------------------------------------------------%
\subsection{Splitting Property}
\label{subsect:split-prop}
As before, we assume to be given some nice 
enlarged
structure data. Moreover, 
we restrict ourselves to the case that 
$\qsf$ and $\Diffeo$ contain at least those hypersurfaces and stratified
isomorphisms, respectively, that are necessary to keep 
Proposition \ref{prop:wind_diffeo} valid. In other words,
one possibility is to choose $\qsf$ to contain at least
all of these ``cubic'' hypersurfaces 
and $\Diffeo$ to contain at least the stratified isomorphisms
described in that Subsubsection \ref{subsubsect:winddiffeo}.
Throughout the whole subsection, let $\pi'$ be some representation 
of $\algdiff$ on $\hilb$ and denote
by $\pi := \pi'\einschr{\alg}$ the corresponding representation of $\alg$.
Additionally, we require $M$ to have at least dimension $3$.

\bprop
\label{prop:reg+inv->einmal_mu_0}
Assume $\pi$ to be regular.
Moreover, let $\EINS_{\nu_0}$ be $\Diffeo$-invariant
for some $\nu_0$. 

Then $\mu_{\nu_0}$ is the Ashtekar-Lewandowski measure $\mu_0$.
\eprop
Recall that regularity always means regularity w.r.t.\ $\epg$, whereas $\epg$
is taken from the nice enlarged structure data.

\bcorr
\label{corr:reg+inv->einmal_mu_0}
Let $\pi$ be regular,
and let there exist a (cyclic) $\Diffeo$-invariant
vector in $\hilb$.

Then there is a first-step decomposition of $\pi'$, such that
$\mu_{\nu_0}$ is the Ashtekar-Lewandowski measure $\mu_0$
for some $\nu_0$ and $\EINS_{\nu_0}$ is (cyclic and) $\Diffeo$-invariant.
\ecorr
\bpf
According to Lemma \ref{lem:canon_decomp:inv+cycl} 
and the agreements thereafter,
we may find some $\nu_0$, such that 
$\EINS_{\nu_0}$ is $\Diffeo$-invariant (and cyclic). 
Now use the proposition above.
\qed
\epf
Note that, as mentioned earlier, we do not distinguish between
the $\Diffeo$-invariances on equivalent representations. More
precisely, we should say in the corollary above: There is an
isomorphism $U : \hilb \nach \hilb'$ with 
$\hilb' = \bigdirsum_{\nu\in\Nu} L_2(\Ab,\mu_\nu)$ for certain
measures $\mu_\nu$ on $\Ab$, 
such that $U \circ \pi' \einschr{C(X)} \circ U^{-1}$
is cyclic on each $L_2(\Ab,\mu_\nu)$ with cyclic vector $\EINS_\nu$; moreover,
$\mu_{\nu_0}$ equals $\mu_0$ for some $\nu_0\in\Nu$ and
$U \pi'(\dwirk\diffeo) U^{-1} \EINS_{\nu_0} = \EINS_{\nu_0}$
for all $\diffeo\in\Diffeo$.

Before we will be able to prove the proposition above, we 
have to provide two estimates.
\blem
\label{lem:weyl_casimir_absch}
Let $T \in \matfkt_\gc$ be a gauge-variant spin network state, and let 
$\darst$ be some representation occurring in $T$. 
Denote the Casimir eigenvalue w.r.t.\ $\darst$ by $\lambda_\darst$
and set $n := \dim\Lieg$. Finally, define $\eta : \R_+ \nach \R_+$ 
according to Lemma \ref{lem:tensor_casimir_absch}.

Then there is a one-parameter group $\weyl_t$ of Weyl operators,
such that, for each $t_0 > 0$ and each even $J \in \N_+$,
there are $(2n)^J$ diffeomorphisms $\diffeo_\varrho$,
such that
\bgl
\Bignorm[\infty]{
  \inv{(2n)^J} \sum_\varrho
     \bigl(\dwirk{\diffeo_\varrho}^{-1} (\weyl_t) 
             - \e^{-\einhalb \lambda_\darst J t^2}\EINS\bigr)
     T} 
& \leq & \supnorm T \: (\e^{\eta(t_0) J t^4} - 1)
\egl
for all $\betrag t < t_0$.
\elem

\bpf
Fix some edge $\gamma\in\gc$, such that $\darst$ is the representation
carried by $\gamma$ in $T$. 
Let, according to Lemma \ref{lem:tensor_casimir_absch},
$\{X_i\}_{i=1}^n$ be a basis of the Lie algebra $\Lieg$ of $\LG$, 
such that $-\inv n\sum_i \darst(X_i) \darst(X_i)$ 
is (up to the prefactor) the Casimir operator $\darst$. Define
\bgl
\casisum\darst\Lieg(t) \breitrel{:=}
   \inv{2n}\sum_{i=1}^{n} \bigl(\darst(\e^{t X_i}) + \darst(\e^{-t X_i})\bigr)
\breitrel\ident  \inv{2n}\sum_{i=1}^{2n}\darst(\e^{t X_i})
\breitrel\ident  \inv{2n}\sum_{i=1}^{2n}\darst(\e^{-t X_i})
\egl
with $X_{i+n} := - X_i$. 
According to Proposition \ref{prop:wind_diffeo},
choose some interval $I\teilmenge [0,1]$, for each $i = 1, \ldots, 2n$ some 
appropriate surface $S_i$ disjoint to $\im\gc$
and some $s\in\Z$.
Moreover, let $\gotd_i : M \nach \Lieg$ be the constant function of
value $\einhalb X_i$, and
fix some strictly increasing sequence 
$(\tau_j)_{j\in\N_+} \teilmenge I$.  

We are now going to consider
the one-parameter group
\bgl
\weyl_t & := & \prod_{i=1}^{2n} \: \weyl_{i,t} 
     \breitrel\ident \prod_{i=1}^{2n} \: \weyl^{S_i,\sigma_{S_i}}_{E_{\gotd_i}(t)}.
\egl
In fact, this is a one-parameter group: All the $S_i$ are
disjoint, whence $\weyl_{i,t}$ and $\weyl_{i',t'}$ commute
by Lemma \ref{lem:disjqusurf->commut}.% 
\footnote{If we choose some
$E(t) : M \nach \LG$ with $E(t) = E_{\gotd_i}(t)$ on each $S_i$ and
define $\sigma_S$ to be the joint intersection function of $S_1,\ldots,S_{2n}$,
we get $\weyl_t = \weyl^{S,\sigma_S}_{E(t)}$.
Recall that we assumed that $\epg$ contains
not only the ``genuine'' subgroups in $\Weyl$,
but also the finite products of such subgroups, provided they mutually commute.
Therefore, 
it is not important that $E(t)$ is possibly not included in $\dsf(S)$.}

Fix now some positive even integer $J$ and some positive $t_0$.
By the choice of $S_i$ and of $(\tau_j)$,  
for each $\varrho : [1,J] \nach [1,2n]$ there is
a diffeomorphism $\diffeo_\varrho\in\Diffeo$
with the properties described in Proposition \ref{prop:wind_diffeo}.
In particular, since 
$\diffeo_\varrho(\gamma)$ intersects $S := \bigcup_i S_i$ 
completely transversally,
the minimal $S$-admissible decomposition of 
$\diffeo_\varrho(\gamma)$ contains $S$-external edges only.
More explicitly,
it equals $\diffeo_\varrho(\gamma_0) \cdots \diffeo_\varrho(\gamma_{J})$ with 
$\gamma_0 = \gamma\einschr{[0,\tau_1]}$,
$\gamma_j = \gamma\einschr{[\tau_j,\tau_{j+1}]}$ 
and $\gamma_{J} = \gamma\einschr{[\tau_{J},1]}$.
By
$\diffeo_\varrho(\gamma(\tau_j)) \in S_{\varrho(j)}$ for all $j$,
we see that $\diffeo_\varrho(\gamma_j)$ starts in $S_{\varrho(j)}$
and ends in $S_{\varrho(j+1)}$ (with the obvious exceptions for $j = 0$ and
$j=J$). 
Since, moreover, by construction,
$\sigma^+(S_{\varrho(j)},\diffeo_\varrho(\gamma_{j-1})) 
 = (-1)^{j + s} = 
 \sigma^-(S_{\varrho(j)},\diffeo_\varrho(\gamma_j))$
for $j = 1,\ldots,J$, we get

\bgl
         \weyl_t (\dwirk{\diffeo_\varrho}(\darst \circ \pi_\gamma)) 
  & = &  \weyl_t (\darst \circ \pi_{\diffeo_\varrho\gamma}) 
  \breitrel=
       \bigtensor_{j=0}^{J} \weyl_t (\darst \circ \pi_{\diffeo_\varrho\gamma_j}) \\
 & = & (\darst \circ \pi_{\diffeo_\varrho\gamma_0}) \: \tensor \:
       \bigtensor_{j=1}^J \Bigl(
        \darst\bigl(\e^{(-1)^{j+s} X_{\varrho(j)} t}\bigr)
        \cdot \: (\darst \circ \pi_{\diffeo_\varrho\gamma_{j}}) \Bigr),
\egl
hence
\bgl\
         [\dwirk{\diffeo_\varrho}^{-1}(\weyl_t)] (\darst \circ \pi_\gamma) 
 & = & (\darst \circ \pi_{\gamma_0}) \: \tensor \:
       \bigtensor_{j=1}^J \Bigl(
        \darst\bigl(\e^{(-1)^{j+s} X_{\varrho(j)} t}\bigr)
        \cdot \: (\darst \circ \pi_{\gamma_{j}}) \Bigr)
\egl
and
\bgl
       \inv{(2n)^J} \sum_\varrho [\dwirk{\diffeo_\varrho}^{-1}(\weyl_t)] (\darst \circ \pi_\gamma) 
 & = & (\darst \circ \pi_{\gamma_0}) \: \tensor \:
       \bigtensor_{j=1}^J \: \inv{2n} \:\sum_{i=1}^{2n}
        \darst(\e^{(-1)^{j+s} X_i t})
        \cdot (\darst \circ \pi_{\gamma_{j}}) \\
 & = & (\darst \circ \pi_{\gamma_0}) \: \tensor \:
       \bigtensor_{j=1}^J \: \casisum\darst\Lieg(t) \:
        \cdot \: (\darst \circ \pi_{\gamma_{j}}).
\egl

Here, we used
\bgl
\sum_{\varrho:[1,J] \nach [1,2n]} \: \bigtensor_{j=1}^J \: a_{\varrho(j),j} & = &
\bigtensor_{j=1}^J \sum_{i=1}^{2n} \: a_{i,j}.
\egl

By assumption, we have
$T = T_\gamma \tensor T_{\gc'} 
   = \sqrt{\dim\darst} (\darst\circ\pi_\gamma)^k_l \tensor T_{\gc'}$
for some matrix indices $k,l$ and some $T_{\gc'} \in \matfkt_{\gc'}$
with $\gc' = \gc \setminus \{\gamma\}$.
Additionally using $S \cap \im (\diffeo_\varrho \gc') = \leeremenge$ and
$\supnorm{\darst^k_l} \leq \supnorm\darst$, we get
\bgl
 &     & 
\Bignorm[\infty]{
  \inv{(2n)^J} \sum_\varrho 
     \bigl(\dwirk{\diffeo_\varrho}^{-1}(\weyl_t) - \e^{-\einhalb \lambda_\darst J t^2}\EINS\bigr)
     T}  \\
 & \leq & 
\Bignorm[\infty]{\sqrt{\dim\darst} \: \Bigl(
     \inv{(2n)^J} \sum_\varrho 
     \bigl(\dwirk{\diffeo_\varrho}^{-1}(\weyl_t) - \e^{-\einhalb \lambda_\darst J t^2}\EINS\bigr)
     (\darst\circ\pi_{\gamma})\Bigr)^k_l \tensor T_{\gc'}}  \\
& \leq & \supnorm{T_\gamma} \: (\e^{\eta(t_0) J t^4} - 1) \: \supnorm{T_{\gc'}} 
\egl
for all $\betrag t < t_0$, 
by Lemma \ref{lem:tensor_casimir_absch}
and the surjectivity \cite{paper3} 
of $\pi_\hyph : \Ab \nach \LG^{\elanz\hyph}$ for every hyph $\hyph$. 
\qed
\epf

\bcorr
\label{corr:weyl+diffeo>eps}
Let $\pi'$ be some representation of $\algdiff$,
let $T \in \matfkt_\gc$ be a nontrivial gauge-variant spin network and let
$\psi\in\hilb$ be $\Diffeo$-invariant with
$\skalprod{\psi}{\pi(T) \psi}_\hilb \neq 0$.

Then there is a one-parameter group $\weyl_t$ of Weyl operators
and $\varepsilon, t_0 > 0$,
such that for all $0 \neq \betrag t < t_0$ 
there is a diffeomorphism $\diffeo$
with 
\bgl
\betrag{\skalprod{\psi}{\pi\bigl((\weyl_t - \EINS)(\dwirk\diffeo T)\bigr) \psi}_\hilb}
 & \geq & \varepsilon.
\egl
\ecorr
\bpf
\bunum
\item
Set 
$\varepsilon := 
    \min\{\einhalb,\inv4\betrag{\skalprod{\psi}{\pi(T) \psi}_\hilb}\} > 0$.
Fix $\tau_0 > 0$ 
and some non-trivial irreducible representation $\darst$ occurring in $T$.
Next, choose positive real $\tau_2$ and $\tau_4$, such that
(using the $\eta(\tau_0)$ as given in Lemma \ref{lem:weyl_casimir_absch})
\bglklein
  \norm\psi_\hilb^2 \: \supnorm T \: 
  (\e^{\eta(\tau_0) \tau} - 1) \breitrel< \varepsilon
    & \breitrel{\text{ for all $\betrag{\tau} < \tau_4$ }} 
\eglklein
and
\bglklein
  \e^{-\einhalb \lambda_\darst \tau} \breitrel< \varepsilon
    & \breitrel{\text{ for all $\betrag{\tau} > \tau_2$. }} 
\eglklein
Define now
\bglklein
J_0 := \einhalb\sqrt{2\tau_2\tau_4} & \breitrel{\text{ and }} & 
t_0 := \min\{\inv{\tau_2} J_0,\sqrt[3]{J_0},\tau_0\}.
\eglklein
We say that $(J,t) \in \N_+ \kreuz \R$ is an {\em admissible}\/ pair
iff 
\bglklein
0 < \betrag{t} < t_0 & \breitrel{\text{ and }} & 
\frac{J_0}{t^3} \leq J \leq 2\frac{J_0}{t^3}
\eglklein
As one checks easily, 
the admissibility of $(J,t)$ implies $Jt^2 > \tau_2$ and $Jt^4 < \tau_4$.
Moreover, for every $0 < \betrag{t} < t_0$, there is an even 
$J(t) \in \N_+$ such that $(t,J(t))$ is admissible.
\item
Choose now a one-parameter subgroup $\weyl_t$ of Weyl operators
and, for all positive integers $J$, some
diffeomorphisms $\diffeo_\varrho$, $\varrho\in\Rho_J$,
as in Lemma \ref{lem:weyl_casimir_absch}. 
Consequently, we have

\bgl[1ex]
 &       & 
           \Bigbetrag{
             \inv{(2n)^J} \sum_{\varrho\in\Rho_J}
             \skalprod{\psi}{\pi\bigl(\bigl(\dwirk{\diffeo_\varrho}^{-1}(\weyl_t) - \e^{-\einhalb\lambda_\darst J t^2} \EINS\bigr)T\bigr) \psi}_\hilb} 
            \\
 & \leq  & \Bignorm[\bound(\hilb)]{
             \pi\Bigl(\inv{(2n)^J} \sum_{\varrho\in\Rho_J}
             (\dwirk{\diffeo_\varrho}^{-1}(\weyl_t) - \e^{-\einhalb\lambda_\darst J t^2} \EINS)T\Bigr) 
           } \:
           \norm\psi_\hilb^2  \\
 & \leq  & \Bignorm[\infty]{
             \inv{(2n)^J} \sum_{\varrho\in\Rho_J}
             (\dwirk{\diffeo_\varrho}^{-1}(\weyl_t) - \e^{-\einhalb\lambda_\darst J t^2} \EINS)T
           } \:
           \norm\psi_\hilb^2  \\
 & \leq  & (\e^{\eta(\tau_0) J t^4} - 1) \: \supnorm T \: \norm\psi_\hilb^2 \s
 &  < & \varepsilon 
\egl
and, using
$(1-\varepsilon) \: \betrag{\skalprod{\psi}{\pi(T) \psi}_\hilb}
 \geq \einhalb \betrag{\skalprod{\psi}{\pi(T) \psi}_\hilb} \geq 2 \varepsilon$,
\bgl
 &   &     \Bigbetrag{
             \inv{(2n)^J} \sum_{\varrho\in\Rho_J} \bigl(1 - \e^{-\einhalb\lambda_\darst J t^2}\bigr)
             \skalprod{\psi}{\pi(T) \psi}_\hilb} \\
 & = &
             \bigl(1 - \e^{-\einhalb\lambda_\darst J t^2}\bigr) \:
             \betrag{\skalprod{\psi}{\pi(T) \psi}_\hilb} 
\breitrel\geq (1 - \varepsilon) \:
             \betrag{\skalprod{\psi}{\pi(T) \psi}_\hilb} 
\breitrel\geq   2  \varepsilon

\egl
for all admissible pairs $(J,t)$.
\item
Altogether, we have
for all admissible pairs $(J,t)$
\bgl[4ex]
 \erstezeile3
\Bigbetrag{
  \inv{(2n)^J} 
          \sum_{\varrho\in\Rho_J}
          \skalprod{\psi}{\pi\bigl((\dwirk{\diffeo_\varrho}^{-1}(\weyl_t) - \EINS)(T)\bigr) \psi}_\hilb} \\
 & \geq  & \Biggl| \:
           \Bigbetrag{
             \inv{(2n)^J} 
              \sum_{\varrho\in\Rho_J} \bigl(1 - \e^{-\einhalb\lambda_\darst J t^2}\bigr)
             \skalprod{\psi}{\pi( T) \psi}_\hilb} 
\\ && \hspace*{2em}{}-
           \Bigbetrag{
             \inv{(2n)^J} 
              \sum_{\varrho\in\Rho_J}
             \skalprod{\psi}{\pi\bigl((\dwirk{\diffeo_\varrho}^{-1}(\weyl_t) - \e^{-\einhalb\lambda_\darst J t^2} \EINS)(T)\bigr) \psi}_\hilb} 
           \:\Biggr| \s
 & >  & \varepsilon.
\egl
\item
Finally, for all $0 < \betrag{t} < t_0$, we may choose a $J(t)$
providing an admissible pair $(J(t),t)$.
By the lines above, there is some 
$\diffeo \in \{\diffeo_\varrho\mid \varrho\in\Rho_{J(t)}\}$, such that
\bgl
\betrag{\skalprod{\psi}{\pi\bigl((\weyl_t - \EINS)(\dwirk\diffeo T)\bigr) \psi}_\hilb} 
 \breitrel= \betrag{\skalprod{\psi}{\pi\bigl((\dwirk{\diffeo}^{-1}(\weyl_t) - \EINS) T \bigr) \psi}_\hilb} 
 & >  & \varepsilon
\egl
using the $\Diffeo$-invariance of $\psi$.
\qed
\eunum
\epf

\bpf[Proposition \ref{prop:reg+inv->einmal_mu_0}]
Corollary \ref{corr:weyl+diffeo>eps} shows that $\Weyl'$ 
splits $\Weyl$ at $\EINS_{\nu_0}$
for every nontrivial gauge-variant spin network state $T$
with $\skalprod{\EINS}{T}_{\nu_0} \ident 
\skalprod{\EINS_{\nu_0}}{\pi(T) \EINS_{\nu_0}}_\hilb \neq 0$.
Hence, $\Weyl'$ splits $\Weyl$ at $\EINS_{\nu_0}$,
by Lemma \ref{lem:sns=cont-erzs}. 
Since
$\pi$ is regular, Proposition \ref{prop:split->eqmeas} 
gives the assertion.
\qed
\epf

%------------------------------------------------------------------------%
%            Abschnitt: Introduction                                     %
%------------------------------------------------------------------------%
\subsection{Naturality}
\label{subsect:rep(algdiff)-reg}
Now we are using nice enlarged structure data and assume additionally
that 
\bunum
\item
$\Diffeo$ contains at least the stratified isomorphisms
described in Subsection \ref{subsect:2types(strat_diffeo)}
and at least those necessary to keep Proposition \ref{prop:locstrat_mfs} valid;
\item
$\qsf$ contains 
at most all $\Diffeo$-orbits of semianalytic subsets in $M$
having a finite wide triangulation and being of lower dimension than 
$M$, but contains at least the $q$-balls and $q$-simplices
with $q < \dim M$.
\eunum

\bprop
\label{prop:reg+inv+typ+fin->fund}
Let $\pi'$ be a $\Diffeo$-natural
representation of $\algdiff$,
such that $\EINS_{\nu_0}$ is a diffeo\-morphism-invariant vector
and $\mu_{\nu_0}$ equals $\mu_0$ for some $\nu_0\in\Nu$.

Then the restriction of $\pi'$ to $\hilb_{\nu_0}$ 
is the fundamental representation $\pi_0$,
i.e., we have $P_\nu \pi'(a) = \pi_0(a) P_\nu$ for
all $a \in \algdiff$. 
\eprop
The proof of the proposition will use several steps we are now going to
write down in separate lemmata. For this, throughout the whole
section, we will assume that 
$\pi'$ is a $\Diffeo$-natural 
representation of $\algdiff$ having some $\EINS_{\nu_0}$
as a $\Diffeo$-invariant vector. Moreover, $\mu_{\nu_0}$ equal $\mu_0$.
Finally, as usual,
we set $\pi := \pi'\einschr{\alg}$.

\blem
\label{lem:diffeo(piunit)=piunit}
Let $S_1$ and $S_2$ be elements in $\qsf$ having orientations
$\sigma_{S_1}$ and $\sigma_{S_2}$. Assume that they are 
oriented-strata equivalent.
Finally, let $g\in\LG$ be some element and 
$d_i : S_i \nach \LG$ for $i=1,2$ be the constant function with value $g$.

Then $\weyl_{d_1}^{S_1,\sigma_{S_1}}$ is a $\pi_{\nu_0}$-unit 
($\pi_{\nu_0}$-scalar) iff
$\weyl_{d_2}^{S_2,\sigma_{S_2}}$ is a $\pi_{\nu_0}$-unit ($\pi_{\nu_0}$-scalar).
\elem
\bpf
Let $\diffeo$ be a product of localized stratified isomorphisms mapping
$S_1$ onto $S_2$ as well as their orientations. Then
\bgl
\dwirk\diffeo(\weyl_{d_1}^{S_1,\sigma_{S_1}})
 & = & \weyl_{\diffeo(d_1)}^{\diffeo(S_1),\diffeo(\sigma_{S_1})}
 \breitrel= \weyl_{d_2}^{S_2,\sigma_{S_2}}.
\egl
Now, the assertion follows from Corollary \ref{corr:scal-krit-nu}.
\qed
\epf

\blem
\label{lem:pi(w)_orthcompl}
Let $S$ some subset of $M$, such that $S$ and $\del S$ are connected
embedded submanifolds in $M$ (without boundary)
and that $\quer S$ is an embedded submanifold in $M$
having boundary $\del S$. Moreover, assume that $S$ has one 
of its natural orientations.
Finally, let $\weyl = \weyl^{S,\sigma_S}_d$ for some constant $d\in\dsf(S)$.

Then $P_{\nu_0} \pi(\weyl) \EINS_{\nu_0} \in \hilb_{\nu_0} \ident L_2(\Ab,\mu_0)$ 
is orthogonal to all non-trivial gauge-variant spin network states 
that are not based on an edge $\gamma$ 
whose image equals $S$, $\del S$ or $\quer S$. 
\elem
Recall that no edge of a gauge-variant spin network is labelled
with the trivial representation.
\bpf
Let $T$ be a gauge-variant spin network state in $\matfkt_\gc$. There
are two main cases:
\bunum
\item
$\im \gc$ neither equals $S$, $\del S$ nor $\quer S$.

By Proposition \ref{prop:generate_indep_locally}, there is an infinite number
of localized stratified diffeomorphisms $\diffeo_i$, 
leaving $S$ (including its orientation and $d$) and each edge of $\gc$ except
for some $\gamma$ invariant, and forming a hyph $\{\diffeo_i(\gamma)\}_i$. 
Consequently, $\alpha_{\diffeo_i} T$ and $\alpha_{\diffeo_j} T$
are orthogonal 
for $i\neq j$.
Moreover, each $\alpha_{\diffeo_i}$ commutes with $\weyl$.
Therefore, by Lemma \ref{lem:pi(w)-orth}, 
$P_{\nu_0} \pi(\weyl) \EINS_{\nu_0}$ is orthogonal 
to $T$. 
\item
$\im\gc$ equals $S$, $\del S$ or $\quer S$.

Assume first $\im\gc = \del S$.
Since $\im\gc$ is compact, $\del S$ has to be 
compact as well. Hence, it is isomorphic to $S^1$. After a possibly
necessary re-orientation of $\gc$, the product
of all paths in $\gc$ is a closed edge $\gamma$ with image $\del S$.
Assume that $T$ is not $(\gamma,\darst)$-based for some $\darst$.
Then there is some vertex $m$ in $\gc$, where
the adjacent edges at $m$ 
are either labelled with different representations or
carry non-matching indices.
Now, as in the previous case, 
but this time by Proposition \ref{prop:generate_indep_locally2},
there are infinitely many localized stratified diffeomorphisms $\diffeo_i$, 
leaving the sets $S$ (including orientation and $d$) and $\del S$
invariant; they simply move $m$ along $\del S$ stretching
$\del S$ a bit. By Lemma \ref{lem:orthrel(sns)},
any two 
$\alpha_{\diffeo_i} T$ and $\alpha_{\diffeo_j} T$
with $i\neq j$ are orthogonal. 
Since each $\alpha_{\diffeo_i}$ commutes with $\weyl$,
Lemma \ref{lem:pi(w)-orth} proves the orthogonality of
$P_{\nu_0} \pi(\weyl) \EINS_{\nu_0}$ and $T$. 

The case of $\im\gc = S$ is completely analogous. For $\im\gc = \quer S$
we may additionally get the case of an embedded interval.
However, this is analogous as well.
\qed
\eunum
\epf
Immediately from the proof of the lemma above and that of 
Proposition \ref{prop:generate_indep_locally}, we get
\bcorr
\label{corr:disj(s,gc)->orth}
Let $S$ be some subset of $M$ and $\weyl^{S,\sigma_S}_d\in\Weyl$ 
be any Weyl operator. Moreover, let $\gc$ be a graph not contained in
the closure of $S$.

Then $P_{\nu_0} \pi(\weyl) \EINS_{\nu_0} \in \hilb_{\nu_0} \ident L_2(\Ab,\mu_0)$ 
is orthogonal to all non-trivial gauge-variant spin network states 
in $\matfkt_\gc$.
\ecorr

We are now going to prove that the Weyl operators to open balls 
given some constant ``labelling'' $d$, are $\pi_{\nu_0}$-units. We start
with the dimensions $0$ and $3+$, but smaller than $\dim M$, proceed
with dimension $1$ and end up with dimension $2$.
\bcorr
\label{corr:pi-unit(0,3+)}
Let $s < \dim M$ be some non-negative integer with $s \neq 1,2$,
and let $S$ be an open or closed $s$-dimensional ball in $M$
given a nice orientation. 

Then $\weyl := \weyl^{S,\sigma_S}_d$ is a $\pi_{\nu_0}$-unit for 
every constant $d\in\dsf(S)$.
\ecorr
\bpf
Let $\gc$ be a non-trivial graph. Since $s \neq 1, 2$,
neither $S$, $\del S$ nor $\quer S$ equals $\im\gc$.
Thus, 
$P_{\nu_0} \pi(\weyl) \EINS_{\nu_0}$ is orthogonal to each non-trivial
$T\in\matfktgsn$. Since $\matfktgsn$ is a continuous $\mu_0$-generating
system, $\weyl$ is 
a $\pi_{\nu_0}$-scalar (see also Lemma \ref{lem:cont_gen_syst:scalar}).

To prove that $\weyl$ is even a $\pi_{\nu_0}$-unit observe first that,
by Propositions \ref{prop:inv_orient_simplball} and \ref{prop:locstrat_mfs}, 
there is a stratified
isomorphism $\diffeo$ mapping $S$ onto itself, 
but reverting its orientation. Thus, 
$\dwirk\diffeo(\weyl) = \weyl^\ast$, whence $\weyl^2$ is 
a $\pi_{\nu_0}$-unit by Corollary \ref{corr:unitaer:scalar-nu->unit}.
Since $\LG$ is compact, there is a square root for any
element. Re-doing the proof for $d_1\in\dsf(S)$ with $d_1\:d_1 = d$
gives the assertion.
\qed
\epf

\blem
\label{lem:weyl+gammabased->scalar}
Let $\weyl \in \Weyl$ be a Weyl operator for some quasi-surface $S$ and some
constant $d\in\dsf(S)$,
and let $\gamma$ be an analytic edge,
such that $P_{\nu_0} \pi(\weyl) \EINS_{\nu_0}$ is contained
in the closure of $\spann\:\bsns\gamma$. 

If the image of $\gamma$ is not completely contained in $S$,
then $\weyl$ is a $\pi_{\nu_0}$-scalar.
\elem
\bpf
Let $m \in \im\gamma \setminus S$. If $\gamma$ is closed,
we may assume that $m$ is not the base point of $\gamma$.
Consider now for each $g\in\LG$
the Weyl operator $\weyl_{g,m}$ given by the quasi-surface $S_m := \{m\}$,
whereas the orientation of $S_m$ is chosen, such that 
the direction of $\gamma$ coincides with the orientation of $S_m$.
Since $S_m$ and $S$ are disjoint, $\weyl_{g,m}$ and $\weyl$ commute.
Moreover, by Corollary \ref{corr:pi-unit(0,3+)}, 
$\weyl_{g,m}$ is a $\pi_{\nu_0}$-unit.
Consequently, by Corollary \ref{corr:commut+pi-unit-nu->inv},
$\weyl_{g,m}$ leaves $P_{\nu_0} \pi(\weyl) \EINS_{\nu_0}$ invariant.
\bunum
\item
Let $m$ be not an endpoint of $\gamma$.

First of all, let 
$T = (T_{\gc,\vekt\darst})^{\vekt i}_{\vekt j} \in \bsns{\gamma,\darst}$ 
for some non-trivial $\darst$ with $\darst_k = \darst$ for all $k$ and with
$m$ being a vertex of $\gc$. One easily checks that
\bgl
\weyl_{g,m}(T) 
 & = & \sum_{r_1, r_2}
           \darst (g^2)^{r_1}_{r_2} \: 
              (T_{\gc,\vekt\darst})^{\vekt i(r_2)}_{\vekt j(r_1)},
\egl
whereas $\vekt i(r)$ is the tuple of all $i_k$ where
the index belonging to the edge leaving at $m$ is replaced by $r$.
Hence,
\bglklein
       \skalprod{P_{\nu_0} \pi(\weyl) \EINS_{\nu_0}}{T}_{\mu_0}
 & = & \skalprod{\weyl_{g,m}^\ast (P_{\nu_0} \pi(\weyl) \EINS_{\nu_0})}{T}_{\mu_0} \\
 & = & \skalprod{P_{\nu_0} \pi(\weyl) \EINS_{\nu_0}}{\weyl_{g,m}(T)}_{\mu_0} \\
 & = & \sum_{r_1, r_2} \darst(g^2)^{r_1}_{r_2} \: 
        \skalprod{P_{\nu_0} \pi(\weyl) \EINS_{\nu_0}}{(T_{\gc,\vekt\darst})^{\vekt i(r_2)}_{\vekt j(r_1)}}_{\mu_0}
\eglklein
for all $g\in\LG$ and therefore, since square roots exist in $\LG$,
\bglklein
\skalprod{P_{\nu_0} \pi(\weyl) \EINS_{\nu_0}}{T}_{\mu_0}
 & = & \int_\LG \skalprod{P_{\nu_0} \pi(\weyl) \EINS_{\nu_0}}{T}_{\mu_0} \:
                \dd\mu_\Haar(g) \\
 & = & \sum_{r_1, r_2} 
        \skalprod{P_{\nu_0} \pi(\weyl) \EINS_{\nu_0}}{(T_{\gc,\vekt\darst})^{\vekt i(r_2)}_{\vekt j(r_1)}}_{\mu_0} \:
           \int_\LG \darst(g)^{r_1}_{r_2} \: \dd\mu_\Haar(g) \\
 & = & 0.
\eglklein
Now, if $T = (T_{\gc,\vekt\darst})^{\vekt i}_{\vekt j} \in \bsns{\gamma,\darst}$
for some non-trivial $\darst$ without $m$ being a vertex of $\gc$,
we may refine $\gc$ by inserting $m$ as a new vertex. Then 
$T$ is a (finite) sum of $(\gamma,\darst)$-based gSNs each
having $m$ as a vertex of the underlying graph. Using the just shown result,
we get 
\bglklein
\skalprod{P_{\nu_0} \pi(\weyl) \EINS_{\nu_0}}{T}_{\mu_0}
 & = & 0
\eglklein
for all $(\gamma,\darst)$-based gauge-variant spin network states.

Altogether, this shows that $P_{\nu_0} \pi(\weyl) \EINS_{\nu_0}$
is orthogonal to all non-trivial gauge-variant spin network states,
i.e., $\weyl$ is a $\pi_{\nu_0}$-scalar.
\item
Let $m$ be an endpoint of $\gamma$.

We argue analogously, using
\bgl
\weyl_{g,m}(T) 
 & = & \sum_r
        \darst (g)^{i_1}_{r} \: (T_{\gc,\vekt\darst})^{\vekt i(r)}_{\vekt j}
\egl
if $m = \gamma(0)$, and similarly for $m = \gamma(1)$.
\qed
\eunum
\epf

\bcorr
\label{corr:pi-unit(1)}
Let $S$ be an open $1$-dimensional ball in $M$ given a nice orientation. 

Then $\weyl := \weyl^{S,\sigma_S}_d$ is a $\pi_{\nu_0}$-unit for 
every constant $d\in\dsf(S)$.
\ecorr
\bpf
Let $\gamma$ be the edge whose interior is $S$ and choose
one of its orientations.
By Lemma \ref{lem:pi(w)_orthcompl},
$P_{\nu_0} \pi(\weyl) \EINS_{\nu_0}$ 
is orthogonal to all non-trivial gauge-variant spin network states 
that are not based on the edge $\gamma$. 
By Corollary \ref{corr:decomp(hilb)-bsns},
$P_{\nu_0} \pi(\weyl) \EINS_{\nu_0}$ is contained
in the closure of the span 
of $\gamma$-based gSNs. Since, however, the endpoints of $\gamma$
are not contained in $S$, Lemma \ref{lem:weyl+gammabased->scalar} implies that 
$\weyl$ is a $\pi_{\nu_0}$-scalar. 
Now, by Proposition \ref{prop:inv_orient_simplball}, 
there is some $\diffeo\in\Diffeo$ being the identity on $S$,
but inverting the orientation
of $S$, i.e., $\dwirk\diffeo(\weyl) = \weyl^\ast$.
Corollary \ref{corr:unitaer:scalar-nu->unit}
implies that $\weyl^2$ is a $\pi_{\nu_0}$-unit. As above,
the assertion follows since square roots exist in $\LG$.
\qed
\epf

\bcorr
\label{corr:pi-unit(2)}
Let $S$ be an open $2$-dimensional ball in $M$ given a nice orientation. 

Then $\weyl := \weyl^{S,\sigma_S}_d$ is a $\pi_{\nu_0}$-unit for 
every constant $d\in\dsf(S)$.
\ecorr
\bpf
The image of an edge $\gamma$ equals $S$, $\del S$ or $\quer S$ iff
$\gamma$ is a closed loop along $\del S \iso S^1$.
By Lemma \ref{lem:pi(w)_orthcompl},
$P_{\nu_0} \pi(\weyl) \EINS_{\nu_0}$ 
is orthogonal to all non-trivial gauge-variant spin network states 
not based on such a $\gamma$. Hence, we have 
$P_{\nu_0} \pi(\weyl) \EINS_{\nu_0} \in \quer{\spann\:\bsns\gamma}$ 
by Corollary \ref{corr:decomp(hilb)-bsns}.
Observe that $\im\gamma \cap S = \leeremenge$.
Now argue as in Corollary \ref{corr:pi-unit(1)}.
\qed
\epf

\bprop
\label{prop:triang->piunit}
Let $S$ be a finitely widely triangulizable subset in $M$ having a natural
orientation with $\dim S < \dim M$. 

Then $\weyl := \weyl^{S,\sigma_S}_d$ is a $\pi_{\nu_0}$-unit for 
every $d\in\dsf(S)$ being constant on $S$.
\eprop
\bpf
\bunum
\item
$S$ is an open $q$-simplex having a nice orientation.

By Corollary \ref{corr:orientstrateq_simplball}
and Proposition \ref{prop:locstrat_mfs}, $S$ is 
oriented-strata equivalent to a nicely oriented $q$-ball. 
So we get the assertion, since
$q$-balls lead to Weyl operators that are $\pi_{\nu_0}$-units 
(Corollaries \ref{corr:pi-unit(0,3+)}, \ref{corr:pi-unit(1)} and \ref{corr:pi-unit(2)}) 
and since this property is inherited
to all oriented-strata equivalent objects 
according to Lemma \ref{lem:diffeo(piunit)=piunit}.
\item
$S$ is finitely widely triangulizable. 

This means, by definition, $S$ is the finite disjoint union of 
nicely oriented simplices. 
Since disjoint unions lead to products of (commuting) 
Weyl operators (see Lemma \ref{lem:disjqusurf->commut}),
we get the assertion as well in the general case.
\qed
\eunum
\epf

\bpf[Proposition \ref{prop:reg+inv+typ+fin->fund}]
Use Proposition \ref{prop:triang->piunit}
and Lemma \ref{lem:pi-unit->pi=pi_0}, observing that each
$\weyl'\in\Weyl'$ is a $\pi'_{\nu_0}$-unit and that $\algdiff$
is generated by $\Weyl$, $\Weyl'$ and $C(X)$.
\qed
\epf

\neueseite

%------------------------------------------------------------------------%
%            Abschnitt: Introduction                                     %
%------------------------------------------------------------------------%
\subsection{Classification}

\bdf
Enlarged structure data are called \df{optimal} iff they are nice
and
\bunum
\item
$M$ is at least three-dimensional;
\item
$\dsf(S)$ contains 
\bunum
\item
at most the constant functions together with their $\gaugetrfs$-orbits;
\eunum
\item
$\Diffeo$ contains 
\bunum
\item
at least the stratified isomorphisms described in 
Subsection \ref{subsect:2types(strat_diffeo)},
\item
at least those necessary to keep Proposition \ref{prop:locstrat_mfs} valid;
\eunum
\item
$\qsf$ contains 
\bunum
\item
at least those hypersurfaces that are necessary to keep 
Proposition \ref{prop:wind_diffeo} valid,
\item
at least the $q$-balls and $q$-simplices for $q < \dim M$,
\item
at most all $\Diffeo$-orbits of semianalytic subsets in $M$
having a finite wide triangulation and being of lower dimension than
$M$.
\eunum
\eunum
\edf

\bthm
\label{thm:main}
Let $\pi''$ be a representation of $\algauto$ on $\hilb$, such that
$\pi := \pi'' \einschr\alg$ is regular and
$\pi' := \pi'' \einschr\algdiff$ is $\Diffeo$-natural.
Moreover, let there exist some $\autos$-invariant vector in $\hilb$
being cyclic for $\pi$. Finally, let optimal enlarged structure data
be given.

Then $\pi''$ is unitarily equivalent to the fundamental representation 
of $\algauto$.
\ethm

\bpf
By Corollary \ref{corr:reg+inv->einmal_mu_0},
regularity and diffeomorphism invariance
imply that there is a first-step decomposition of $\pi'$,
such that some $\mu_{\nu_0}$ is the Ashtekar-Lewandowski measure $\mu_0$ 
and that $\EINS_{\nu_0}$ is $\Diffeo$-invariant.
Naturality, diffeomorphism invariance and
Proposition \ref{prop:reg+inv+typ+fin->fund}
imply that each $\weyl\in\Weyl$ w.r.t.\ a constant $d \in \dsf(S)$ is a $\pi_{\nu_0}$-unit. 
As $\EINS_{\nu_0}$ is also $\gaugetrfs$-invariant, $\weyl\in\Weyl$ is 
a $\pi_{\nu_0}$-unit for every $d \in \dsf(S)$ 
by Lemma \ref{lem:inv=pi_nu-unit} and Proposition \ref{prop:weyl-gauge-covariance}.
Cyclicity gives the proof by Corollary \ref{corr:unit-nu->cyclic}.
\qed
\epf

We remark that the results above can be directly extended
to semianalytic sets having the same dimension as $M$. Of course,
the triangulizability has to be guaranteed and the intersection functions
have to be adjusted. The latter one can be done, e.g., by setting
$\sigma_S(\gamma)$ for closed $S$ to be one iff $\gamma$ starts at the boundary
of $S$ and then leaves $S$ non-tangentially. The proofs, however, 
have to be modified accordingly. In particular, there is no longer an
extra dimension available to mirror simplices and balls.
Instead, we now use that there are diffeomorphisms mapping
simplices (enlarged by one of its faces) onto two other, disjoint simplices
whose union is the original simplex again 
(Proposition \ref{prop:two-in-one-simplex}).
The proofs that the corresponding Weyl operators are $\pi_\nu$-units,
should now 
use the first statement of Corollary \ref{corr:unitaer:scalar-nu->unit}
and proceed inductively on the dimension.

%------------------------------------------------------------------------%
%            Abschnitt: Introduction                                     %
%------------------------------------------------------------------------%
\subsection{Discussion}
Having now obtained the desired uniqueness theorem, we might ask, whether
the assumptions for it are reasonable. 
\subsubsection{Structure Data}
First of all, let us consider 
the enlarged structure data.
\blem
The following enlarged structure data are optimal:
\bunum
\item
$M$ is an at least three-dimensional analytic manifold;
\item
$\LG$ is a nontrivial connected compact Lie group;
\item
$\Pf$ consists of all piecewise analytic paths in $M$;
\item
$\gaugetrfs$ contains the generalized gauge transforms;%
\footnote{The statement remains true if we assume $\gaugetrfs$ to be any
subset of $\Gb$, e.g., just the (stratified analytic) gauge transforms 
of a particular principal fibre bundle.}
\item
$\Diffeo$ contains the stratified analytic diffeomorphisms in $M$;
\item
$\qsf$ contains the semianalytic sets in $M$
(together with their $\Diffeo$-orbits)
having lower dimension than $M$ and having a finite wide triangulation;
\item
$\isf(S)$ contains the natural%
\footnote{To be precise, $\isf(S)$ should contain the natural intersection
functions of $S$, if $S$ is a submanifold. In the general case,
include all intersection functions that are joint intersection
functions given by the nice intersection functions
for the submanifolds forming the respective triangulation. Finally,
if necessary, collect all intersection functions generated by the
action of $\Diffeo$ on stratified sets of the types mentioned previously.}
intersection functions of $S$;
\item
$\dsf(S)$ contains the constant functions on $M$ (together with their $\gaugetrfs$-orbits);
\item
$\epg$ contains the one-parameter subgroups of Weyl operators 
consistent
with $\qsf$, $\{\isf(S)\}$, $\{\dsf(S)\}$.
\eunum
\elem
From our point of view (see also the discussion
in Subsection \ref{subsect:struct_data}), 
all ingredients are natural up to the 
restrictions on $\qsf$ and, maybe, on $M$ and $\dsf(S)$. 
The inclusion of semianalytic sets is reasonable, since 
the stratified diffeomorphisms map analytic hypersurfaces to semianalytic
sets anyway. At the same time, the inclusion of lower-dimensional
surfaces becomes natural. But, it would be desirable
to at least replace the condition of wide triangulizability by
the ``standard'' triangulizability, since in this case it is known
that any semianalytic set is triangulizable. 
The requirement that each simplex in the triangulation is
nicely oriented, is not too restrictive, since every naturally
oriented, embedded surface is at least locally nicely oriented.
The finiteness, on the
other hand, cannot simply be dropped. This may at most be possible
for compact $M$. In fact, then every semianalytic set has a 
compact closure and compact boundary. Then we may triangulize them
finitely, by local finiteness. Redoing the procedure with the 
(lower-dimensional) semianalytic
set given by the intersection of the original one with its boundary,
we may successively get a finite decomposition of the original
set into simplices. For non-compact $M$, this is no longer true. Simply 
take a hyperplane in $\R^3$ being triangulizable, of course, but
not finitely. Well, although our proofs above have aimed at the finite case,
we may extend the uniqueness result immediately to this example.
Simply use that a hyperplane can be rotated onto itself inverting
its orientation, and argue as in Corollary \ref{corr:pi-unit(0,3+)}. 
In other words, it may be,
as already mentioned above, that every analytic manifold
is widely triangulizable; but even if not, 
there seems to be still some leeway in our argumentation above
to keep the uniqueness given in the more general context.
However, to explore this, several technical investigations
in the field of semianalytic sets are necessary that go much beyond the scope
of this paper.

We mentioned also the restriction that $M$ has to be at least
three-dimensional. Well, for quantum gravity this is no problem at
all, since the space-like hypersurfaces are three-dimensional.
The space-time is even four-dimensional, although this does not seem
relevant here, since we work with compact structure groups
excluding the full covariant formulation of general relativity
in four dimensions using structure group $SO(3,1)$ or $Sl(2,\C)$.
Nevertheless, we expect our result to be true in dimension $2$ 
as well. In dimension $1$, one should check it by hand -- $M$ can only be
a line or a circle.

Another issue concerns the choice of functions $d\in\dsf(S)$ to label 
the stratified sets. Constant functions mark some minimal condition.
On the other hand, our proofs in Subsection \ref{subsect:rep(algdiff)-reg}
only go through 
for constant labellings. In fact, only these guarantee that diffeomorphisms
mapping some $S$ onto itself preserve even its labelling.
The most obvious way out might be to add some stronger notion of 
regularity. In particular, we might reuse the idea of step functions
for the definition of integrals. This means, we should approximate 
an arbitrary (sufficiently ``smooth'') function by simple functions,
i.e.\ by sums of step functions,
having sufficiently nice, disjoint supports. These sums now correspond to
products of Weyl operators with constant labellings. Since 
these are represented identically, we would get the desired uniqueness
for representations that are in this sense regular and if 
each $d$ can be approximated this way. However, this approximation 
again may be in conflict with the triangulation problem above. 
Therefore, at this point, 
we state only the directly given

\blem
Besides nice enlarged structure data, assume that each
$\dsf(S)$ consists of some subset of continuous functions $d : M \nach \LG$.
Equip $\dsf(S)$ with the supremum norm on $S$ induced by some
fixed norm on $\LG$. Assume there is some sequence
$(d_i)_{i\in\N}$ with $d_i \gegen d$ in $\dsf(S)$,
such that for all $i$ there are finitely many $S_{i,k_i}$ 
forming a decomposition of $S$ and each having
a finite wide triangulation, whereas $d_i$ is constant on each $S_{i,k_i}$.

Then, given the assumptions of Theorem \ref{thm:main},
$\pi'$ is equivalent to $\pi_0$, provided $\pi'$ is 
$\mpsg^{0,S,\sigma_S}$-regular for all $S\in\qsf$ and $\sigma_S \in \isf(S)$.
\elem
Recall that $\pi_0$ itself is always
$\mpsg^{0,S,\sigma_S}$-regular, i.e., if $d_i$ converges
pointwise on $S$ to $d$, then the corresponding Weyl operators 
converge weakly.
\bpf
Let $d$, $S$ and $\sigma_S$ be fixed.
The Weyl operators corresponding to $S_{i,k_i}$ and $d_i \einschr{S_{i,k_i}}$
are even $\pi_{\nu_0}$-units according to the proof of Theorem \ref{thm:main},
hence each $\weyl^{S,\sigma_S}_{d_i}$ as well, 
by Lemma \ref{lem:disjqusurf->commut}. 
Proposition \ref{prop:dense_lambda-reg} and 
the $\mpsg^{0,S,\sigma_S}$-regularity imply that $\weyl^{S,\sigma_S}_d$
is a $\pi_{\nu_0}$-unit as well. 
Corollary \ref{corr:unit-nu->cyclic} gives the proof.
\qed
\epf

\subsubsection{Further Assumptions}
Let us now say a few words about the other assumptions of 
Theorem \ref{thm:main}. That we restrict ourselves to cyclic
representations, is no restriction at all, since any 
(non-degenerate) representation
can be decomposed into cyclic ones. Rather, the assumption that 
there is a cyclic vector being at the same time diffeomorphism invariant,
is a restriction. This means that we only consider theories having a
diffeomorphism invariant ``vacuum''. Well, this may be justified by
the corresponding invariance of general relativity leading to
some special kind of quantum geometry. Next, we assumed at least 
the ``standard'' regularity mapping weakly continuous
one-parameter subgroups into weakly continuous ones. It may be desirable
to drop this assumption; however, even in the classical theory of
quantum mechanics, the Stone-von Neumann theorem relies on the
regularity assumption. Indeed, it is very difficult to prove 
results without referring to it. However, in our case, there may be some hope,
since the diffeomorphism group is that large and may thence identify so
many objects in order to, possibly, 
replace some or all of the regularity assumptions.
The naturality of the action of diffeomorphisms is discussed below.

\subsubsection{Improvements}
Finally, 
we would like to emphasize that we were able to drop a crucial
assumption and to weaken another 
made in the paper \cite{d60} by Sahlmann and Thiemann:

First of all, we did not need any assumptions about the domains
of the operators. This was possible, since we are working 
with the exponentiated Weyl operators from the very beginning. The
only point, where we went down to the non-exponentiated
regime, was in Subsection \ref{subsect:split-prop} 
(and Appendix \ref{app:estimate}). 
But even there, we did not do this
for generators of the {\em represented}\/ Weyl operators.
In fact, we did only use results for the convergence of the
genuine Weyl operators w.r.t.\ the supremum norm. This way,
we get some {\em ``analytic''}\/ convergence at the exponentiated level 
that, afterwards, leads to the emergence of the Ashtekar-Lewandowski
measure by splitting and regularity.

Secondly, we significantly weakened the assumptions 
on the representation of the diffeomorphisms.
Although we re-used the name ``natural'' representation, 
our definition imposes much less restrictions than that in \cite{d60}.
There, the action of diffeomorphisms is said to be natural if
it is the pull-back representation of $\Diffeo$ on each 
$L_2(\Ab,\mu_\nu)$. This, however, is well-defined only
if the pull-back action of $\Diffeo$ is well-defined on 
$L_2(\Ab,\mu_\nu)$.
In fact, in general, it is not. To see this, we use again the 
general notation of Section \ref{sect:gen_set}.
Namely, let $\mu_\nu$ be the Dirac measure at some $x \in X$
being not invariant w.r.t.\ $\Weyl'$.
Then $\hilb_\nu = L_2(X,\mu_\nu)$ is isomorphic to $\C$ by
$\psi \auf \psi(x)$ for any measurable function $\psi$ on $X$.
Take some $\weyl'\in\Weyl'$ with $\xi_{\weyl'}(x) \neq x$, and
let the even continuous function $\psi$ be one at $\xi_{\weyl'}(x)$, 
but zero at $x$.
Therefore, $\psi = 0$ in $\hilb_\nu$,
but, at the same time,
$(\weyl'(\psi))(x) = \psi(\xi_{\weyl'}(x)) = 1$,
hence $\weyl'(\psi) = 1$ in $\hilb_\nu$. In other words,
the extension of the pull-back mapping from $C(X)$ to $L_2(X,\mu_\nu)$
is ill defined.
Additionally, one sees immediately, that, even if the pull-back representation
is well defined, it is unitary only if 
$\mu_\nu$ is $\Weyl'$-invariant.
This, of course, restricts the possible measures
drastically. We, instead, defined naturality (see Definition \ref{def:natural})
much less restrictive. Firstly, we do not refer
to the pull-back representation at all. Secondly, before we impose conditions
on the projection of $\pi'$ to certain $\hilb_\nu$, we check whether 
$\hilb_\nu$ is invariant w.r.t.\ $\Weyl'$. Only then and only 
if $\mu_{\nu_1}$ and $\mu_{\nu_2}$ coincide, we, thirdly,
require that the respective projections of $\pi'$ coincide.
This way, the problems indicated above are circumvented.

Nevertheless, one should think why one required naturality
at all in our case $X = \Ab$ and $\Weyl' \iso \Diffeo$.
Recall that there are three different objects to be considered:
the continuous functions on $\Ab$, the Weyl operators $\weyl\in\Weyl$
and the diffeomorphisms $\dwirk\diffeo\in\Weyl'$. The first two of them
are dynamical, the last one is just a constraint. Therefore, it is reasonable
to distinguish between them. For instance, it is not required that 
$\pi'$ is regular. In fact, the diffeomorphisms act arbitrarily
non-continuously on $\hilb$ already in the fundamental representation: 
Given some $\psi\in L_2(\Ab,\mu_0)$, say a spin network state on some graph
$\gc$, then $\dwirk\diffeo(\psi)$ is orthogonal to $\psi$ provided
$\gc$ is not preserved by $\diffeo$ (actually being a negligible restriction).
Moreover, since $C(\Ab)$ is continuous, in any case, we may 
decompose the restriction of any representation to $C(\Ab)$
into canonical representations on $\Ab$ w.r.t.\ certain measures.
It now may be conceivable that, if the continuous functions
on $\Ab$ cannot distinguish between two of these addends, then the
purely kinematical, constraining part cannot either. In other
words, if two measures in the first-step decomposition coincide,
then the induced representations should be identified.
There is no obvious reason why diffeomorphisms should not
keep the addends of the first-step decomposition invariant -- but, there
is also no reason why they should. Therefore, although it might be reasonable 
to restrict oneself to natural representations of diffeomorphisms,
this assumption does not seem to be absolutely desirable.
If we do this, however,
observe that the arguments above should not be applied to the Weyl operators. 
Indeed, first of all, these are dynamical objects, and, secondly,
they act on a higher level, namely, on $\Ab$ affected by the dynamics 
and not on the paths being the ultimate constituents of the theory
and being the domain for the homomorphisms in $\Ab$.

\subsubsection{Main Open Issues}

Of course, the remarks above are not at all final answers why to
consider just these assumptions. At least from the mathematical
point of view, it would be highly desirable to have more general results
available. We have given some hints here for direct extensions, however,
the field is still open, in particular:

\bquest
Is naturality w.r.t.\ diffeomorphisms necessary?
\equest

\bquest
Is regularity w.r.t.\ Weyl operators necessary?
\equest

\neueseite
%------------------------------------------------------------------------%
%            Abschnitt: Introduction                                     %
%------------------------------------------------------------------------%
\section{Acknowledgements}
The author thanks Abhay Ashtekar, Jerzy Lewandowski, 
Stefan M\"uller, Andrzej Oko\l\'ow, Hans-Bert Rademacher,
Hanno Sahlmann, Konrad Schm\"udgen,
Matthias Schwarz, Thomas Thiemann, Rainer Verch and Elmar Wagner 
for fruitful discussions. 
Moreover, the author is very grateful to Garth Warner for his remarks and,
in particular, for pointing out a mistake in
an earlier version of this article. Fortunately, all the results
kept valid.
Additionally, the author thanks the three anonymous referees for their
very valuable comments and suggestions helping him to improve the article.
The author has been supported in part by the Emmy-Noether-Programm 
(grant FL~622/1-1) of the Deutsche Forschungsgemeinschaft.
%------------------------------------------------------------------------%
%            Abschnitt: Introduction                                     %
%------------------------------------------------------------------------%
\anhangengl
\section{Continuity Criterion}
\blem
\label{lem:strongcont_crit}
Let $Y$ be some sequential topological space. 
Let $X$ be a Banach space and let $\mps : Y \nach \bound(X)$ be some 
map. Moreover, let $\norm{\mps(\cdot)} : Y \nach \R$
be locally bounded. 
Assume, finally, that there is some subset $E \teilmenge X$, such that 
$y \auf \mps(y) e$ is continuous for all $e\in E$ and
that $\spann\:E$ is dense in $X$. 

Then $y \auf \mps(y) x$ is continuous for all $x\in X$.
\elem
\bpf
Fix some $y'\in Y$ and choose a neighbourhood $U$ of $y'$,
such that $\norm{\mps(\cdot)}$ is bounded on $U$, 
say, by $c$. 
Let now $\varepsilon > 0$ and $x \in X$. Then there are $x_1, x_2\in X$
with $x_1 \in \spann\:E$ and $\norm{x_2} \leq \varepsilon$,
such that $x = x_1 + x_2$. Since $y \nach \mps(y) e$ is continuous
for $e\in E$, so it is for $e\in\spann\:E$. Hence, there is 
some neighbourhood $U' \teilmenge U$ of $y'$, such that
$\norm{\mps(y) x_1 - \mps(y') x_1} \leq \varepsilon$ for all $y \in U'$.
Consequently,
\bgl
\norm{\mps(y) x - \mps(y') x} 
 & \leq & \norm{\mps(y) x_1 - \mps(y') x_1} + \norm{\mps(y) x_2 - \mps(y') x_2} \\
 & \leq & \norm{\mps(y) x_1 - \mps(y') x_1} + (\norm{\mps(y)} + \norm{\mps(y')}) \norm{x_2} \\
 & \leq & (2c+1) \varepsilon
\egl
for all $y \in U'$. 
Hence, $y \auf \mps(y) x$ is continuous in $y'$ for all $x\in X$.
Since $y'$ was arbitrary, we get the proof.
\qed
\epf

%------------------------------------------------------------------------%
%                                                                        %
%------------------------------------------------------------------------%
\section{Two Estimates}
\label{app:estimate}
\blem
\label{lem:opprod_hilb_absch}
Let $H$ be some Hilbert space and $N \in \N$.
Moreover, let $A$, $A_i$ and $B_i$ 
be linear continuous operators on $H$,
such that $\norm{A} \leq 1$ and $\norm{B_i} \leq 1$ for all $i = 1,\ldots,N$.
Then
\bglklein 
\bignorm{\prod_{i=1}^N A_i B_i - \prod_{i=1}^N A B_i} 
  & \leq & \prod_{i=1}^N \bigl(1 + \norm{A_i - A} \bigr) - 1.
\eglklein
\elem
\bpf
We have 
\bglklein
          \bignorm{\prod_{i=1}^N A_i B_i - \prod_{i=1}^N A B_i} 
 &  =   & \bignorm{\prod_{i=1}^N (A + [A_i - A]) B_i - \prod_{i=1}^N A B_i} \\[1ex]
 & \leq & \prod_{i=1}^N \bigl(\norm{A B_i} + \norm{(A_i - A) B_i} \bigr) -
          \prod_{i=1}^N \norm{A B_i} \\[1ex]
 & \leq & \prod_{i=1}^N \bigl(\norm{A} + \norm{A_i - A} \bigr) -
          \prod_{i=1}^N \norm{A} \\[1ex]
 & \leq & \prod_{i=1}^N \bigl(1 + \norm{A_i - A} \bigr) - 1.
\eglklein
\qed
\epf

\neueseite

\blem
\label{lem:tensor_casimir_absch}
Let $\LG$ be a connected, compact (hence linear) Lie group and 
let $\darst$ be an irreducible representation of $\LG$ on $V_\darst$.
Moreover, 
let $\{X_i\}_{i=1}^n$ be a basis of the Lie algebra $\Lieg$ of $\LG$, such that 
$-\inv n\sum_i \darst(X_i) \darst(X_i)$ 
is (up to the prefactor) the (quadratic) Casimir operator $C_\darst$ for $\darst$.
Set 
\zgl{
\casisum\darst\Lieg(t) \breitrel{:=} 
    \inv{2n}\sum_{i=1}^{n} \bigl(\darst(\e^{t X_i}) + \darst(\e^{-t X_i})\bigr).
}

Then, 
for all $t_0 > 0$, there is some $\eta(t_0) > 0$ with
\bglklein
\bignorm[\infty]{ 
 \darst \tensor \bigtensor_{j=1}^J \bigl(
      \casisum\darst\Lieg(t) \cdot \darst\bigr) 
\: - \: \e^{-\einhalb\lambda_\darst J t^2} \: \bigtensor_{j=0}^J \darst} 
& \leq &  \e^{\eta(t_0) J t^4} - 1
\eglklein

for all $\betrag t < t_0$ and all positive integers $J$.
Here, $\lambda_\darst$ is the Casimir eigenvalue w.r.t.\ $\darst$,
and $\supnorm\cdot$ denotes the supremum norm in $\LG^{J+1}$
induced by the standard operator norm $\norm\cdot$ on $V_\darst$.
\elem
\bpf
Let 
\bgl
f_1(t) := \casisum\darst\Lieg(t) 
\breitrel{\breitrel{\text{ and }}} 
f_2(t) := \e^{-\einhalb\lambda_\darst t^2} \darst(\EINS). 
\egl
We have $f_1(0) = \darst(\EINS) = f_2(0)$.
Next, 
$f''_1(0) = \inv n \sum_{i=1}^n \darst(X_i) \darst(X_i) = - C_\darst$. Since
$C_\darst = \lambda_\darst \darst(\EINS)$, we have 
$f''_1(0) = - \lambda_\darst \darst(\EINS) = f''_2(0)$.
Since, moreover, the derivatives of odd degree vanish for both $f_1$ and $f_2$,
the derivatives of $f_1$ and $f_2$ coincide up to degree $3$.
Hence,
for every $t_0 > 0$ there is some $\eta(t_0)>0$, such that
$\norm{f_1(t) - f_2(t)} < \eta(t_0) t^4$ for all $\betrag{t} < t_0$.
Here, we used the analyticity of $f_1$ and $f_2$ on full $\R$.

Using Lemma \ref{lem:opprod_hilb_absch} and $\norm{\darst(g)} = 1$ (by unitarity of $\LG$),
we get
\bgl
      \erstezeile4 \Bignorm{
         \darst(g_0) \:
         \prod_{j=1}^{J} \bigl(\casisum\darst\Lieg(t) \: \darst(g_j)\bigr)
         \: - \: \e^{-\einhalb\lambda_\darst J t^2} \prod_{j=0}^{J} \darst(g_j)}  \\
 &\leq& \prod_{j=1}^{J} \Bigl(1 + \bignorm{
            \casisum\darst\Lieg(t) 
         \: - \: \e^{-\einhalb\lambda_\darst t^2} \darst(\EINS)}\Bigr) - 1 \\
 &\leq& \prod_{j=1}^{J} \Bigl(1 + \eta(t_0) t^4 \Bigr) - 1 \\
 &\leq& \e^{\eta(t_0) J t^4} - 1
\egl
for all $g_0,\ldots,g_J\in\LG$ and all $\betrag t < t_0$.
\qed
\epf

%------------------------------------------------------------------------%
%                                                                        %
%------------------------------------------------------------------------%
\section{``Bumpy'' Stratified Isomorphisms}
\blem
\label{lem:winddiffeo-bumps}
Let $\tau_1$ and $\tau_2$ be real numbers with $\tau_1 < \tau_2$.
Moreover, let $0 < \varepsilon < \einhalb(\tau_2 - \tau_1)$ and $a > 0$.
Finally, let $n\geq 2$ be an integer and define
\zgl{
C := [\tau_1 - \varepsilon,\tau_2 + \varepsilon] 
 \kreuz [-2\varepsilon,2a+2\varepsilon] \kreuz B_{2\varepsilon}^{n-2}
 \teilmenge \R^n,}
where $B_r^m$ is the ball around the origin in $m$ dimensions with radius
$r$.

Then there is a stratified analytic isomorphism $\diffeo$ of $\R^n$
with the following properties (see also Figure \ref{fig:wind_diffeo}
on page \pageref{fig:wind_diffeo}):
\bunum
\item
$\diffeo$ is the identity outside $C$;
\item
$\diffeo$ changes the second (i.e.\ $y$-)coordinate only;
\item
$\diffeo$ maps the first (i.e.\ $x$-)coordinate axis
(restricted to $[\tau_1 - \varepsilon,\tau_2 + \varepsilon]$)
to the union of straight lines connecting the
points 
\zgl{\text{
$(\tau_1 - \varepsilon, 0, \vec 0)$, $(\tau_1 + \varepsilon, 2a, \vec 0)$,
$(\tau_2 - \varepsilon, 2a, \vec 0)$, and $(\tau_2 + \varepsilon, 0, \vec 0)$.}}
\eunum
\elem

\neueseite
\bpf
W.l.o.g.\ we may assume that $\tau := \tau_2 = - \tau_1$.
Decompose $C$ into the 18 subsets%
\footnote{Of course, if $n=3$, there are 27 connected components, and
for $n=2$ there are only 9. We drop the corresponding cases here.}
\zgl{G_{ij0} := G_{ij} \kreuz B_\varepsilon^{n-2}}
and 
\zgl{G_{ij+} := G_{ij} \kreuz 
   \bigl(B_{2\varepsilon}^{n-2} \setminus \inter B_{\varepsilon}^{n-2}\bigr)}
having overlapping boundaries 
(for the definition of $G_{ij}$ see Figure \ref{fig:wind_diffeo}).
We are going to explicitly construct a diffeomorphism $\diffeo$ 
mapping $G_{ij\ast}$ onto some $H_{ij\ast}$. 
Before stating the explicit formulae,
we explain them verbally for $n=2$. $G_{11}$ is mapped to $H_{11}$, such that 
lines parallel to the $x$-axis are mapped to lines through $(x_0,y_0)$,
whereas the line $x = -\tau - \varepsilon$ is preserved pointwise.
The mapping between $G_{12}$ and $H_{12}$ simply makes (mutually
parallel) sloped lines out of lines parallel to the $x$-axis.
$G_{13}$ is mapped to $H_{13}$ similarly as $G_{11}$ to $H_{11}$.
The maps $G_{2i} \nach H_{2i}$ map a line parallel to the $x$-axis again
to such a line. The shift is completely determined by the shift on the
left boundaries of the $G_{2i}$. These, of course, are already 
given by maps of the right boundaries of $G_{1i}$. The maps for $G_{3i}$
will not be given explicitly. They just follow by the reflection symmetry
w.r.t.\ $x = 0$. The ideas above widely fix $\diffeo$. We only have to
take care of the matching conditions in the $\vec z$-directions.
Here, we introduce a ``fall-off'' when $\norm{\vec z}$ is in
$[\varepsilon,2\varepsilon]$. For this, we define 
$g(\vec z) := \einhalb\bigl(1-\cos(\frac{\pi}{\varepsilon}\norm{\vec z} )\bigr)$.

In the following, we will use that for 
any analytic function $h:\R \kreuz \R^{n-2}$ and any $y_0 \in \R$,
\fktdefabgesetzt{\diffeo_\aux}{\R^n}{\R^n}{(x,y,\vec z)}%
   {\bigl(x,y + h(x,\vec z)(y_0 - y),\vec z\bigr)}
is invertible analytically on 
\zgl{U_h := \{(x,y,\vec z) \mid h(x,\vec z) \neq 1\}} by 
\fktdefabgesetzt{\diffeo_\aux^{-1}}{U_h}{\R^n.}{(x,y,\vec z)}%
   {\bigl(x,\frac{y-h(x,\vec z) y_0}{1-h(x,\vec z)},\vec z\bigr)}
   
Let us now state the diffeomorphism setting $y_0 := a + 2 \varepsilon$:
\bunum
\item
Define
\fktdefabgesetzt{\diffeo_{110}}{\R^n}{\R^n}{(x,y,\vec z)}%
   {\bigl(x,y+ \frac{a}{\varepsilon}\frac{x + \tau + \varepsilon}{2a+\varepsilon}(y_0 - y),\vec z\bigr)}
and 
\fktdefabgesetzt{\diffeo_{11+}}{\R^n}{\R^n}{(x,y,\vec z)}%
   {\bigl(x,y+ g(\vec z)\:\frac{a}{\varepsilon}\frac{x + \tau + \varepsilon}{2a+\varepsilon}(y_0 - y),\vec z\bigr)}
Both $\diffeo_{110}$ and $\diffeo_{11+}$ are analytically invertible on 
$x < -\tau+\varepsilon+\frac{\varepsilon^2}a$
and are the identity on $x = -\tau-\varepsilon$.
Moreover, they 
coincide on $G_{110} \cap G_{11+}$. Finally, 
$\diffeo_{11+}$ is the identity on $\norm{\vec z} = 2\varepsilon$.
\item
Define
\fktdefabgesetzt{\diffeo_{120}}{\R^n}{\R^n}{(x,y,\vec z)}%
   {\bigl(x,y+ \frac{a}{\varepsilon}(x + \tau + \varepsilon),\vec z\bigr)}
and 
\fktdefabgesetzt{\diffeo_{12+}}{\R^n}{\R^n.}{(x,y,\vec z)}%
   {\bigl(x,y+ g(\vec z)\:\frac{a}{\varepsilon}(x + \tau + \varepsilon),\vec z\bigr)}
Both $\diffeo_{120}$ and $\diffeo_{12+}$ are analytically invertible on 
full $\R^n$, coincide on $G_{120} \cap G_{12+}$
and are the identity on $x = -\tau-\varepsilon$.
In particular, observe that 
\zgl{\diffeo_{120}(-\tau-\varepsilon,y,\vec 0) = (-\tau-\varepsilon,y,\vec 0)}
and
\zgl{\diffeo_{120}(-\tau+\varepsilon,y,\vec 0) = (-\tau+\varepsilon,y+2a,\vec 0),}
i.e.\ 
$\diffeo_{120}(-\tau,0,\vec 0) = (-\tau,a,\vec 0)$,
\item
The maps $\diffeo_{13\ast} : \R^n \nach \R^n$ are defined 
analogously to the case of $\diffeo_{11\ast}$.
\item
The maps $\diffeo_{2i\ast}$ are given by 
\fktdefabgesetzt{\diffeo_{2i\ast}}{\R^n}{\R^n,}{(x,y,\vec z)}%
   {\bigl(x,\pr_y \diffeo_{1i\ast}(-\tau+\varepsilon,y,\vec z), \vec z\bigr)}
where $\pr_y$ is the projection to the $y$-component.
\item
The remaining maps $\diffeo_{3i\ast}$ are defined using the
reflection symmetry w.r.t.\ $x=0$.
\eunum
One immediately checks that $\diffeo:\R^n \nach \R^n$ defined by 
$\diffeo\einschr{G_{ij\ast}} := \diffeo_{ij\ast}$ and 
$\diffeo\einschr{\R^n\setminus C} := \ido$ is a well-defined stratified
analytic isomorphism with the desired properties.
\qed
\epf

%------------------------------------------------------------------------%
%            Literaturverzeichnis                                        %
%------------------------------------------------------------------------%

\begin{thebibliography}{10}

\bibitem{a117}
{Abhay Ashtekar: New Hamiltonian formulation of general relativity. {\it Phys.
  Rev.} {\bf D36} (1987) {1587--1602}.}

\bibitem{a72}
{Abhay Ashtekar and Chris J. Isham: Representations of the holonomy algebras of
  gravity and nonabelian gauge theories. {\it Class. Quant. Grav.} {\bf 9}
  (1992) {1433--1468}. \\ {\sf e-print:\ hep-th/9202053}.}

\bibitem{a28}
{Abhay Ashtekar and Jerzy Lewandowski: Differential geometry on the space of
  connections via graphs and projective limits. {\it J. Geom. Phys.} {\bf 17}
  (1995) {191--230}. \\ {\sf e-print:\ hep-th/9412073}.}

\bibitem{a30}
{Abhay Ashtekar and Jerzy Lewandowski: Projective techniques and functional
  integration for gauge theories. {\it J. Math. Phys.} {\bf 36} (1995)
  {2170--2191}. {\sf e-print:\ gr-qc/9411046}.}

\bibitem{a13}
{Abhay Ashtekar and Jerzy Lewandowski: Quantum theory of geometry {I}: Area
  operators. {\it Class. Quant. Grav.} {\bf 14} (1997) {A55--A82}. {\sf
  e-print:\ gr-qc/9602046}.}

\bibitem{a48}
{Abhay Ashtekar and Jerzy Lewandowski: Representation theory of analytic
  holonomy {$C^*$} algebras. In: {\it Knots and Quantum Gravity} (Riverside,
  CA, 1993), edited by John C. Baez, pp. 21--61, Oxford Lecture Series in
  Mathematics and its Applications~1 (Oxford University Press, Oxford, 1994).
  {\sf e-print:\ gr-qc/9311010}.}

\bibitem{d17}
{John C. Baez and Stephen Sawin: Diffeomorphism-invariant spin network states.
  {\it J. Funct. Anal.} {\bf 158} (1998) {253--266}. {\sf e-print:\
  q-alg/9708005}.}

\bibitem{d3}
{John C. Baez and Stephen Sawin: Functional integration on spaces of
  connections. {\it J. Funct. Anal.} {\bf 150} (1997) {1--26}. {\sf e-print:\
  q-alg/9507023}.}

\bibitem{d70}
{Benjamin Bahr and Thomas Thiemann: Automorphisms in Loop Quantum Gravity. \\ {\sf
  e-print:\ 0711.0373 [gr-qc]}.}

\bibitem{m6}
{Edward Bierstone and Pierre D. Milman: Semianalytic and subanalytic sets. {\it
  Publ. Math. IHES} {\bf 67} (1988) {5--42}.}

\bibitem{BratRob2}
{Ola Bratteli and Derek W. Robinson: {\it Operator Algebras and Quantum
  Statistical Mechanics, vol. 2 (Equilibrium States, Models in Quantum
  Statistical Mechanics)}. Springer-Verlag, New York, 1996.}

\bibitem{BratRob1}
{Ola Bratteli and Derek W. Robinson: {\it Operator Algebras and Quantum
  Statistical Mechanics, vol. 1 ($C^\ast$- and $W^\ast$-Algebras, Symmetry
  Groups, Decomposition of States)}. Springer-Verlag, New York, 1987.}

\bibitem{paper20}
{Christian Fleischhack: Construction of Generalized Connections. \\ {\sf e-print:\ math-ph/0601005}.}

\bibitem{paper3}
{Christian Fleischhack: Hyphs and the Ashtekar-Lewandowski Measure. {\it J.
  Geom. Phys.} {\bf 45} (2003) {231--251}. {\sf e-print:\ math-ph/0001007}.}

\bibitem{paper21}
{Christian Fleischhack: Irreducibility of the Weyl Algebra in Loop Quantum
  Gravity. {\it Phys. Rev. Lett.} {\bf 97} (2006) {061302}.}

\bibitem{diss}
{Christian Fleischhack: Mathematische und physikalische Aspekte
  verallgemeinerter Eichfeldtheorien im Ashtekarprogramm (Dissertation).
  Universit{\"a}t Leipzig, 2001.}

\bibitem{paper15}
{Christian Fleischhack: Proof of a Conjecture by Lewandowski and Thiemann. {\it
  Commun. Math. Phys.} {\bf 249} (2004) {331--352}. {\sf e-print:\
  math-ph/0304002}.}

\bibitem{paper10}
{Christian Fleischhack: Regular Connections among Generalized Connections. {\it
  J. Geom. Phys.} {\bf 47} (2003) {469--483}. {\sf e-print:\ math-ph/0211060}.}

\bibitem{paper2+4}
{Christian Fleischhack: Stratification of the Generalized Gauge Orbit Space.
  {\it Commun. Math. Phys.} {\bf 214} (2000) {607--649}. {\sf e-print:\
  math-ph/0001006, math-ph/0001008}.}

\bibitem{m3}
{Patrick X. Gallagher: Zeros of group characters. {\it Math. Zeitschr.} {\bf
  87} (1965) {363--364}.}

\bibitem{m4}
{R. M. Hardt: Stratification of Real Analytic Mappings and Images. {\it Invent.
  Math.} {\bf 28} (1975) {193--208}.}

\bibitem{m8}
{William Huebsch and Marston Morse: Diffeomorphisms of manifolds. {\it Rend.
  Circ. Mat. Palermo} {\bf 11} (1962) {291--318}.}

\bibitem{d67}
{Wojciech Kami{\'n}ski, Jerzy Lewandowski, and Marcin Bobie{\'n}ski: Background
  independent quantizations: the scalar field I. {\it Class. Quant. Grav.} {\bf
  23} (2006) {2761--2270}. \\ {\sf e-print:\ gr-qc/0508091}.}

\bibitem{d68}
{Wojciech Kami{\'n}ski, Jerzy Lewandowski, and Andrzej Oko\l\'ow: Background
  independent quantizations: the scalar field II. {\it Class. Quant. Grav.}
  {\bf 23} (2006) {5547--5586}. \\ {\sf e-print:\ gr-qc/0604112}.}

\bibitem{d66}
{Jerzy Lewandowski and Andrzej Oko\l\'ow: Automorphism covariant
  representations of the holonomy-flux $*$-algebra. {\it Class. Quant. Grav.}
  {\bf 22} (2005) {657--680}. \\ {\sf e-print:\ gr-qc/0405119}.}

\bibitem{lost}
{Jerzy Lewandowski, Andrzej Oko\l\'ow, Hanno Sahlmann, and Thomas Thiemann:
  Uniqueness of diffeomorphism invariant states on holonomy-flux algebras. {\it
  Commun. Math. Phys.} {\bf 267} (2006) {703--733}. {\sf e-print:\
  gr-qc/0504147}.}

\bibitem{e46}
{Jerzy Lewandowski and Thomas Thiemann: Diffeomorphism invariant quantum field
  theories of connections in terms of webs. {\it Class. Quant. Grav.} {\bf 16}
  (1999) {2299--2322}. {\sf e-print:\ gr-qc/9901015}.}

\bibitem{m9}
{Stanis\l aw \L ojasiewicz: On semi-analytic and subanalytic geometry. In: {\it
  Panoramas of mathematics} (Warszawa, 1992/1994), pp. 89--104, Banach Center
  Publications 34 (Polish Academy of Sciences, Warszawa, 1995).}

\bibitem{m5}
{Stanis\l aw \L ojasiewicz: Triangulation of semi-analytic sets. {\it Ann.
  Scuola Norm. Sup. Pisa} {\bf 18} (1964) {449--474}.}

\bibitem{d61}
{Andrzej Oko\l\'ow and Jerzy Lewandowski: Diffeomorphism covariant
  representations of the holonomy-flux $\ast$-algebra. {\it Class. Quant.
  Grav.} {\bf 20} (2003) {3543--3568}. \\ {\sf e-print:\ gr-qc/0302059}.}

\bibitem{m7a}
{Mario Pezzana: Strong analytic triangulation for manifolds (Review of an
  article by Massimo Ferrarotti ({\em Boll. Un. Mat. Ital. A} {\bf 17} (1980)
  79--84)). {\it Math. Rev.} {\bf 81e} (1981) {32012}.}

\bibitem{d64}
{Hanno Sahlmann: Some Results Concering the Representation Theory of the
  Algebra Underlying Loop Quantum Gravity. {\sf e-print:\ gr-qc/0207111}.}

\bibitem{d65}
{Hanno Sahlmann: When Do Measures on the Space of Connections Support the Triad
  Operators of Loop Quantum Gravity? {\sf e-print:\ gr-qc/0207112}.}

\bibitem{d63}
{Hanno Sahlmann and Thomas Thiemann: Irreducibility of the
  Ashtekar-Isham-Lewan\-dow\-ski Representation. {\it Class. Quant. Grav.} {\bf
  23} (2006) {4453--4472}. \\ {\sf e-print:\ gr-qc/0303074}.}

\bibitem{d60}
{Hanno Sahlmann and Thomas Thiemann: On the Superselection Theory of the Weyl
  Algebra for Diffeomorphism Invariant Quantum Gauge Theories. {\sf e-print:\
  gr-qc/0302090}.}

\bibitem{EMS124}
{Masamichi Takesaki: {\it Theory of Operator Algebras I (Encyclopaedia of
  Mathematical Sciences 124)}. Springer-Verlag, Berlin, 2002.}

\bibitem{Whitehead}
{J. H. C. Whitehead: On $C^1$-complexes. {\it Annals of Math.} {\bf 41} (1940)
  {809--824}.}

\bibitem{Whitney}
{Hassler Whitney: {\it Geometric Integration Theory}. Princeton University
  Press, Princeton, New Jersey, 1957.}

\bibitem{Z1}
{Eberhard Zeidler \mbox{}(Hrsg.): {\it Taschenbuch der Mathematik, Bd. 1}.
  Teubner-Verlag Leipzig, Stuttgart, 1996.}

\end{thebibliography}
\end{document}